\documentclass[a4paper,11pt]{article}

\providecommand{\href}[2]{#2}\begingroup\raggedright\endgroup

\usepackage{jheppub}
\usepackage[T1]{fontenc}
\usepackage{graphicx}
\usepackage{grffile}
\usepackage{amsmath,amssymb}
\usepackage{mathrsfs}
\usepackage{bm}
\usepackage{amsthm,amsfonts,srcltx}

\theoremstyle{plain}
\newtheorem{prob}{Problem}[section]

\newtheorem{prop}[prob]{Proposition}

\title{\boldmath 
Janossy densities for chiral random matrix ensembles
and their applications to two-color QCD
}

\author[a,b]{Hiroyuki Fuji,}
\author[c,d]{Issaku Kanamori,}
\author[e]{Shinsuke M. Nishigaki}

\affiliation[a]{Faculty of Education, Kagawa University,
1-1 Saiwai-cho, Takamatsu 760-8521, Japan}
\affiliation[b]{Centre for Quantum Geometry of Moduli Spaces, 
Aarhus University, 
Ny Munkegade 118, DK-8000 Aarhus C, Denmark}
\affiliation[c]{Department of Physical Science, Hiroshima University,
Higashi-hiroshima 739-8526, Japan}
\affiliation[d]{\footnote{Present affiliation}RIKEN Center for Computational Science, Kobe 650-0047, Japan}
\affiliation[e]{Department of Physics and Materials Science,
Shimane University, Matsue 690-8504 Japan}

\emailAdd{fuji@ed.kagawa-u.ac.jp}
\emailAdd{kanamori-i@riken.jp}
\emailAdd{mochizuki@riko.shimane-u.ac.jp}

\abstract{We compute individual distributions of low-lying eigenvalues of massive chiral random matrix
ensembles by the Nystr\"om-type quadrature method
for evaluating the Fredholm determinant and Pfaffian
that represent the analytic continuation of the Janossy densities (conditional gap probabilities).
A compact formula for individual eigenvalue distributions suited for precise numerical evaluation by the Nystr\"om-type method
is obtained in an explicit form, and  the $k^{\text{\tiny th}}$ smallest eigenvalue distributions
are numerically evaluated for chiral unitary and symplectic ensembles in the microscopic limit.
As an application of our result, the low-lying Dirac spectra of the SU(2) lattice gauge theory
with $N_F=8$ staggered flavors are fitted to
the numerical prediction from the chiral symplectic ensemble,
leading to
a precise determination of
the chiral condensate
of a two-color QCD-like system in the future.
}

\preprint{%
{\flushright
HUPD-1904\\
}}
\begin{document} 
\maketitle
\flushbottom

\section{Introduction}
\label{sec:intro}

Random matrix theory (RMT) has served as fundamental tool for 
analysing quantum spectra of classically chaotic systems.
Universality of the level statistics of invariant RMTs provides a basis upon which the system-specific information,
due e.g.\ to the presence of short periodic orbits or to the weak localization effect, may be encoded \cite{Berry_Keating}.
In the application of RMT to QCD or gauge theories in general, 
the focus is on the distributions of several smallest eigenvalues of chiral RM ensembles,
as they describe the spectral statistics of gauge-covariant Dirac operators in the broken phase of chiral symmetry.
(Examples of such applications are found in \cite{Edwards:1999ra,DeGrand:2005vb,Fukaya:2007fb,Buividovich:2008ip,Lehner:2011km}.)
This relation is particularly useful with lattice simulations.
If a gauge theory is in the chirally broken phase and not in the conformal window, 
its low-energy excitations are unambiguously described by 
the chiral Lagrangian on one of the Riemannian symmetric spaces (Nambu-Goldstone manifolds) $\mathcal{M}$.
In that case, (i) the low-lying Dirac eigenvalues $0\leq \lambda_1\leq \lambda_2\leq \cdots$ 
measured on lattices of different volumes $V$
will, after
prescribed unfolding $x_k=\Sigma V\lambda_k$ and scaling of quark masses
$\mu_f=\Sigma V m_f$,
with a constant $\Sigma$ independent of the volumes,
obey a single statistical distribution 
$p_k(x; \{\mu_f\})=\langle \delta(x-x_k) \rangle$, and (ii) this distribution will be identical to the one
from the RMT that is equivalent to the zero-momentum part of the chiral Lagrangian on $\mathcal{M}$ \cite{Shuryak:1992pi}.
If the theory is in the symmetric phase of the chiral symmetry,
no such scaling with the volume, which collapses the distributions of
$\lambda_k$'s 
from different volumes onto a single function, would appear.
Previously this criterion was applied to QCD around the critical temperature,
and the inconsistency with RMT (including non-scaling of unfolded Dirac eigenvalues 
with volumes) was considered as a sign of chiral symmetry
restoration \cite{Damgaard:2000}.
In addition, if the theory is conformal, no scale should appear so that
the chiral condensate 
$\Sigma$
should disappear in the chiral limit and description with RMT is not applicable.

In the proposal of the walking technicolor model \cite{Appelquist:1986}, 
the choice of the gauge group of techni-gluons and the representation of techni-quarks
are rather open (as long as the one-loop beta function coefficient is negative and small), 
since these particles would be confined under the energy scale of several hundred TeV and would escape direct detection.
This spurred extensive numerical searches of the conformal window (where $\beta(g_*)=0$) and
the walking regime (where $\beta(g)<0$ but small) on various lattice settings with choices of colors/flavors/representations.
Summaries of 
recent activities with lattice simulations are found in
\cite{Pica:2017gcb, Svetitsky:2017xqk, Witzel:2019jbe}.
In an attempt to identify the chirally broken phase below the conformal window for the SU(3) $N_F=4$ and $8$ systems,
Fodor et al.\ \cite{Fodor:2009wk} 
fitted the Dirac spectra of these gauge theories to the analytic results
from the chiral GUE (Dyson index $\beta=2$).
Subsequently, one of the present author (I.K.) and others
tried a similar comparison of the Dirac spectrum of the SU(2) $N_F=8$ system
 (see e.g. \cite{Leino:2017lpc, Leino:2018qvq, Leino:2018yfd}
 for the current situation
of this system) to the chiral GSE ($\beta=4$) \cite{Huang:2015vkr}.

\begin{figure}[h] 
\begin{center}
\vspace{1.5cm}
\hspace{-2cm}
 \includegraphics[bb=0 0 360 223,width=65mm]{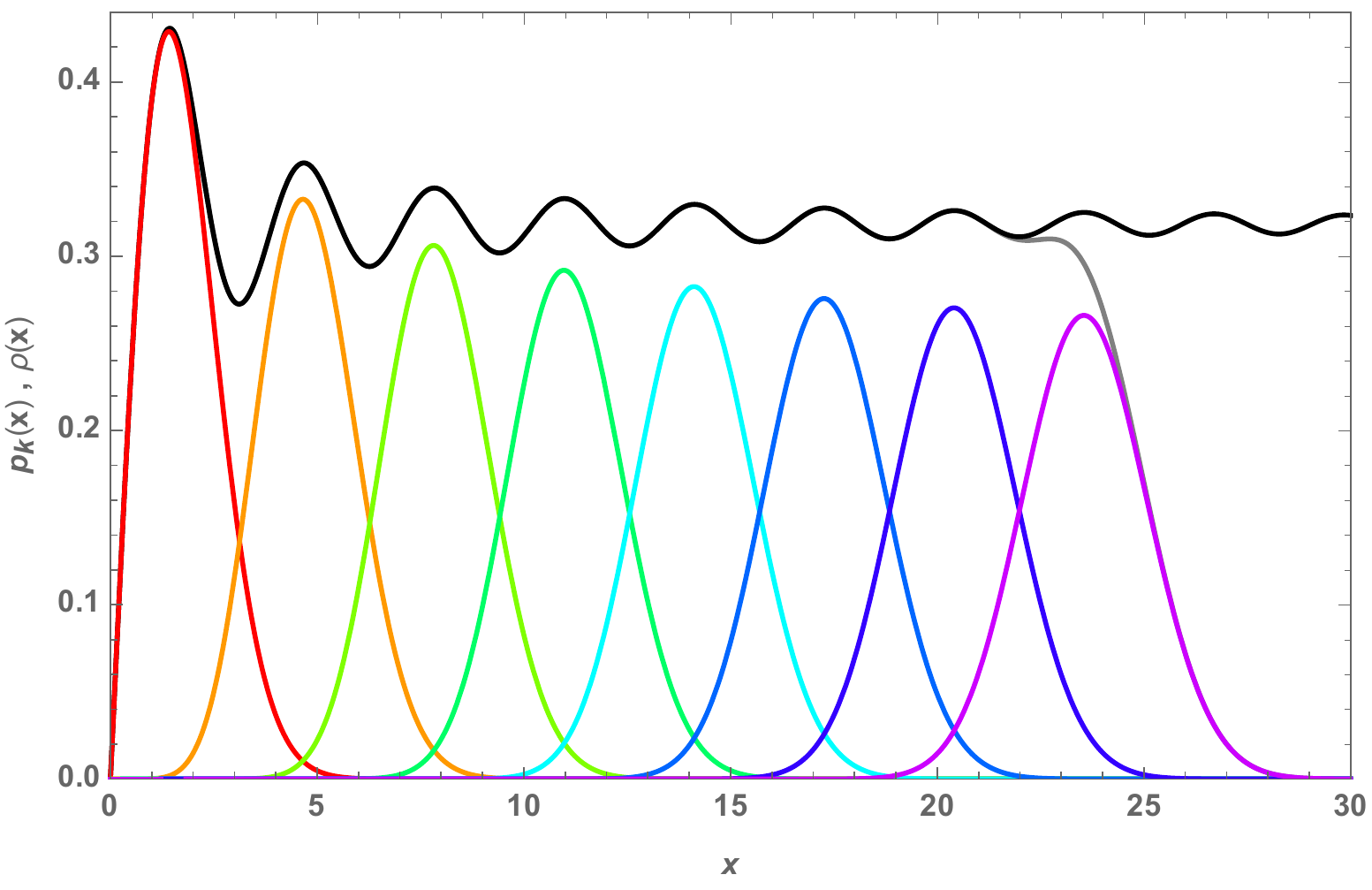}
\caption{\label{Fig1}
First eight eigenvalue distributions $p_1(x), \ldots, p_8(x)$ (red to purple),
their sum (gray),
and the microscopic spectral density $\rho(x)$ (black, normalized as $\rho(\infty)=1/\pi$)
of the quenched ($N_F=0$) chiral Gaussian unitary ensemble.
}
\end{center}
\end{figure}

For the above approach of fitting Dirac spectra to the corresponding RMT predictions
to be practically useful,
it is highly desirable to single out individual distributions of
each of the ordered RM eigenvalues $p_k(x)$ from the spectral density 
$\rho(x; \{\mu\})=\langle \sum_{k} \delta(x-x_k) \rangle =\sum_{k\geq 1}p_k(x; \{\mu\})$,
as the latter becomes rather structureless after a couple of oscillations (Fig.~\ref{Fig1}).

The standard technique to access such individual eigenvalues is to use
the spectral kernel.
Once the spectral kernel is obtained, one can give an analytic expression
of the distribution.
Moreover, by combining 
Nystr\"{o}m-type (quadrature) 
evaluation of Fredholm determinants and Pfaffians,
one can numerically evaluate the distribution of individual eigenvalues.
Damgaard and one of the authors (S.M.N.) have previously derived
analytic expressions of such individual eigenvalue distributions
for chiral RM ensembles at three Dyson indices $\beta$
and with scaled quark mass parameters $\{\mu_f\}$, initially by the shift-of-variable method 
\cite{Nishigaki:1998is,Damgaard:2000ah} and
later by the Nystr\"{o}m-type 
evaluation of Fredholm determinants and Pfaffians of the spectral kernels 
\cite{Nishigaki:2016nka}.
There, technical difficulties have prevented us from obtaining analytic formulas
for the chiral GSE ($\beta=4$) with even numbers of massless flavors and
for the chiral GOE ($\beta=1$) with even values of the topological charge.
Especially, the former restriction is frustrating, as it obstructs applications 
to the SU(2) systems with $N_F=8$ and $12$ staggered flavors
that are popular lattice settings of walking technicolor candidates.
Because of this reason, the Monte Carlo method with finite-size
matrices was used in \cite{Huang:2015vkr} to generate
the spectral distribution of the RM side in their analysis of SU(2)
$N_F=8$ system.
The purpose of this paper is to lift this restriction by providing an analytic formula for
the conditional gap probability,
a.k.a.~the Janossy density,
that interpolates the ordinary determinantal or Pfaffian formula
for the $k$-point correlation function and the Fredholm determinant/Pfaffian expression for the
gap probability.
Then our formula is numerically evaluated very efficiently by the Nystr\"{o}m-type  method.
As an application of our result, the low-lying Dirac spectra of the SU(2) lattice gauge theory
with $N_F=8$ staggered flavors are fitted to the derived RM prediction.\footnote{
We shall use the same data as \cite{Huang:2015vkr} in this paper, but
there are major differences in our analysis from \cite{Fodor:2009wk}.
Our lattice data is obtained with the unimproved staggered fermion action
and suffers from large taste breaking effects.
Consequently, we do not observe the 4-fold degeneracy characteristic of the staggered tastes,
and the lightest of these corresponds to 2 flavors.
Moreover, due to the Kramers degeneracy of the SU(2) Dirac operator,
the degeneracy of the lightest fermion modes are 4-fold, 
to which we must compare the prediction of RMT with $N_F=4$ instead of $N_F=8$.}

This paper is organized as follows.
In Sect.\ 2 we start by reviewing known formulas on the spectral-statistical distributions
of chiral RMTs and their Janossy densities.
In Sect.\ 3 we present a formula for the individual eigenvalue distributions suited for 
precise numerical evaluation by the Nystr\"{o}m-type method.
Specifically, we shall provide numerical data of $p_{1}(x), \ldots, p_{4}(x)$ 
for the chiral GSE with $N_F=4$ and $8$ degenerate massive flavors.
In Sect.\ 4 we determine the values of chiral condensate of the SU(2) system with $N_F=8$
 the first eigenvalue distribution of the corresponding chiral GSE.
Conclusions and discussions on feasible applications of our results 
are presented in Sect.\ 5.
In order to avoid plethora of formulas in RMT and of lattice details in the main text,
some of them are relocated to the Appendices.

\section{Fredholm determinants and Pfaffians for chiral Gaussian random matrix ensembles}
In this section, we will summarize some necessary ingredients about the chiral random matrix ensembles, and  
derive our main formulae for the Fredholm determinants and Pfaffians of Gaussian chiral random matrix ensembles.

\subsection{Gaussian chiral random matrix ensembles and the microscopic limit}
Consider $N\times (N+\nu)$ matrices $W$ with $W\in \mathbb{R}^{N\times (N+\nu)}$, 
$W\in \mathbb{C}^{N\times (N+\nu)}$, or $W\in \mathbb{H}^{N\times (N+\nu)}$. 
Each ensemble is labelled by the Dyson index $\beta=1,2,4$, respectively.
The non-negative integer $\nu$ denotes
the corank of the matrix
$
H=
{\small 
\left(\begin{array}{cc}
0 & W \\
W^{\dagger} & 0
\end{array}
\right)
}
$
and will correspond to the topological charge when $H$ is interpreted as modelling Dirac operator of a gauge theory
\cite{Verbaarschot:1994qf}.
Let $Z_{N,\beta,\nu}(\{m_a\})$ be the partition function for the Gaussian chiral random matrix ensembles with $\alpha$
parameters $m_a$ ($a=1,\ldots,\alpha$),
which will correspond to quark masses,
such that
\begin{align}
Z_{N,\beta,\nu}(\{m_a\})=\int dW\mathrm{e}^{-\beta\,\mathrm{tr}(W^{\dagger}W)}\prod_{a=1}^{\alpha}\mathrm{det}
\left(\begin{array}{cc}
m_a & \mathrm{i}W \\
\mathrm{i}W^{\dagger} & m_a
\end{array}
\right),
\end{align}
where det stands for the determinant for $\beta=1,2$ and the quaternionic determinant (qdet) for $\beta=4$.
In particular for $\beta=4$ it is understood that twofold degenerated eigenvalues in the determinant are only counted once.
In terms of eigenvalues $\{x_i\}$ for the Wishart matrix $W^{\dagger}W$, i.e.\  
the squares of non-zero eigenvalues $\{\pm \lambda_i\}$ of the Hermitian matrix $H$,
$Z_{N,\beta,\nu}(\{m_a\})$ 
is expressed as follows:
\begin{align}
&Z_{N,\beta,\nu}(\{m_a\})
=\left(\prod_{a=1}^{\alpha}m_a^{\nu}\right)
\int_0^{\infty}\cdots\int_0^{\infty}\prod_{i=1}^{N}\left(dx_i\,x_i^{\frac{\beta(\nu+1)}{2}-1}\mathrm{e}^{-\beta x_i}\prod_{a=1}^{\alpha}(x_i+m_a^2)\right)
\prod_{i>j}^N|x_i-x_j|^{\beta}.
\label{eq:partition_function}
\end{align}
Likewise the $p$-level correlation function $R_{N,\beta,\nu}^{(p)}(\lambda_1,\ldots,\lambda_p;\{m_a\})$ of 
the Hermitian matrix $H$
is defined by
\begin{align}
&R_{N,\beta,\nu}^{(p)}(\lambda_1,\ldots,\lambda_p;\{m_a\})=\left(2^p\prod_{j=1}^p|\lambda_j|\right)\sigma_{N,\beta,\nu}^{(p)}(\lambda_1^2,\ldots,\lambda_p^2;\{m_a\}),
\\
&\sigma_{N,\beta,\nu}^{(p)}(x_1,\ldots,x_p;\{m_a\})=\frac{\Sigma_{N,\beta,\nu}^{(p)}(x_1,\ldots,x_p;\{m_a\})}{\Sigma_{N,\beta,\nu}^{(0)}(\{m_a\})}, \\
&\Sigma_{N,\beta,\nu}^{(p)}(x_1,\ldots,x_p;\{m_a\})
\nonumber \\
&=\frac{1}{(N-p)!}\int_0^{\infty}dx_{p+1}\cdots\int_0^{\infty}dx_N\prod_{i=1}^{N}\left(x_i^{\frac{\beta(\nu+1)}{2}-1}\mathrm{e}^{-\beta x_i}\prod_{a=1}^{\alpha}(x_i+m_a^2)\right)\prod_{i>j}^N|x_i-x_j|^{\beta}.
\end{align}
Here we introduce variables $z_j$'s such that
\begin{align}
z_j=\left\{
\begin{array}{cl}
-m_{j}^2, &\quad j=1,\ldots ,\alpha  \\
x_{j-\alpha}, &\quad j=\alpha+1,\ldots \alpha+p 
\end{array}
\right..
\end{align}
The $p$-level correlation functions for $\beta=2$ are rewritten as the determinant of the spectral kernel $K(z_i,z_j)$ \cite{Mehta,Nagao,Forrester_book,Nishigaki}:
\begin{align}
\sigma_{N,\beta=2,\nu}^{(p)}(x_1,\ldots,x_{p};\{m_a\})=\frac{1}{N!\,\Sigma_{N,\beta=2,\nu}^{(0)}(\{m_a\})}\det[K(z_i,z_j)]_{i,j=1}^{p+\alpha}.
\label{eq:correlator_GUE}
\end{align}
$R^{(p)}_N(x_1,\ldots,x_{p-\alpha};\{m_a\})$ is given by the determinant of the scalar kernel \cite{Damgaard:1997ye,Wilke:1997gf}.
For $\beta=1,4$, 
the skew-orthogonal polynomial method involves
the quaternionic determinant $\mathrm{qdet}$ \cite{Dyson:1962es} of the quaternionic kernel \cite{Nagao:2000qn,Nagao:2000cb,Akemann:2000ze,Akemann:2000yq}.
In particular, $p$-level correlation functions are given by
$(p+\alpha)\times (p+\alpha)$ 
quaternionic determinants of the quaternionic kernel,
which is rewritten by
a $2(p+\alpha)\times 2(p+\alpha)$ Pfaffian of its $\mathbb{C}$-number $2\times 2$ representative
(denoted by the same $K(z_i,z_j)$ for notational simplicity),
\begin{align}
\sigma_{N,\beta=(1,4),\nu}^{(p)}(x_1,\ldots,x_{p};\{m_a\})&=\frac{1}{N!\,\Sigma_{N,\beta=1,4,\nu}^{(0)}(\{m_a\})}\mathrm{qdet}[K(z_i,z_j)]_{i,j=1}^{p+\alpha}
\nonumber \\
&
=\frac{1}{N!\,\Sigma_{N,\beta=1,4,\nu}^{(0)}(\{m_a\})}\mathrm{Pf}\left(Z[K(z_i,z_j)]_{i,j=1}^{p+\alpha}\right),
\label{eq:correlator_GSE}
\end{align}
where $Z=\mathrm{i}\sigma_2\otimes\mathbb{I}_{p+\alpha}$ stands for the skew-unit matrix $Z^2=-\mathbb{I}_{2(p+\alpha)}$.

Now we will consider the asymptotic limit:
\begin{align}
N\to\infty,\quad x_i,\;m_a\to 0,\quad \zeta_i=\sqrt{8Nx_i},\;\mathrm{and} \;\mu_a=\sqrt{8N}m_a:\mathrm{fixed}.
\label{eq:asymptotic_limit}
\end{align}
This limit corresponds
the microscopic limit of the QCD-like theory on a box of volume $V$ such that
\begin{align}
V\to\infty,\quad m_a\to 0,\quad \mu_a=\Sigma V m_a:\mathrm{fixed},
\end{align}
where $\Sigma$ stands for the chiral condensate in the chiral limit.

In this asymptotic limit (\ref{eq:asymptotic_limit}), the scaled $p$-level correlation function $R_{\beta=2,\nu}^{(p)}(\zeta_1,\ldots,\zeta_p;\{\mu_a\})$ for the chiral GUE with $2\alpha$ dynamical quarks whose masses are doubly degenerated $\mu_{a}=\mu_{a+\alpha}$ ($a=1,\ldots,\alpha$)\footnote{
In \cite{Damgaard:1997ye}, an alternative representation of the $p$-level correlation function $R^{(p)}(\zeta_1,\ldots,\zeta_p;\{\mu_a\})$
is also found for general mass parameters. (See eq.~(\ref{eq:massive_correlator}) in Appendix \ref{app:confluent_GUE}.)
} is found as follows 
\cite{Damgaard:1997ye,Wilke:1997gf,Akemann:2000ze,Akemann:2000yq,Jackson:1996jb}:
\begin{align}
&R_{\beta=2,\nu}^{(p)}(\zeta_1,\ldots,\zeta_p;\{\mu_a\})
=\frac{1}{Z_{\beta=2,\nu}(\{\mu_a\})}\det\left(
\begin{array}{cc}
\left[K_{--}(\mu_a,\mu_b)\right]_{a,b=1,\ldots,\alpha}  & \left[K_{-+}(\zeta_i,\mu_b)\right]_{\substack{i=1,\ldots,p\\ b=1,\ldots,\alpha}} 
\\
 \left[K_{+-}(\mu_a,\zeta_j)\right]_{\substack{a=1,\ldots,\alpha\\ j=1,\ldots,p}} 
& 
\left[K_{++}(\zeta_i,\zeta_j)\right]_{i,j=1,\ldots,p}
\label{eq:decompostion_kernel}
\end{array}
\right), 
\\
&K_{++}(\zeta,\zeta')=\frac{\sqrt{\zeta\zeta'}}{\zeta^{\prime\,2}-\zeta^2}[J_{\nu}(\zeta)\zeta'J_{\nu+1}(\zeta')-J_{\nu}(\zeta')\zeta J_{\nu+1}(\zeta)], 
\nonumber \\
&
K_{++}(\zeta,\zeta)=\frac{\zeta}{2}[J_{\nu}(\zeta)^2+J_{\nu+1}(\zeta)^2],  \nonumber
\\
&K_{+-}(\zeta,\mu')=\frac{-\sqrt{\zeta\mu'}}{-\mu^{\prime\,2}-\zeta^2}[J_{\nu}(\zeta)(-\mu')I_{\nu+1}(\mu')-I_{\nu}(\mu')\zeta J_{\nu+1}(\zeta)],\nonumber \\
&K_{-+}(\mu,\zeta')=\frac{-\sqrt{\mu\zeta'}}{\zeta^{\prime\,2}+\mu^2}[I_{\nu}(\mu)\zeta'J_{\nu+1}(\zeta')-J_{\nu}(\zeta')(-\mu) I_{\nu+1}(\mu)],\nonumber \\
&K_{--}(\mu,\mu')=\frac{\sqrt{\mu\mu'}}{\mu^{\prime\,2}-\mu^2}[I_{\nu}(\mu)\mu'I_{\nu+1}(\mu')-I_{\nu}(\mu')\mu I_{\nu+1}(\mu)],
\nonumber \\
&K_{--}(\mu,\mu)=\frac{\mu}{2}[I_{\nu}(\mu)^2-I_{\nu+1}(\mu)^2],\quad  Z_{\beta=2,\nu}(\mu_1,\ldots,\mu_{\alpha})=\det\left(\left[K_{--}(\mu_a,\mu_b)\right]_{a,b=1,\ldots,\alpha}\right),
\label{eq:decompostion_kernel2}
\end{align}
where 
$J_{\nu}(x)$ and $I_{\nu}(x)$ denote the Bessel and the modified Bessel functions, respectively,
\begin{align}
I_{\nu}(x)=\mathrm{i}^{-\nu}J_{\nu}(\mathrm{i}x)=\sum_{m=0}^{\infty}\frac{1}{m!\Gamma(m+\nu+1)}\left(\frac{x}{2}\right)^{2m+\nu}.
\end{align}

For $\beta=4$, the scaled $p$-level correlation function in the asymptotic limit (\ref{eq:asymptotic_limit})
is found 
for $N_F=4\alpha$ quadruply degenerated flavors $\mu_a=\mu_{a+\alpha}=\mu_{a+2\alpha}=\mu_{a+3\alpha}$ ($a=1,\ldots,\alpha$) and $N_F=2\alpha$ doubly degenerated flavors $\mu_a=\mu_{a+\alpha}$ ($a=1,\ldots,\alpha$), manifestly in \cite{Nagao:2000cb}.
\begin{align}
&R_{\beta=4,\nu}^{(p)}(\zeta_1,\ldots,\zeta_p;\{\mu_a\})=\frac{1}{Z_{\beta=4,\nu}(\{\mu_a\})}\mathrm{Pf} [ZK_{ij}].
\end{align}
Explicit expressions of matrix elements of the spectral kernels\footnote{
An explicit formula for the $p$-level correlation function is known as well for the chiral GOE ($\beta=1$) \cite{Nagao:2000cb},
but the convergence of the Nystr\"om-type discretization of the Fredholm Pfaffian
is not guaranteed due to the discontinuity of $\mathrm{sgn}(\zeta-\zeta')$ in its kernel elements.
To avoid such analytical difficulty, we will focus on the study of the Fredholm Pfaffian for the chiral GSE, 
and leave discussions of the chiral GOE for the future work.
} $ZK_{ij}$  are summarized in Appendix \ref{app:kernel}.

\subsection{Individual eigenvalue distributions}

We now focus on the individual distribution  of the $k^{\text{\tiny th}}$ smallest eigenvalue for the chiral  random matrix ensembles \cite{Forrester:1993vtx}.
There are various techniques to analyze the gap probabilities \cite{Gaudin,Mehta:1970zz}
such as linear differential equations \cite{Edelman,Edelman2} or Painlev\'e transcendental equations \'a la  Tracy-Widom \cite{TW_airy,Tracy:1993xj}.
An alternative method to find individual distribution  of the $k^{\text{\tiny th}}$ smallest eigenvalue in the asymptotic limit (\ref{eq:asymptotic_limit})
has also been developed in \cite{FH,Damgaard:2000ah}. (See also \cite{Akemann:2003tv,Akemann:2007yj,Akemann:2008va,Akemann:2009gsa,Akemann:2011up,Akemann:2012pn}.)
The procedure of this method consists of three steps \cite{Nishigaki:2016nka}:
\begin{enumerate}
\item Relate the joint distribution of the first $k$ eigenvalues to the partition function with $\beta k+\beta (\nu+1)/2-1$
additional masses and a fixed topological charge $2/\beta +1$.
\item Replace the partition function by the microscopically-scaled form \cite{Guhr:1996vx,Jackson:1996jb,Nagao:2000qn,Nagao:2000cb}
by taking the asymptotic limit (\ref{eq:asymptotic_limit}).
\item Integrate over the scaled variables $\zeta_i$ ($i=1,\ldots,k$) in a cell $0\le \zeta_1\le\cdots\le\zeta_{k-1}\le\zeta_k$.
\end{enumerate}

On actual implementation of the above method,
the numerical integration over $k$ scaled variables in the third step becomes resource-consuming.
To circumvent such technical issue,
we will consider Fredholm determinants and Pfaffians 
for the chiral random matrix ensembles with $\alpha$ mass parameters  
as the generating function of the joint distribution of the first $k$ eigenvalues, and 
utilize the quadrature method \cite{wolfram_GL} to evaluate them numerically \cite{Nishigaki:2012rn,Nishigaki:2012jw,Nishigaki:2015qfa,Yamamoto:2017isf}.
In this section, we will derive a compact formula\footnote{In \cite{FW,WBF}, 
what we call $E(k;I;{m_a})$ with $k=0$, $\alpha=1$ for the chiral GUE ($\beta=2$) has essentially been worked out.
We would like to thank P.~Forrester for kindly reminding us of their works.} 
of Fredholm determinants and Pfaffians which will be efficient for numerical computations.

Let $P_{N+\alpha,\beta,\nu}(x_1,\ldots,x_{N+\alpha})$ be the distribution of the probability for all eigenvalues of the rank $N$ matrix,
\begin{align}
P_{N+\alpha,\beta,\nu}(x_1,\ldots,x_{N+\alpha})=\frac{1}{N!\,C^{(0)}_{N,\beta,\nu}(\{m_a\})}\prod_{i=1}^{N+\alpha} x_i^{\frac{\beta(\nu+1)}{2}-1}\mathrm{e}^{-\beta x_i}\prod_{i>j}^{N+\alpha}|x_i-x_j|^{\beta}.
\end{align}
The $x_i$-independent prefactor $C^{(0)}_{N,\beta,\nu}(\{m_a\})$ is defined 
so that $P_{N+\alpha,\beta,\nu}(x_1,\ldots,x_{N+\alpha})$ obeys the normalization condition.
\begin{align}
\int_{-\infty}^{\infty}\cdots\int_{-\infty}^{\infty}dx_1\cdots dx_{N+\alpha}\,
P_{N+\alpha,\beta,\nu}(x_1,\ldots,x_{N+\alpha})\prod_{a=1}^{\alpha}\chi_{{\{-m_a^2\}}}(x_a)\prod_{i=\alpha+1}^{N+\alpha}\chi_{[0,\infty]}(x_i)=1,
\end{align}
where $\chi_I(x)$ stands for the characteristic function on $I\subset \mathbb{R}$. 
If $I$ is a line segment $[a,b]$ ($a<b$) or a semi-infinite line, the characteristic function is given by
\begin{align}
\chi_I(x)=\left\{
\begin{array}{cc}
1 & (x\in I) \\
0& (x\not\in I) 
\end{array}
\right. .
\end{align} 
If $I$ consists of one point $\{y\}$, 
\begin{align}
\chi_{\{y\}}(x)=\delta(x-y).
\end{align}

Consider the joint probability $E(k;I;\{m_a\})$ that one finds exactly
$k$ eigenvalues on an interval $I$ along the real axis and $\alpha$ eigenvalues in $\mathbb{R}_{<0}$ such that
\begin{align}
E(k;I;\{m_a\})=&
\frac{(N+\alpha)!}{k!\alpha!(N-k)!}
\int_{-\infty}^{\infty}\cdots\int_{-\infty}^{\infty}dx_1\cdots dx_{N+\alpha}\,P_{N+\alpha,\beta,\nu}(x_1,\ldots,x_{N+\alpha})
\nonumber \\
&
\times\prod_{a=1}^{\alpha}\chi_{\{-m_{a}^2\}}(x_a)\prod_{j=\alpha+1}^{\alpha+k}\chi_{I}(x_j)\prod_{l=\alpha+k+1}^{N+\alpha}(1-\chi_I(x_l)).
\label{eq:E_k}
\end{align}
Such a joint probability $E(k;I;\{m_a\})$ is known as an analytic continuation of the Janossy density \cite{Janossy1,Janossy2,Forrester_book}.
(See Appendix \ref{app:Janossy} for the definition of the Janossy density.)
The cumulative distribution $F_k(s)$ and the probability distribution $p_k(s)$
of the $k^{\text{\tiny th}}$ smallest positive eigenvalue
are expressed by
\begin{align}
F_k(s)=1-\sum_{\ell=0}^{k-1}E(\ell;[0,s];\{m_a\}),\qquad
p_k(s)=\frac{\partial}{\partial s}F_k(s).
\end{align}
In the next subsections, we shall show that the generating function $\tau(z;I;\{m_a\})$ of the probability $E(k;I;\{m_a\})$ given by
\begin{align}
\tau(z;I;\{m_a\})&= 
\sum_{k= 0}^{N}(1-z)^kE(k;I;\{m_a\})
\nonumber\\
&=
\left\langle \sum_{i=1}^{N+\alpha} \Bigl(\prod_{a=1}^{\alpha}\delta(x_i+m_a^2) \prod_{j(\neq i)}(1-z\chi_I(x_j)\,) \Bigr)\right\rangle
\nonumber \\
&=\frac{(N+\alpha)!}{\alpha!N!}
\int_{-\infty}^{\infty}\cdots\int_{-\infty}^{\infty}dx_1\cdots dx_{N+\alpha}\,P_{N+\alpha,\beta,\nu}(x_1,\ldots,x_{N+\alpha})
\nonumber \\
&\qquad\qquad\qquad\qquad\qquad\qquad
\times
\prod_{a=1}^{\alpha}\chi_{\{-m_{a}^2\}}(x_a)\prod_{j=\alpha+1}^{N+\alpha}(1-z\chi_{I}(x_j))
\label{eq:tau_E_k}
\end{align}
is rewritten as a 
block-decomposed Fredholm determinant or Pfaffian of the spectral kernels in (\ref{eq:correlator_GUE}) or (\ref{eq:correlator_GSE}).

\newpage
\subsection{Fredholm determinant for chiral Gaussian unitary ensemble}
We start by sketching the proof
for the simplest case
$\beta=2$, $\alpha=1$, $m_1^2=-y$:
\begin{align}
&\tau(z; I; \sqrt{-y})
\nonumber \\
&=(N+1)\left(\int-z\int_I dx_2\right)\cdots \left(\int-z\int_I dx_{N+1}\right) P_{N+1,\beta=2,\nu}(y, x_2,\ldots,x_{N+1})
\nonumber \\
&=(N+1)\int dx_2\cdots dx_{N+1} \, P_{N+1,\beta=2,\nu}(y,x_2,\cdots,x_{N+1})
\nonumber \\
&\quad
-(N+1)N z\int_I dx_2 \int dx_3\cdots dx_{N+1} \, P_{N+1,\beta=2,\nu}(y,x_2,x_3,\cdots,x_{N+1}) 
\nonumber \\
&\quad +(N+1)\frac{N(N-1)}{2!} z^2 \int_I dx_2 dx_3 \int dx_4\cdots dx_{N+1} P_{N+1,\beta=2,\nu}(y,x_2,x_3,x_4,\cdots,x_{N+1})
\nonumber \\
&\quad-\cdots
\nonumber \\
&=\sigma^{(0)}_{N,\beta=2,\nu}(\sqrt{-y})-z\int_I dx_2\, \sigma^{(1)}_{N,\beta=2,\nu}(x_2;\sqrt{-y})
\nonumber \\
&\quad+\frac{z^2}{2!} \int_I dx_2 dx_3 \,\sigma^{(2)}_{N,\beta=2,\nu}(x_2,x_3;\sqrt{-y})
\nonumber \\
&\quad
-\frac{z^3}{3!} \int_I dx_2 dx_3 dx_4\,\sigma^{(3)}_{N,\beta=2,\nu}(x_2,x_3,x_4;\sqrt{-y}) +\cdots.
\label{eq:rewrite_fredholm1}
\end{align}
To rewrite correlation functions $\sigma^{(k)}_{N,\beta=2,\nu}$ in terms of the spectral kernel (\ref{eq:kernel_notation}), 
we will prepare some notations such as\footnote{
It is noted that $K\circ K= K$ holds on $\mathbb{R}_+$, but $K\circ K\ne K$ on the interval $I$.
}
\begin{align}
 (f\circ g)(x,x')=\int_I dx''\, f(x,x'') g(x'',x'),\quad  \mathrm{tr}\, f=\int_I dx\,f(x,x),\quad \overbrace{K\circ K\circ \cdots \circ K}^n =K^n.
\end{align}
In addition, we assume that the quadrature discretization of the Riemann integral on $I$ to be well-defined in the continuum limit $M\to\infty$ 
(which is always implicit below),
\begin{align}
&\{x_1,\ldots,x_M\}\in I, \ \ 
dx_1,\ldots,dx_M> 0,\qquad
\sum_{i=1}^M f(x_i) dx_i 
\stackrel{M\to\infty}{\longrightarrow} \int_I f(x) dx .
\end{align}
We further introduce following notations for the block decomposition of the spectral kernel 
integrated over $I$.
\begin{align}
&\kappa=K(-y,-y),\quad \bm{k}=\left[\sqrt{d x_i}\,K(x_i,-y)\right]_{i=1,\ldots,M},\nonumber \\
&\bm{k}^{\mathrm{T}}=\left[\sqrt{d x_i}\,K(-y,x_j)\right]_{j=1,\ldots,M},\quad
\bm{K}=\left[\sqrt{d x_i}\, K(x_i,x_j)\sqrt{d x_j}\right]_{i,j=1,\ldots,M}.
\label{eq:kernel_notation}
\end{align}
Adopting eq.~(\ref{eq:correlator_GUE}) and these notations, one can rewrite the Fredholm determinant $\tau(z; I; \sqrt{-y})$
in terms of the block-decomposed scalar kernel as follows:
\begin{align}
&\tau(z; I; \sqrt{-y})\cdot Z_{N,\beta=2,\nu}(\sqrt{-y})
\nonumber \\
&=K(-y,-y)-z\int_I dx_2
\det\left|
\begin{array}{cc}
K(-y,-y) & K(-y,x_2)\\
K(x_2,-y) & K(x_2,x_2)
\end{array}
\right|
\nonumber \\
& \quad+\frac{z^2}{2!} \int_I dx_2 dx_3 
\det\left|
\begin{array}{ccc}
K(-y,-y) & K(-y,x_2) & K(-y,x_3)\\
K(x_2,-y) & K(x_2,x_2) & K(x_2,x_3)\\
K(x_3,-y) & K(x_3,x_2) & K(x_3,x_3)
\end{array}
\right|
\nonumber \\ 
&\quad -\frac{z^3}{3!} \int_I dx_2 dx_3 dx_4
\det\left|
\begin{array}{cccc}
K(-y,-y) & K(-y,x_2) & K(y,x_3) & K(-y,x_4)\\
K(x_2,-y) & K(x_2,x_2) & K(x_2,x_3) & K(x_2,x_4)\\
K(x_3,-y) & K(x_3,x_2) & K(x_3,x_3) & K(x_3,x_4)\\
K(x_4,-y) & K(x_4,x_2) & K(x_4,x_3) & K(x_4,x_4)
\end{array}
\right|
+\cdots
\nonumber
\\
&=
\kappa-z\left\{ \kappa \,\mathrm{tr} \bm{K} -\bm{k}^{\mathrm{T}}\bm{k} \right\}
+\frac{z^2}{2!} \left\{ 
\kappa(\mathrm{tr} \bm{K})^2-\kappa\,\mathrm{tr} \bm{K}^2
-2\bm{k}^{\mathrm{T}}\bm{k}\,\mathrm{tr} \bm{K}
+2\bm{k}^{\mathrm{T}}\bm{K}\bm{k}
\right\}  \nonumber \\
&\quad 
-\frac{z^3}{3!}  \bigl\{ 
\kappa(\mathrm{tr} \bm{K})^3
-3\kappa\,\mathrm{tr} \bm{K}\,\mathrm{tr} \bm{K}^2
+2\kappa\,\mathrm{tr} \bm{K}^3
-3\bm{k}^{\mathrm{T}} \bm{k}(\mathrm{tr} \bm{K})^2
+3\bm{k}^{\mathrm{T}} \bm{k}\,\mathrm{tr} \bm{K}^2
\nonumber \\
&\quad\quad\quad\quad
-6\bm{k}^{\mathrm{T}} \bm{K} \bm{k}\,\mathrm{tr} \bm{K}
+6\bm{k}^{\mathrm{T}} \bm{K}^2 \bm{k}
\bigr\} 
+\cdots\ .
\nonumber
\end{align}
Reorganizing summations, one finds
\begin{align}
&\tau(z; I; \sqrt{-y})\cdot Z_{N,\beta=2,\nu}(\{-y\})
\nonumber \\
&=
\kappa\left\{1-\mathrm{tr}\, z \bm{K} +\frac{1}{2!} (\mathrm{tr}\, z \bm{K})^2-\frac{1}{3!} (\mathrm{tr}\,z \bm{K})^3+\cdots\right\}
\nonumber \\
&\quad\quad \times
\left\{1- \frac12 \mathrm{tr} (z \bm{K})^2+\cdots\right\}\left\{1-\frac13 \mathrm{tr} (z \bm{K})^3+\cdots\right\}\cdots
\nonumber \\
&\quad
+z \bm{k}^{\mathrm{T}} \bm{k}\left\{1- \mathrm{tr}\,z \bm{K}+\frac1{2!} (\mathrm{tr} \,z \bm{K})^2-\cdots\right\}
\left\{1-\frac12 \mathrm{tr} (z \bm{K})^2+\cdots\right\}\cdots
\nonumber \\
&\quad
+z^2 \bm{k}^{\mathrm{T}} \bm{K} \bm{k}\left\{1-\mathrm{tr}\,z \bm{K}+\cdots\right\}\cdots
\nonumber \\
&\quad
+z^3 \bm{k}^{\mathrm{T}} \bm{K}^2 \bm{k}\left\{1-\cdots\right\}\cdots
\nonumber \\
&\quad
+\cdots
\nonumber
 \\
&=\left\{\kappa+z \bm{k}^{\mathrm{T}} (\mathbb{I}+z \bm{K}+(z \bm{K})^2+\cdots) \bm{k}\right\}
\nonumber \\
&\quad
\times\exp\left(-\mathrm{tr} \,z \bm{K}- \frac12 \mathrm{tr} (z \bm{K})^2-\frac13 \mathrm{tr} (z \bm{K})^3-\frac14 \mathrm{tr} (z \bm{K})^4-\cdots\right)
\nonumber \\
&=
\left\{\kappa+z \bm{k}^{\mathrm{T}} (\mathbb{I}-z \bm{K})^{-1} \bm{k}\right\}
\det(\mathbb{I}-z \bm{K})
=
-\det
\left|
\begin{array}{cc}
-\kappa & -\sqrt{z}\bm{k}^{\mathrm{T}} \\
-\sqrt{z}\bm{k} & \mathbb{I}-z\bm{K}
\end{array}
\right|.
\label{eq:rewrite_fredholm2}
\end{align}
Thus we obtain a compact expression of $\tau(z; I; \sqrt{-y})$ in terms of the Fredholm determinant.

The generalization to the case with $\alpha$ eigenvalues lying at $y_a$ ($a=1,\ldots,\alpha$)
proceeds in the same way as the derivation of eq.~(\ref{eq:rewrite_fredholm2}), leading to
\begin{align}
\tau(z; I;\{\sqrt{-y_q}\})
&=
\frac{\det
\left|
\begin{array}{cc}
-\kappa & -\sqrt{z}\bm{k}^{\mathrm{T}} \\
-\sqrt{z}\bm{k} & \mathbb{I}-z\bm{K}
\end{array}
\right|}{\det(-\kappa)}:=\frac{\det\mathcal{K}(z)}{\det(-\kappa)},
\label{eq:det_formula}
\end{align}
where the notation for the block decomposition of kernels (\ref{eq:kernel_notation}) is generalized as
\begin{align}
&\kappa=\left[ K(-y_a,-y_b)\right]_{a,b=1,\ldots,\alpha}, \quad 
\bm{k}=\left[\sqrt{d x_i}\,K(x_i,-y_b)\right]_{\substack{i=1,\ldots,M\\b=1,\ldots,\alpha}},
\nonumber \\
&\bm{k}^{\mathrm{T}}=\left[K(-y_a,x_j)\sqrt{d x_j}\right]_{\substack{a=1,\ldots,\alpha \\j=1,\ldots,M}}, \quad
\bm{K}=\left[\sqrt{d x_i}\, K(x_i,x_j)\sqrt{d x_j}\right]_{i,j=1,\ldots,M} .
\label{eq:kernel_notation2}
\end{align}
The numerator $\mathcal{K}(z)$ clearly interpolates the (ordinary) determinantal form for the $k$-level correlation function $\det \kappa$
in the case $I\to \emptyset$  (for which $\bm{k}, \bm{K}\to 0$) and the Fredholm determinantal form $\det\left(\mathbb{I}-z \bm{K}\right)$
for the generating function of the gap probability in the `quenched' limit $y_a\to \infty$ (for which $\kappa\to\mathbb{I}$ and $\bm{k}\to 0$).

For $y_a>0$ and $y_a\in I$, $\tau(1; I; \{\sqrt{-y_a}\})$ represents
the Janossy density $J_{\alpha,I}(\{y_a\})$ defined as the probability of finding no eigenvalue in the interval $I$ 
except for the ones at designated points $y_a\in I$ ($a=1,\ldots,\alpha$),
for the (classical) Laguerre unitary ensemble.
On the other hand, after an analytic continuation to $y_a=-m_a^2<0$ and setting $I=[0,s]$,
$\tau(1; I; \{m_a\})$ represents
the probability $E(0;[0,s];\{m_a\})$ of finding no eigenvalue smaller than $s$ 
for the massive Laguerre unitary ensemble (see discussions in Appendix \ref{app:Janossy2}).

Finally, changing the eigenvalue variables back to the chiral Gaussian and
taking the asymptotic limit (\ref{eq:asymptotic_limit}),
eq.~(\ref{eq:det_formula}) leads to
\begin{align}
\tau(z; I;\{\mu_a\})
&=
\frac{\det
\left|
\begin{array}{cc}
-\kappa & -\sqrt{z}\bm{k}^{\mathrm{T}} \\
-\sqrt{z}\bm{k} & \mathbb{I}-z\bm{K}
\end{array}
\right|}{\det(-\kappa)},
\label{eq:det_formula_scaled}
\end{align}
with the kernel elements given by their scaled forms (\ref{eq:decompostion_kernel2}),
\begin{align}
&\kappa=\left[K_{--}(\mu_a,\mu_b)\right]_{a,b=1,\cdots,\alpha},\quad 
\bm{k}=\left[\sqrt{d \zeta_i}\,K_{+-}(\zeta_i,\mu_b)\right]_{\substack{i=1,\ldots,M\\b=1,\ldots,\alpha}},\nonumber \\
&\bm{k}^{\mathrm{T}}=\left[\sqrt{d \zeta_i}\,K_{-+}(\mu_a,\zeta_j)\right]_{\substack{a=1,\ldots,\alpha \\j=1,\ldots,M}},\quad 
\bm{K}=\left[\sqrt{d \zeta_i}\, K_{++}(\zeta_i,\zeta_j)\sqrt{d \zeta_j}\right]_{i,j=1,\ldots,M}.
\label{eq:kernel_notation_scaled}
\end{align}

\subsection{Fredholm Pfaffian for chiral Gaussian symplectic ensemble}
Generalization of the result of the previous subsection to the chiral GOE and GSE is straightforward:
one finds the quaternionic determinant formula simply by replacing $K$ with the quaternionic kernel and ``det'' with  ``qdet'' simultaneously, 
because the quaternionic determinant shares the same linear algebraic properties which are utilized in the derivation of the determinant formula (\ref{eq:det_formula_scaled}). 
In particular for the chiral GSE, one can use the explicit formulae of the correlation functions and spectral kernels for $N_F=4\alpha$ and $N_F=2\alpha$ in \cite{Nagao:2000cb}. (See Appendix \ref{app:kernel}.)
Indeed, applying the correlation functions $R^{(p)}$ in Appendix \ref{app:kernel} to eq.~(\ref{eq:rewrite_fredholm1}) and repeating the same steps
leading to zeq.~(\ref{eq:det_formula_scaled}), one finds the following Pfaffian formula
\begin{align}
\tau(z; I;\{\mu_a\})&=\frac{\mathrm{qdet}\left|\begin{array}{cc}-\kappa &-\sqrt{z}\bm{k}^{\mathrm{T}} \\
 -\sqrt{z}\bm{k} &
\mathbb{I}-z\bm{K}
\end{array}
\right|}{\mathrm{qdet}(-\kappa)}
=\frac{\mathrm{Pf}\left[Z\mathbb{J}_{2\alpha}-Z(z\circ K)\right]}{\mathrm{Pf}\left[-ZK^{(0)}\right]}
\nonumber \\
&=\frac{\sqrt{\mathrm{det}\left[\mathbb{J}_{2\alpha}-z\circ K\right]}}{\sqrt{\mathrm{det}K^{(0)}}}
:=\frac{\sqrt{\det \mathcal{K}(z)}}{\sqrt{\det K^{(0)}}},
\label{eq:Pfaffian_formula}
\end{align}
where 
\begin{align}
\mathbb{J}_{2\alpha}=
\mathrm{diag}(\overbrace{0,\cdots,0}^{2\alpha},1,1,\cdots).
\end{align}
The matrix elements $S_{AB}$, $D_{AB}$, and $I_{AB}$ ($A,B=\pm$) of the quaternionic kernel $K$ are given in eqs.~(\ref{eq:S})--(\ref{eq:I}).

For the quadruply degenerated case $N_F=4\alpha$, $z\circ K$ with $\mu_a$ ($a=1,\ldots,\alpha$) is given by
\begin{align}
z\circ K=\left(
\begin{array}{cc}
\left[K_{--}(\mu_a,\mu_b)\right] & \sqrt{z}\left[K_{+-}(\mu_a,\zeta_j)\sqrt{d\zeta_j}\right] \\
\sqrt{z}\left[\sqrt{d\zeta_i}K_{-+}(\zeta_i,\mu_b)\right]&
z\left[\sqrt{d\zeta_i}K_{++}(\zeta_i,\zeta_j)\sqrt{d\zeta_j}\right]
\end{array}
\right),
\end{align}
where $Z=\mathrm{i}\sigma_2\otimes\mathbb{I}_{\alpha+M}$, and 
\begin{align}
&K_{AB}
=\left(\begin{array}{cc}
[-S_{AB}(\xi_A,\xi_B)] & [-I_{AB}(\xi_A,\xi_B)]  \\
{[D_{AB}(\xi_A,\xi_B)]} & {[-S^{\mathrm{T}}_{AB}(\xi_A,\xi_B)]}
\end{array}
\right),\nonumber \\
& (\xi_+,d\xi_+)=(\zeta,d\zeta),\quad (\xi_-,d\xi_-)=(\mu,1).
\end{align}
For the doubly degenerated case $N_F=2\alpha$, $z\circ K$ for even $\alpha$ with $y_a=-\mu_a^2$ ($a=1,\ldots,\alpha$)  is given by
\begin{align}
Z(z\circ K)=\left(
\begin{array}{ccc}
[I_{--}(\mu_a,\mu_b)]
& 
\sqrt{z}[I_{-+}(\mu_a,\zeta_j)\sqrt{d\zeta_j}]& 
\sqrt{z}\left[S_{-+}(\mu_a,\zeta_j)\sqrt{d\zeta_j}\right]
\\
-\sqrt{z}[\sqrt{d\zeta_i}I_{-+}^{\mathrm{T}}(\mu_b,\zeta_i)] &
z[\sqrt{d\zeta_i}I_{++}(\zeta_i,\zeta_j)\sqrt{d\zeta_j}] 
&
z[\sqrt{d\zeta_i}S_{++}(\zeta_i,\zeta_j)\sqrt{d\zeta_j}]
\\
-\sqrt{z}[\sqrt{d\zeta_i}S_{-+}^{\mathrm{T}}(\mu_b,\zeta_i)]
& 
-z[\sqrt{d\zeta_i}S^{\mathrm{T}}_{++}(\zeta_j,\zeta_i)\sqrt{d\zeta_j}]
&
z[\sqrt{d\zeta_j}D_{++}(\zeta_i,\zeta_j)\sqrt{d\zeta_i}]
\end{array}
\right),
\end{align}
and $z\circ K$ for odd $\alpha$ is by
\begin{align}
&Z(z\circ K)
 \\
&{\small 
=\left(
\begin{array}{cccc}
[I_{--}(\mu_a,\mu_b)]
&
[Q_{-}(\mu_a)]
&
\sqrt{z}[I_{-+}(\mu_a,\zeta_j)\sqrt{d\zeta_j}]
&
\sqrt{z}\left[S_{-+}(\mu_a,\zeta_j)\sqrt{d\zeta_j}\right]
\\
-[Q_{-}^{\mathrm{T}}(\mu_b)]
&
0
&
-\sqrt{z}[Q_{+}^{\mathrm{T}}(\zeta_j)\sqrt{d\zeta_j}]
&
-\sqrt{z}[P^{\mathrm{T}}_+(\zeta_j)\sqrt{d\zeta_j}] 
\\
-\sqrt{z}[\sqrt{d\zeta_i}I^{\mathrm{T}}_{-+}(\mu_b,\zeta_i)]
&
\sqrt{z}[\sqrt{d\zeta_i}Q_{+}(\zeta_i)]
&
z[\sqrt{d\zeta_i}I_{++}(\zeta_i,\zeta_j)\sqrt{d\zeta_j}]
&
z[\sqrt{d\zeta_i}S_{++}(\zeta_i,\zeta_j)\sqrt{d\zeta_j}]
\\
-\sqrt{z}[\sqrt{d\zeta_i}S_{-+}^{\mathrm{T}}(\mu_b,\zeta_i)]
&
\sqrt{z}\left[\sqrt{d\zeta_i}P_{+}(\zeta_i)\right]
&
-z[\sqrt{d\zeta_i}S^{\mathrm{T}}_{++}(\zeta_j,\zeta_i)\sqrt{d\zeta_j}]
&
z[\sqrt{d\zeta_i}D_{++}(\zeta_i,\zeta_j)\sqrt{d\zeta_j}]
\end{array}
\right).
}
\nonumber
\end{align}
Matrix elements of $z\circ K$ in the asymptotic limit (\ref{eq:asymptotic_limit}) are summarized in 
eqs.~(\ref{eq:S})--(\ref{eq:I}), and (\ref{eq:Q})--(\ref{eq:P}).

In case that some of the masses $\mu_a$'s are degenerated, one should adopt the confluent limit of the spectral kernel.
Some details of the confluent limit of the spectral kernel is discussed in Appendix \ref{app:confluent}.

\section{Numerical evaluation of the Janossy density via the Nystr\"om-type discretization}\label{section3}

In evaluating the Fredholm determinant (\ref{eq:det_formula}) and Pfaffian (\ref{eq:Pfaffian_formula}) numerically, 
the Nystr\"om-type discretization proves to be a highly efficient method\footnote{
We will compare our results with the Monte Carlo simulations to examine the efficiency of this method.
See Appendix \ref{app:lattice_data} on details of the Monte Carlo simulation.}
 \cite{Bornemann1,Bornemann2}.This numerical method is based on the quadrature rule (see a brief summary in Appendix \ref{app:nystrom}), and
in seminal works by F.~Bornemann,  it is shown that the Nystr\"om-type discretization
of Fredholm determinants of 
integral operators
of trace class 
(i.e.~for unitary and symplectic ensembles)
converges exponentially as the order of the discretization grows. 
In the following, 
we employ the Gauss-Legendre quadrature rule of order $M$ with
the nodes $\zeta_i$ and the weights $w_i=d\zeta_i$ ($i=1,\ldots,M$) given in eq.~(\ref{eq:node_weight}).\\

\subsection{Chiral GUE with doubly degenerated masses $N_F=2\alpha$}

The Nystr\"om-type discretization of the Fredholm determinant
for the individual eigenvalue distribution with $\beta=2$ and $N_F=2\alpha$ is 
given as follows:
\begin{align}
&\tau(z;[0,s];\{\mu_a\})
\nonumber \\
&=\det\left(
\begin{array}{cc}
-[K_{--}(\mu_a,\mu_b)]_{a,b=1,\ldots\alpha} & -\sqrt{z}\left[K_{-+}(\mu_a,\zeta_j)\sqrt{w_j}\right]_{\substack{a=1,\ldots,\alpha \\ j=1,\ldots,M}} \\
-\sqrt{z}\left[\sqrt{w_i}K_{+-}(\zeta_i,\mu_b)\right]_{\substack{i=1,\ldots,M \\ b=1,\ldots,M}} &\mathbb{I}_M-z\left[\sqrt{w_i}K_{++}(\zeta_i,\zeta_j)\sqrt{w_j}\right]_{i,j=1,\ldots,M} 
\end{array}
\right)
\nonumber \\
&\qquad
\bigg/\det(-[K_{--}(\mu_a,\mu_b)]_{a,b=1,\ldots\alpha}),
\label{eq:tau_chGUE}
\end{align}
where the matrix elements $K_{AB}$ are found in eq.~(\ref{eq:decompostion_kernel2}).

\begin{figure}[htbp]
 \begin{minipage}{0.5\hsize}
 \vspace{-0.5cm}
  \begin{center}
   \hspace*{0.5cm}
   \includegraphics[bb=0 0 360 223,width=90mm]{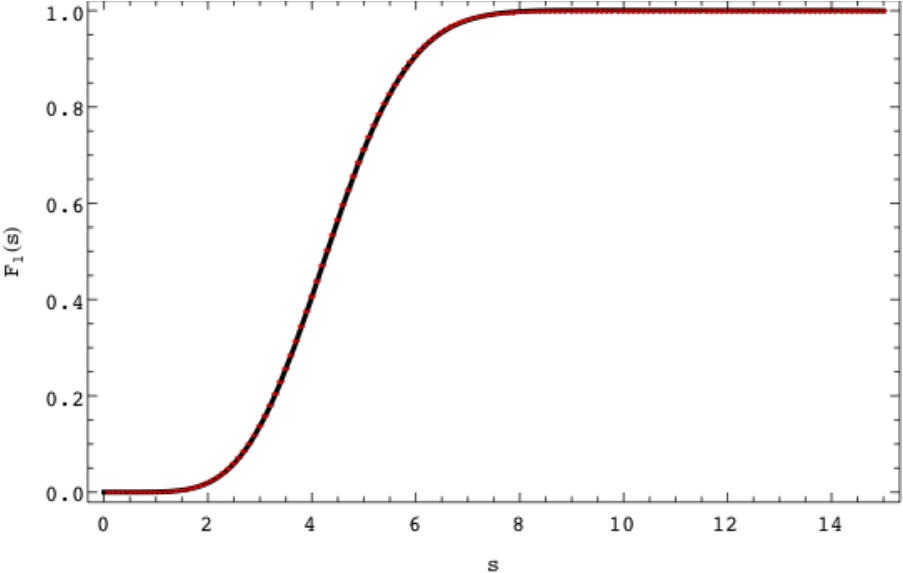}
    \end{center}
 \end{minipage}
 \begin{minipage}{0.5\hsize}
 \begin{center}
    \includegraphics[bb=0 0 360 223,width=90mm]{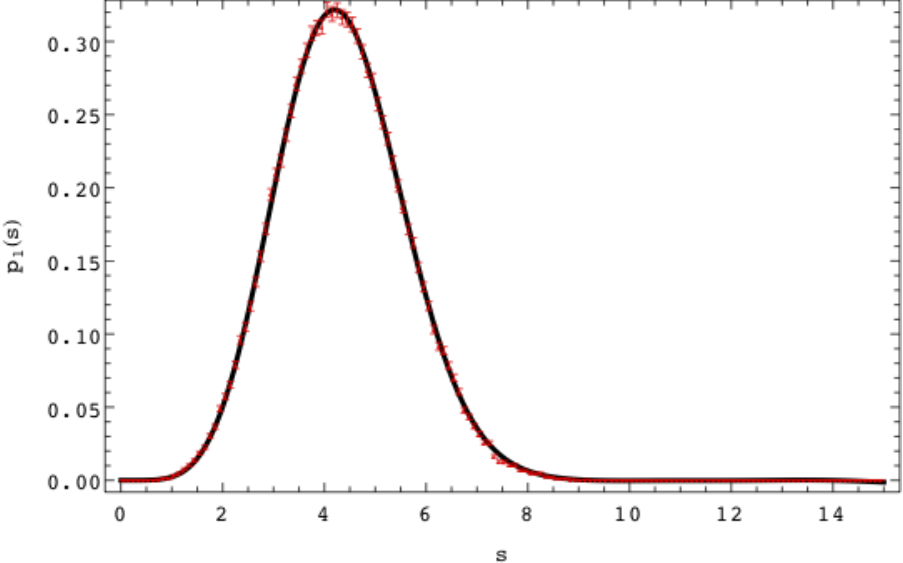}
 \end{center}
  \label{fig:beta21_3}
 \end{minipage}
\caption{\label{fig:beta22}Plots of $F_1(s)$ (left) and $p_1(s)$ (right) for 
the chiral GUE
with $N_f=2$ doubly-degenerated masses $\mu_1=0.1$ and the topological charge $\nu=0$ are depicted. In this computation, 
the quadrature of order $M=5$ is used to discretize the Fredholm
 determinant (\ref{eq:tau_chGUE}).  On both plots, data obtained with
 Monte Carlo simulation with matrix rank $N=1000$ is overlaid (red symbol with error bar, though
 the error in the left panel is hard to recognize by eye).}
\end{figure}

We will evaluate $F_1(s)=1-\tau(1;[0,s];\mu_1)$ and $p_1(s)=\partial_s F_1(s)$ 
using the expression (\ref{eq:tau_chGUE}) and compare with the Monte Carlo simulation.
For $\alpha=1$ with $\mu_1=0.1$ and the topological charge $\nu=0$, we obtain the numerical plots of  $F_1(s)$ and $p_1(s)$ in Fig.~\ref{fig:beta22}
for the rank $M=5$ of the Gaussian quadrature and find a good agreement with the Monte Carlo simulation with the matrix rank $N=1000$.  
\\
\vspace{-1cm}

\subsection{Chiral GSE with quadruply degenerated masses $N_F=4\alpha$}

For $\beta=4$ and with $N_F=4\alpha$, the Nystr\"om-type discretization of the Fredholm Pfaffian is 
given by
\begin{align}
&\tau(z;[0,s];\{\mu_a\})=\frac{|\det^{1/2}(\mathcal{K}(z))|}{|\det^{1/2}(K_{--})|},
\nonumber \\
&\mathcal{K}(z)=\left(
\begin{array}{cc}
-\left[K_{--}(\mu_a,\mu_b)\right] & -\sqrt{z}\left[K_{+-}(\mu_a,\zeta_j)\sqrt{w_j}\right] \\
-\sqrt{z}\left[\sqrt{w_i}K_{-+}(\zeta_i,\mu_b)\right]&
\mathbb{I}_{2M}-z\left[\sqrt{w_i}K_{++}(\zeta_i,\zeta_j)\sqrt{w_j}\right]
\end{array}
\right),
\end{align}
where the matrix elements $K_{AB}$ are found in eqs.~(\ref{eq:S})--(\ref{eq:I}). \\
\vspace{-0cm}
\begin{figure}[htbp]
 \hspace{0.0cm}
 \noindent\hfil
 \begin{minipage}{0.50\hsize}
 \begin{center}
   \includegraphics[bb=0 0 360 223,width=100mm]{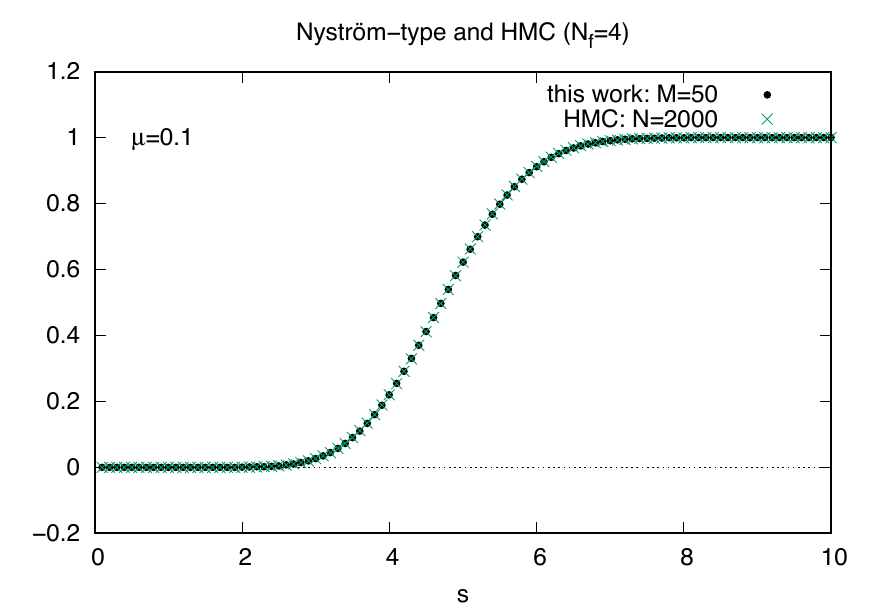}
 \end{center}
  \label{fig:beta44_test_HMC}
 \end{minipage}\\
 \begin{minipage}{0.50\hsize}
 \vspace{-0.4cm}
 \begin{center}
   \includegraphics[bb=0 0 360 223,width=100mm]{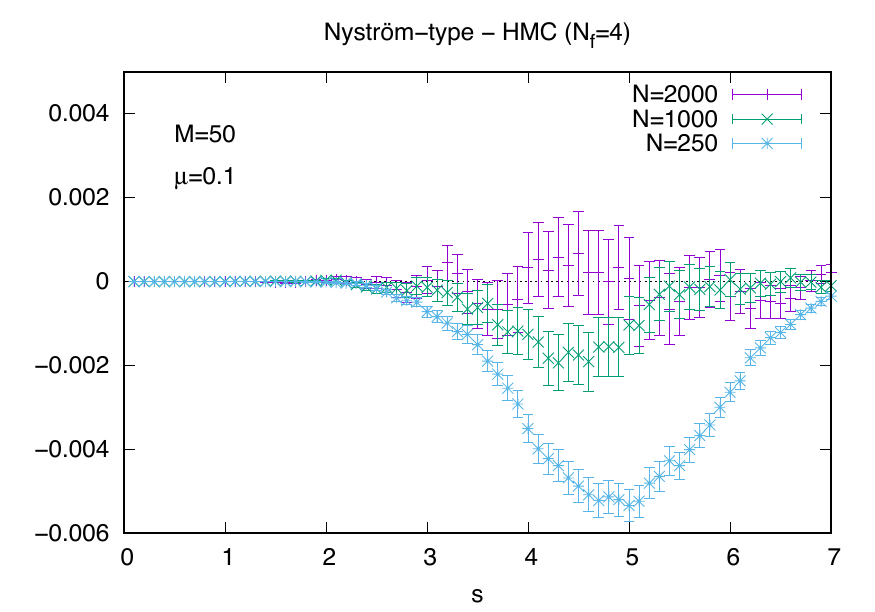}
 \end{center}
 \end{minipage}
 \begin{minipage}{0.50\hsize}
 \vspace{-0.4cm}
 \begin{center}
   \includegraphics[bb=0 0 360 223,width=100mm]{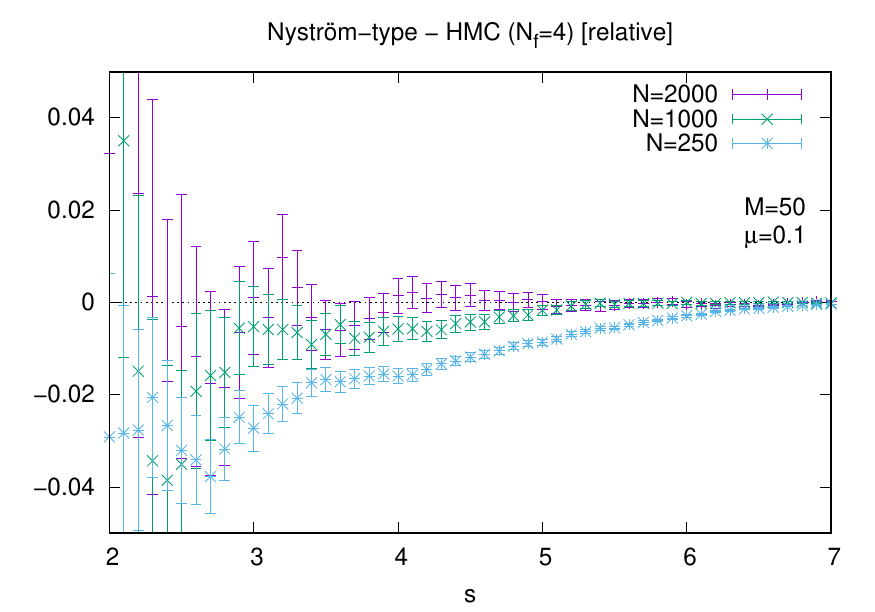}
 \end{center}
 \end{minipage}

 \caption{\label{fig:beta44}
 $F_1(s)$ is computed for
 the chiral GSE
 with $N_F=4\ (\alpha=1)$ quadruply-degenerated masses $\mu_1=0.1$ and the topological charge $\nu=0$ in two ways.
 In the top panel, the
 Nystr\"om-type discretization  of order $M=50$ is applied (black dot)
 and the hybrid Monte Carlo simulation is applied with
 the random rank $N=2000$, $F_1^{\mathrm{HMC}}(s)$ (green cross).
 The error of the HMC result, which is not shown in
 the top panel, is smaller than the symbols.
 The bottom left panel  shows the difference of these two methods,
 $F_1(s)-F_1^{\mathrm{HMC}}(s)$ with $N=250, 1000,
 2000$.  The computational results of the hybrid Monte
 Carlo simulation converges to that of the Nystr\"om-type discretization
 as $N$ grows. The errors plotted come from the Monte Carlo result.
 The relative difference normalized by the 
 Nystr\"om-type is also plotted in in the bottom right panel.  
 Note that the relative difference looses its meaning for $s\lesssim 2$ as 
 the Nystr\"om-type result becomes smaller than the Monte Carlo error.
 }
\end{figure}

\noindent
$\bullet\  N_F=4$

The numerical plot of $F_1(s)$ for $N_F=4$ with the quadruply degenerated mass 
$\mu_1=0.1$ and the topological charge $\nu=0$ is depicted in 
black dots in
Fig.~\ref{fig:beta44} top.
In this computation we have chosen $M=50$. 
In the same Figure,
the result of the hybrid Monte Carlo simulation $F_1^{\mathrm{HMC}}(s)$ of the chiral random matrix ensemble (\ref{eq:partition_function}) with the matrix rank $N=2000$ is shown in green dots as an overlay.

In order to verify the numerical computation with the Nystr\"om-type discretization,
we closely looked at the difference $F_1(s)-F_1^{\mathrm{HMC}}(s)$ for matrix ranks $N=250, 1000, 2000$. 
Fig.~\ref{fig:beta44} bottom shows that the computational results of the hybrid Monte Carlo simulation 
indeed converge to the Nystr\"om-type discretization as $N$ grows, confirming
that the numerical evaluation of $F_1(s)$ by the Nystr\"om-type discretization at $M=50$ is good enough.

\vspace{0cm}
\begin{figure}[htbp]
  \begin{center}
  \hspace*{3cm}
    \includegraphics[bb=0 0 360 223,width=120mm]{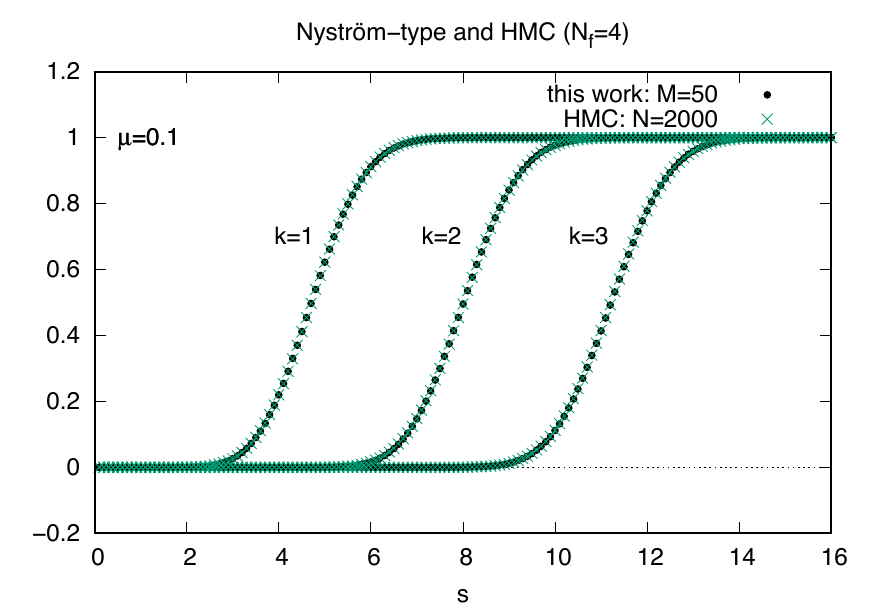}
  \end{center}
 \caption{\label{fig:beta44E_all}
 $F_k(s)$ ($k=1,2,3$) for $N_F=4$ quadruply-degenerated mass parameters
 $\mu_1=0.1$ and the topological charge $\nu=0$ together with Monte Carlo result.
 Black dot: Nystr\"om-type discretization of order $M=50$.
 Green cross: hybrid Monte Carlo simulation with the random matrix rank $N=2000$.}
 \end{figure}

The generalized gap probability 
$E_k(s):=E(k;[0,s];\{\mu_a\})$ in eq.~(\ref{eq:E_k}) is given by
\begin{align}
E_k(s)=\frac{(-1)^k}{k!}\frac{\partial^k}{\partial z^k}\tau(z;[0,s];\{\mu_a\})\Big|_{z=1}.
\end{align}
The $z$-derivatives of $\tau(z;[0,s];\{\mu_a\})$ are evaluated directly by the Taylor expansion
of the determinant, and 
the explicit expressions of $E_k(s)$'s as the sum of trace factors are listed in Appendix \ref{app:trace}.
Numerical computations for $F_k(s)$ for $0\leq s \leq 16$ are depicted in
Fig.~\ref{fig:beta44E_all}, and a good agreement is observed with the
computations of the hybrid Monte Carlo simulation with $N=2000$.

\noindent
$\bullet\  N_F=8$

\begin{figure}[htbp]
\begin{minipage}{0.5\hsize}
\begin{center}
 \includegraphics[bb=0 0 360 223,width=70mm]{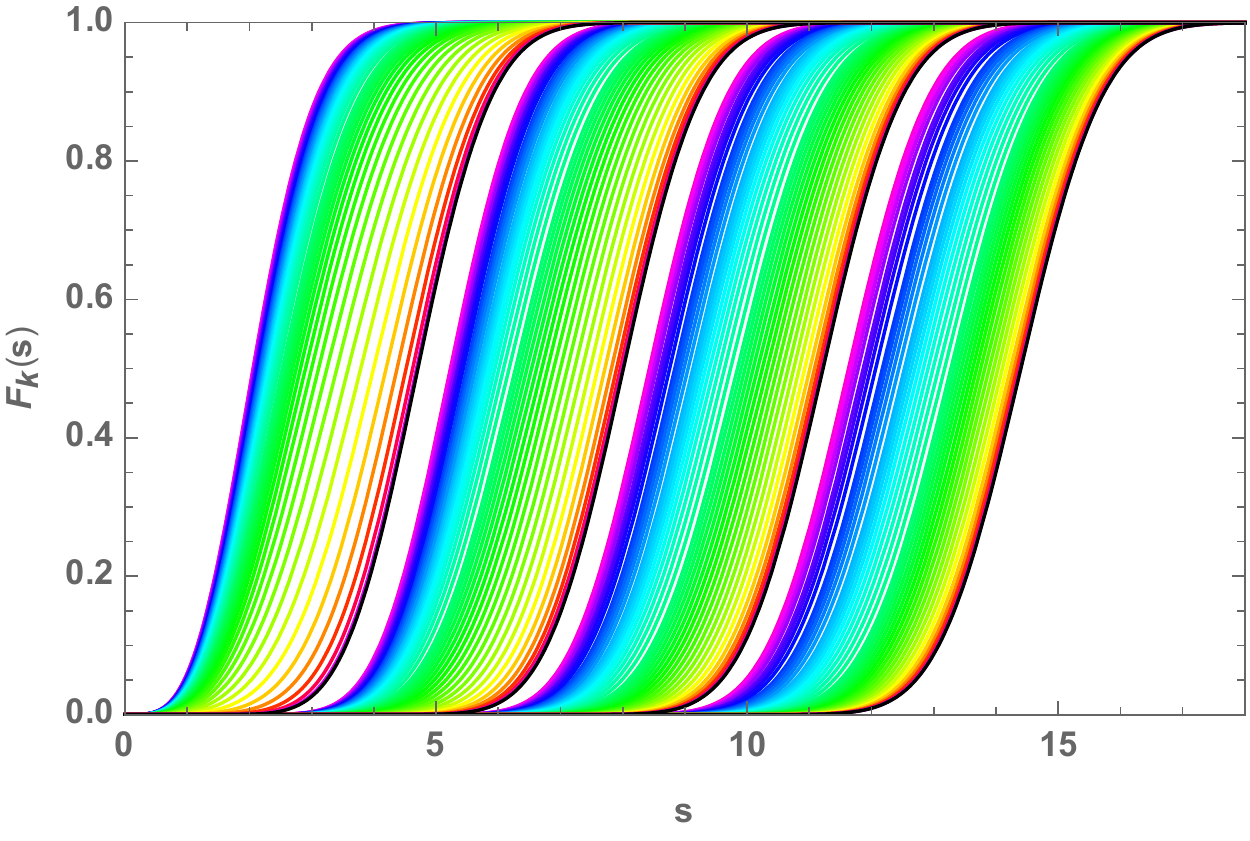}
\end{center}
\end{minipage}
\begin{minipage}{0.5\hsize}
\begin{center}
 \includegraphics[bb=0 0 360 223,width=70mm]{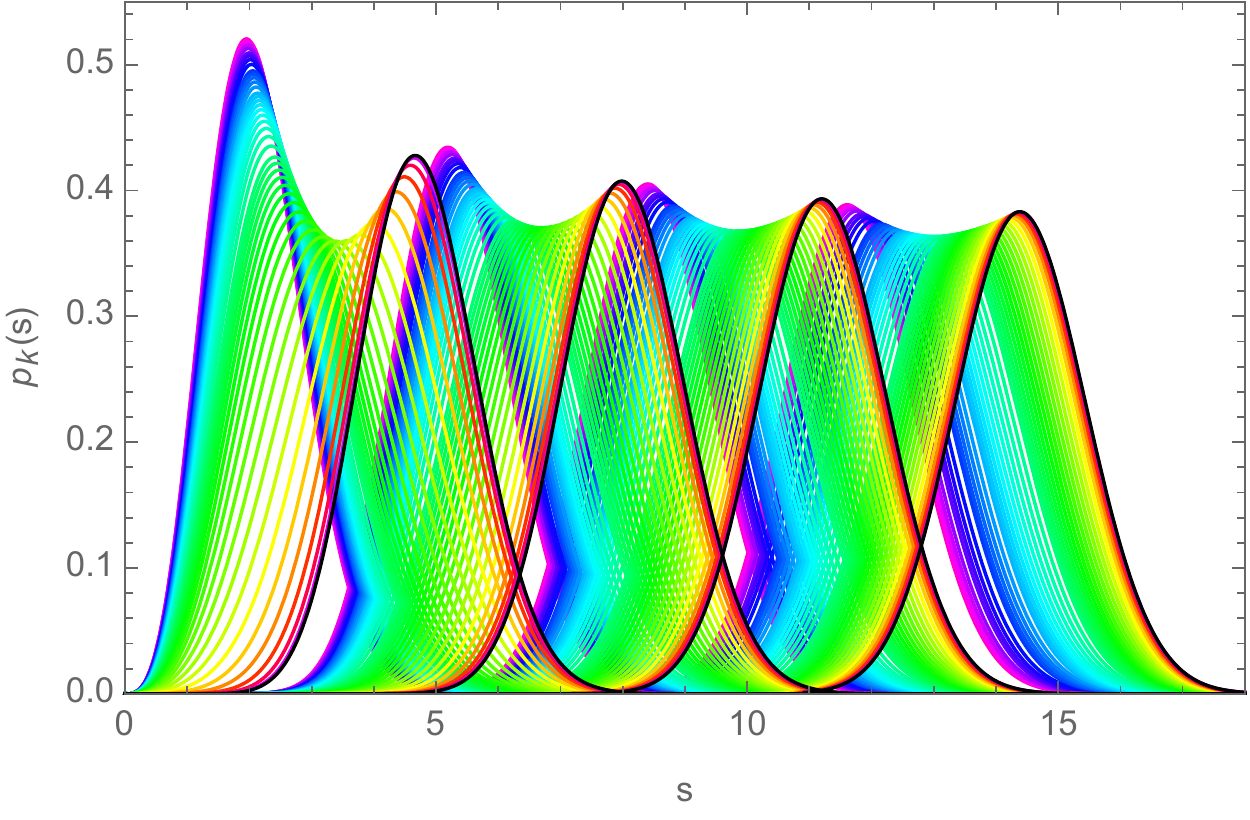}
\end{center}
\end{minipage}
\vspace{1cm}
\\
\begin{minipage}{0.5\hsize}
\begin{center}
 \includegraphics[bb=0 0 360 223,width=65mm]{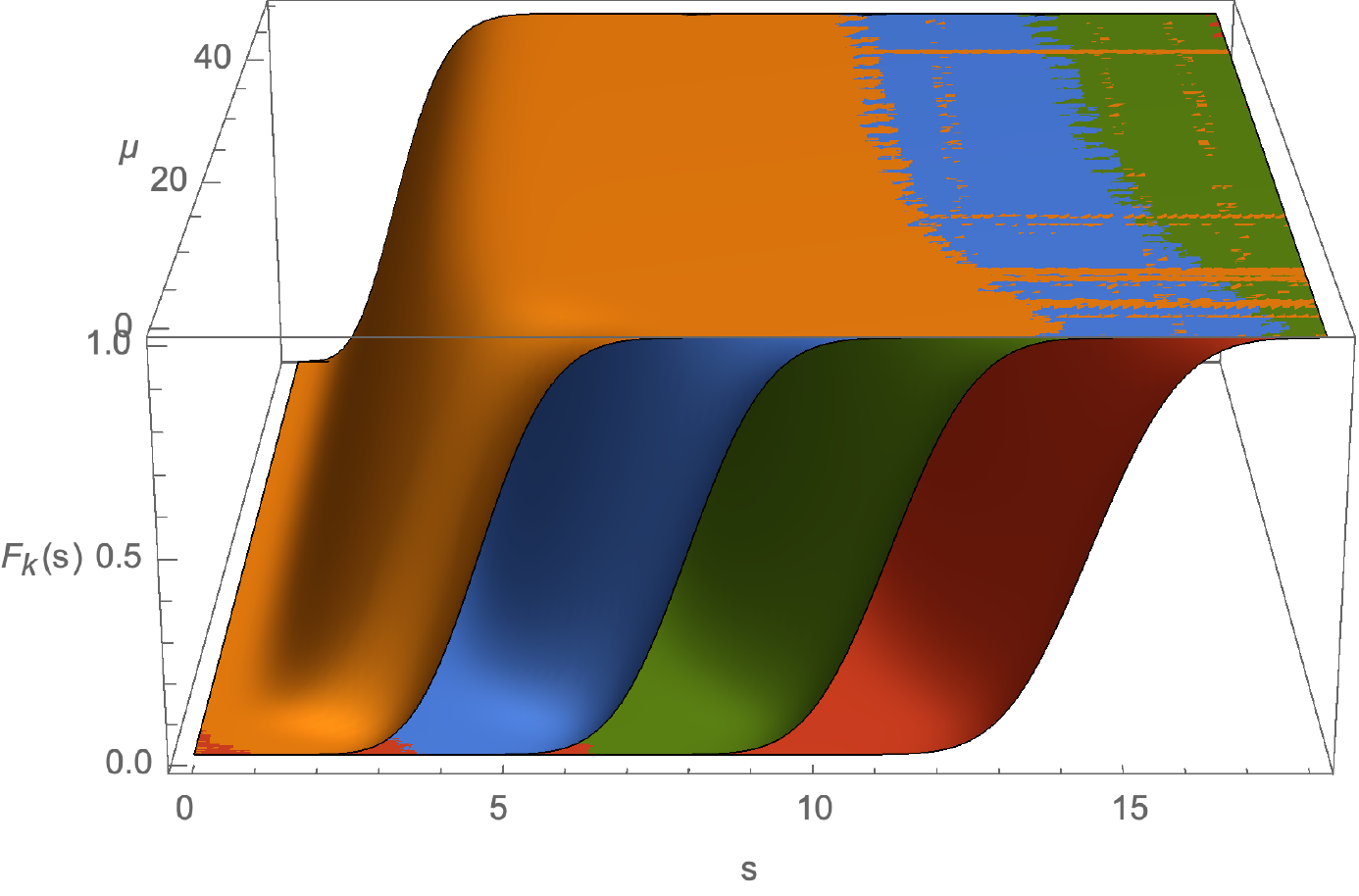}
\end{center}
\end{minipage}
\begin{minipage}{0.5\hsize}
\begin{center}
 \includegraphics[bb=0 0 360 223,width=65mm]{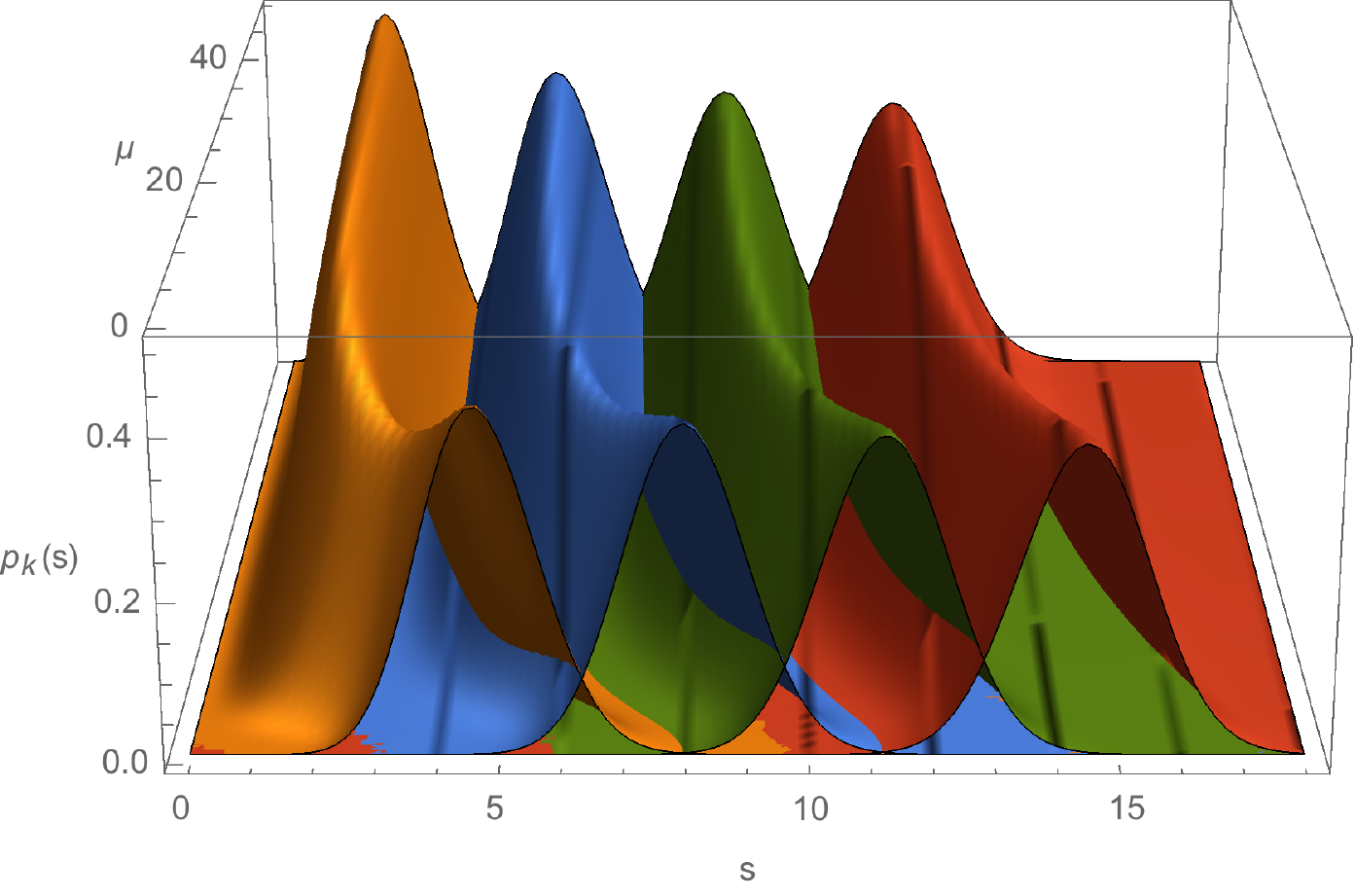}
\end{center}
\end{minipage}
 \caption{\label{fig:beta48_2}
Scaled mass parameter ($\mu$) dependence of [left panels] cumulative distributions $F_{k}(s)$
and [right panels] probability distributions $p_k(s)$ of the four smallest unfolded eigenvalues
$k=1,2,3,4$ of the massive chiral GSE with $N_F=8$ degenerate flavors and the topological charge $\nu=0$, computed at $M=128$.
In the upper panels, mass parameters are chosen at $\mu=0$ (black, $N_F=0$ with $\nu=4$),
$0.5, \cdots (\mbox{step}\;0.5), 10, \cdots (\mbox{step}\;1), 20, \cdots (\mbox{step}\;2), 30,  \cdots (\mbox{step}\;5), 60, \cdots (\mbox{step}\;10), 100, 200$
 (red to purple), 
 $\infty$ 
 (gray, $N_F=0$ with $\nu=0$). 
The lower panels are interpolations of the upper ones.
}
 \end{figure}

Numerical plots of $F_k(s)$ and $p_k(s)=\partial_s F_{k}(s)$ for $N_F=8\ (\alpha=2)$ with 8-fold degenerated mass
$\mu=\mu_1=\mu_2$ and the topological charge $\nu=0$, computed at $M=128$, are depicted in Figs.~\ref{fig:beta48_2}.
The computed numerics of $F_k(s)$ are appended as Supplementary Material because
this case is practically important within our application to the two-color lattice QCD with staggered quarks;
the case with $N_F=4\ (\alpha=1)$ doubtlessly has its chiral symmetry broken as in the ordinary QCD,
and those with $N_f\geq 12\ (\alpha\geq 3)$ have negative 1-loop $\beta$-function coefficients
$\beta_0=(11N_C-2N_F)/3$ and are IR free.
Accordingly, $N_F=8$ is the only case 
which evokes the question of whether its nature is either QCD-like, conformal, or 
walking (which would nominate the model as a possible candidate for the technicolor scenario),
and motivates us to compare its Dirac spectrum to the massive chiral GSE prediction so as to 
confirm or exclude if it is QCD-like.\\

\subsection{Chiral GSE with doubly degenerated masses $N_F=2\alpha$}

For $\beta=4$ and with $N_F=2\alpha$, 
the quaternionic kernel for Janossy density of the $\beta=4$ ensemble is treated independently for even and odd $\alpha$.\\

\noindent{$\bullet$ {$N_F=2$}}

For the case of odd $\alpha$, the Nystr\"om-type discretization of $\tau(z;[0,s];\{\mu_a\})$ yields
\begin{align}
&\tau(z;[0,s];\{\mu_a\})=\frac{\left|\det^{1/2}\left(\mathcal{K}(z)\right)\right|}{\left|\det^{1/2}\left(\mathcal{K}^{(0)}\right)\right|},
\nonumber \\
&{\scriptsize
\mathcal{K}(z)
=\left(
\begin{array}{cccc}
\mathbb{I}_M-z[\sqrt{w_iw_j}S_{++}(x_i,x_j)]
& -z[\sqrt{w_iw_j}I_{++}(x_i,x_j)]
&
\sqrt{z}[\sqrt{w_j}Q_{+}(x_j)]
& \sqrt{z}[I_{-+}(\mu_a,x_j)]
\\
z[\sqrt{w_iw_j}D_{++}(x_i,x_j)]
&\mathbb{I}_M-z[\sqrt{w_iw_j}S^{\mathrm{T}}_{++}(x_j,x_i)]
& \sqrt{z}[\sqrt{w_j}P_{+}(x_j)]
&
\sqrt{z}[\sqrt{w_j}S_{-+}(\mu_a,x_j)]
\\
-\sqrt{z}\left[\sqrt{w_i}S^{\mathrm{T}}_{-+}(\mu_b,x_i)\right]
&
-\sqrt{z}[\sqrt{w_i}I^{\mathrm{T}}_{-+}(\mu_b,x_i)]
&
-[Q_{-}(\mu_b)]
&
-[I_{--}(\mu_a,\mu_b)]
\\
-\sqrt{z}\left[\sqrt{w_i}P_{+}^{\mathrm{T}}(x_i)\right] &
-[\sqrt{w_i}Q_{+}^{\mathrm{T}}(x_i)]
&
0
&
-[Q_{-}^{\mathrm{T}}(\mu_a)]
\end{array}
\right),
}
\nonumber 
\\
&\mathcal{K}^{(0)}
=\left(
\begin{array}{cc}
-[Q_{-}(\mu_b)]
&
-[I_{--}(\mu_a,\mu_b)]
\\
0
&
-[Q_{-}^{\mathrm{T}}(\mu_a)]
\end{array}
\right),
\end{align}
where the matrix elements are found in eqs.~(\ref{eq:S})--(\ref{eq:I}) and (\ref{eq:Q})--(\ref{eq:P}).

\begin{figure}[htbp]
 \hspace{0cm}
 \noindent\hfil
 \begin{minipage}{0.50\hsize}
 \begin{center}
 \includegraphics[bb=0 0 360 223,width=100mm]{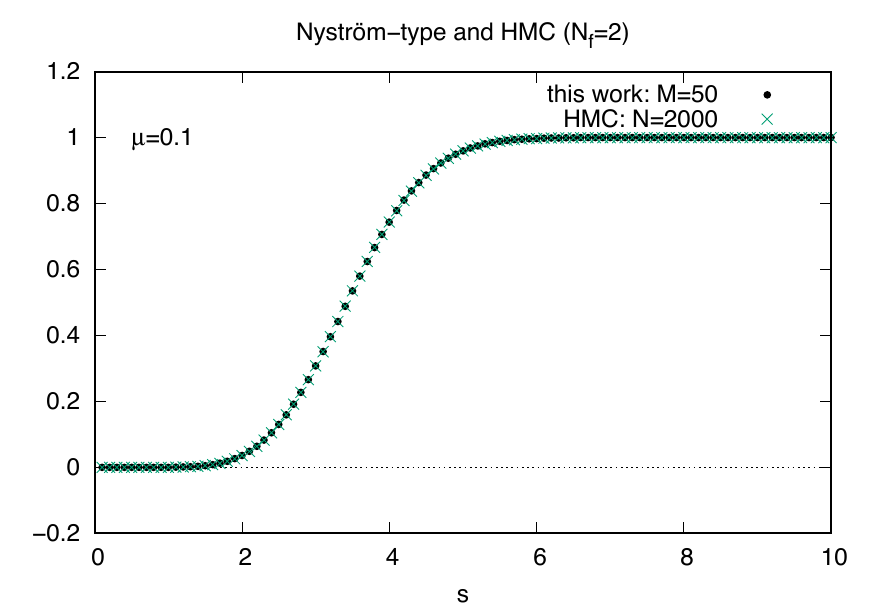}
 \end{center}
  \label{fig:beta42_test_HMC}
 \end{minipage}\\
 \begin{minipage}{0.50\hsize}
  \vspace{-0.4cm}
 \begin{center}
 \includegraphics[bb=0 0 360 223,width=100mm]{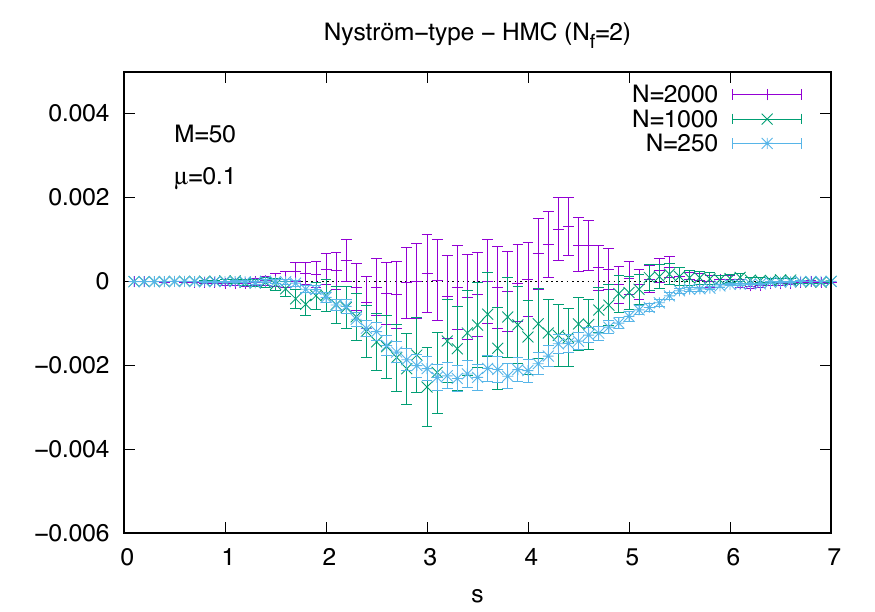}
 \end{center}
 \end{minipage}
 \begin{minipage}{0.50\hsize}
  \vspace{-0.4cm}
 \begin{center}
 \includegraphics[bb=0 0 360 223,width=100mm]{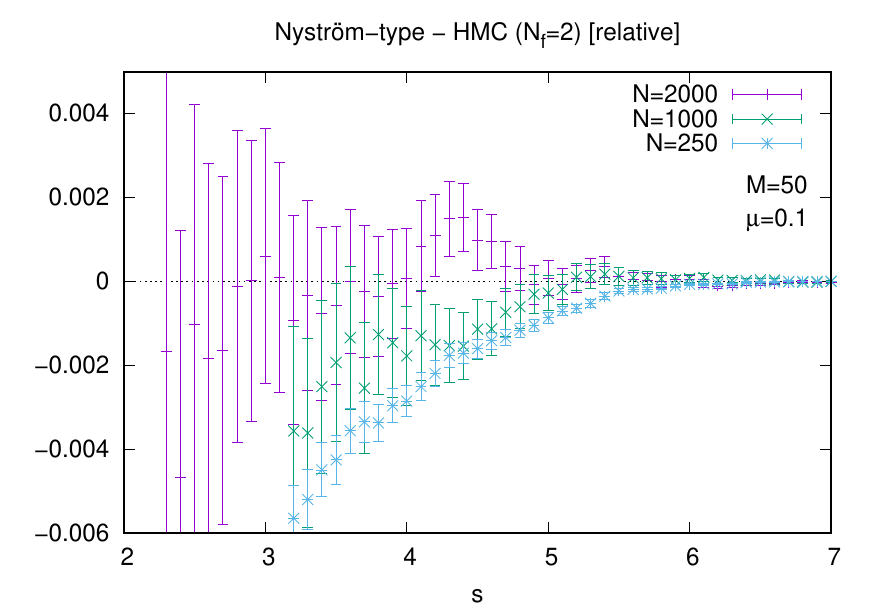}
 \end{center}
 \end{minipage}
 \caption{\label{fig:beta42}
The same plot as Fig.~\ref{fig:beta44} but with doubly-degenerated
 masses $\mu_1=0.1$ ($N_F=2$) and the topological charge $\nu=0$.  
 $F_1(s)$ for the $\beta=4$ ensemble  is evaluated in two ways.
 Top panel: The Nystr\"om-type discretization is applied
of order $M=50$ (black dots) and the hybrid Monte Carlo simulation is
applied with the random matrix rank $N=2000$ (green cross),
 for which the statistical errors are smaller than the symbols and not
 shown in the plot. Bottom panels: Difference between 
 Nystr\"om-type discretization and hybrid Monte Carlo with several
 values of matrix rank $N$.}
\end{figure}

$F_1(s)$ for $\alpha=1$ (i.e. $N_F=2$) with doubly degenerated mass $\mu_1=0.1$ and the topological charge $\nu=0$ is evaluated numerically by the Nystr\"om-type discretization of order  $M=50$. Plots are depicted in Fig.~\ref{fig:beta42} (black dots). The hybrid Monte Carlo simulation $F_1^{\mathrm{HMC}}(s)$ of  the chiral random matrix ensemble (\ref{eq:partition_function}) with the matrix rank $N=1000$ is depicted  (green dots) in Fig.~\ref{fig:beta42} as an overlay.

\noindent{$\bullet$ {The confluent limit for $N_F=2+2+2+2$}}

The next example is the chiral GSE for $N_F=2+2+2+2$ (i.e. $\alpha=4$) in the complete confluent limit. 
The Nystr\"om-type discretization of $E_0(s)$ of order $M$ is given by
\begin{align}
&E_0(s)=\frac{\left|\det^{1/2}\left(\mathcal{K}(z=1)\right)\right|}{|\det^{1/2}\mathcal{K}^{(0)}|},
\qquad \mathcal{K}(z=1)=\left(\begin{array}{cc}
\mathcal{S}_1 & -\mathcal{I} \\
{-\mathcal{D}} & \mathcal{S}_2
\end{array}
\right),\\
& \mathcal{S}_1=\left(\begin{array}{ccc}
\mathbb{I}_M-[\sqrt{w_iw_j}S_{++}(\zeta_i,\zeta_j)]& [\sqrt{w_j}I_{-+}^{(3,0)}(\mu,\zeta_j)] & [\sqrt{w_j}I_{-+}^{(2,0)}(\mu,\zeta_j)] \\
{-[\sqrt{w_i}S_{-+}^{(2,0)}(\mu,\zeta)] } & -I_{--}^{(2,3)}(\mu,\mu) &  0 \\
{[\sqrt{w_i}S_{-+}^{(3,0)}(\mu,\zeta)]} & 0 & -I_{--}^{(2,3)}(\mu,\mu)
\end{array}
\right), \nonumber \\
& \mathcal{I}=\left(\begin{array}{ccc}
[\sqrt{w_iw_j}I_{++}(\zeta_i,\zeta_j)]& -[\sqrt{w_j}I_{-+}^{(1,0)}(\mu,\zeta_j)] & -[\sqrt{w_j}I_{-+}^{(0,0)}(\mu,\zeta_j)] \\
{[\sqrt{w_i}I_{-+}^{(2,0)}(\mu,\zeta)]} &   -I_{--}^{(1,2)}(\mu,\mu)  &  -I_{--}^{(0,2)}(\mu,\mu) \\
{-[\sqrt{w_i}I_{-+}^{(3,0)}(\mu,\zeta)]} &  I_{--}^{(1,3)}(\mu,\mu)  &  I_{--}^{(0,3)}(\mu,\mu) 
\end{array}
\right), \nonumber \\
& \mathcal{D}=\left(\begin{array}{ccc}
-[\sqrt{w_iw_j}D_{++}(\zeta_i,\zeta_j)]& [\sqrt{w_j}S_{-+}^{(1,0)}(\mu,\zeta_j)]& [\sqrt{w_j}S_{-+}^{(0,0)}(\mu,\zeta_j)] \\
{[\sqrt{w_i}S_{-+}^{(0,0)}(\mu,\zeta)]} &  I_{--}^{(0,3)}(\mu,\mu)  &  I_{--}^{(0,2)}(\mu,\mu) \\
{-[\sqrt{w_i}S_{-+}^{(1,0)}(\mu,\zeta)]} & -I_{--}^{(1,3)}(\mu,\mu)  &  -I_{--}^{(1,2)}(\mu,\mu) 
\end{array}
\right),\nonumber 
\end{align}
\begin{align}
& \mathcal{S}_2=\left(\begin{array}{ccc}
\mathbb{I}_M-[\sqrt{w_iw_j}S_{++}(\zeta_i,\zeta_j)]& -[\sqrt{w_j}S_{-+}^{(1,0)}(\mu,\zeta_j)] & -[\sqrt{w_j}S_{-+}^{(0,0)}(\mu,\zeta_j)]\\
{-[\sqrt{w_i}I_{-+}^{(0,0)}(\mu,\zeta)]} & -I_{--}^{(0,1)}(\mu,\mu) &  0 \\
{[\sqrt{w_i}I_{-+}^{(1,0)}(\mu,\zeta)]} & 0 & -I_{--}^{(0,1)}(\mu,\mu)
\end{array}
\right), \nonumber 
\\
&
\mathcal{K}^{(0)}=\left(
\begin{array}{cccc}
0 & I_{--}^{(0,1)}(\mu,\mu) & I_{--}^{(0,2)}(\mu,\mu) &I_{--}^{(0,3)}(\mu,\mu) \\
-I_{--}^{(0,1)}(\mu,\mu) & 0 & I_{--}^{(1,2)}(\mu,\mu) & I_{--}^{(1,3)}(\mu,\mu) \\
-I_{--}^{(0,2)}(\mu,\mu) &  -I_{--}^{(1,2)}(\mu,\mu)  &0 & I_{--}^{(2,3)}(\mu,\mu) \\
-I_{--}^{(0,3)}(\mu,\mu) & - I_{--}^{(1,3)}(\mu,\mu)  &- I_{--}^{(2,3)}(\mu,\mu) &0
\end{array}
\right),
\nonumber
\end{align}
where the matrix elements $S_{AB}$, $D_{AB}$, and $I_{AB}$ ($A,B=\pm$) are found in eqs.~(\ref{eq:S})--(\ref{eq:I}), 
and $S^{(a,b)}_{AB}$, $D^{(a,b)_{AB}}$, and $I^{(a,b)}_{AB}$ ($A,B=\pm$)  are found in eqs.~(\ref{eq:S_confl})--(\ref{eq:I_confl}).

\begin{figure}[htbp]
\noindent\hfil
\hspace*{0cm}
 \begin{minipage}{0.50\hsize}
  \begin{center}
 \includegraphics[bb=0 0 360 223,width=100mm]{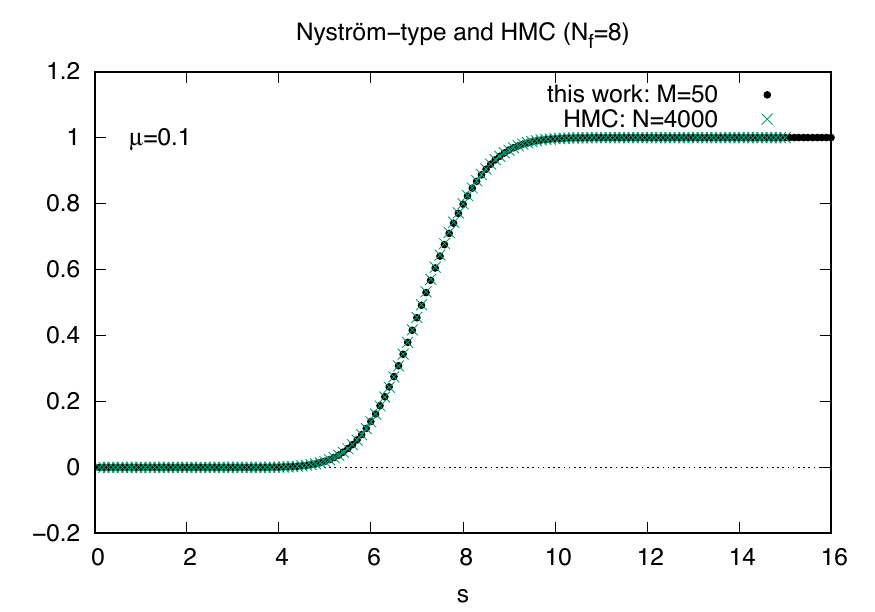}
  \end{center}
 \end{minipage}\\
 \begin{minipage}{0.50\hsize}
 \begin{center}
 \includegraphics[bb=0 0 360 223,width=100mm]{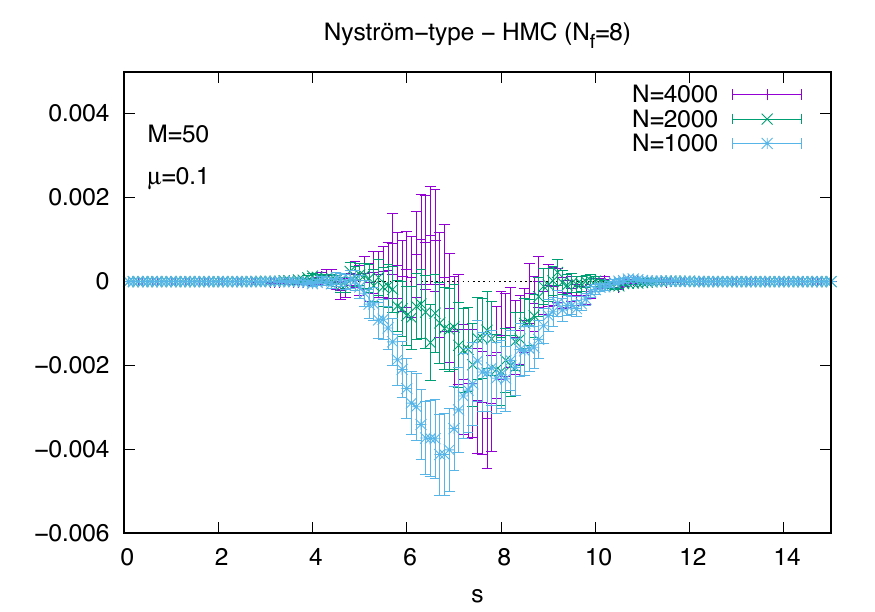}
 \end{center}
 \end{minipage}
 \begin{minipage}{0.50\hsize}
 \begin{center}
 \includegraphics[bb=0 0 360 223,width=100mm]{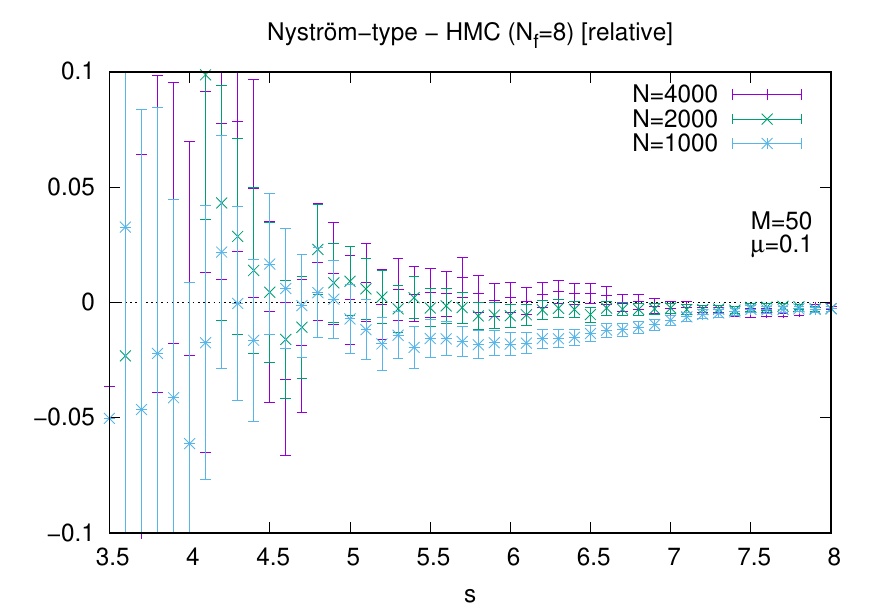}
 \end{center}
 \end{minipage}
 \caption{\label{fig:beta48}
 The same plot as Fig.~\ref{fig:beta44} but with $N_F=8$.
 $F_1(s)$ is computed in two ways for the chiral GSE with
 $N_F=8$ doubly-degenerated masses in the complete confluent limit with
 $\mu_1=0.1$ and the topological charge $\nu=0$.
 Nystr\"om-type discretization of order $M=50$ (black dot)
 and hybrid Monte Carlo simulation with the random rank $N=4000$
 are applied (green cross) in the top panel,
 and  the random rank $N=4000$ (green dots in the top figure),
 respectively.  The errors for HMC is smaller than the symbols and not
 shown in the plot.
 The bottom panels show
 the difference $F_1(s)-F_1^{\mathrm{HMC}}(s)$ for
 $N=1000, 2000, 4000$.
 As the $N$ grows, the HMC results converge to the result from
 Nystr\"om-type discretization.  Compared with the previous cases,
 however,
 its convergence is slower.}
\end{figure}

For the degenerated mass $\mu=0.1$ and the topological charge $\nu=0$, we find a good agreement of $F_1(s)=1-E_0(s)$
with the hybrid Monte Carlo simulation $F_1^{\mathrm{HMC}}(s)$.
The numerical plots are shown in Fig.~\ref{fig:beta48} (left panel) for $M=50$ of the Nystr\"om discretization of the Fredholm Pfaffian  and the hybrid Monte Carlo simulation of the rank $N=2000$. (Black dots: Nystr\"om discretization, Green dots: hybrid Monte Carlo simulation.)
The difference $F_1(s)-F_1^{\mathrm{HMC}}(s)$ in Fig.~\ref{fig:beta48}
(right panel) confirms us that the difference reduces as the rank of matrix grows, and these results confirm us that these numerical computations are consistent and valid.

\begin{figure}[h]
  \begin{center}
  \hspace*{3cm}
 \includegraphics[bb=0 0 360 223,width=120mm]{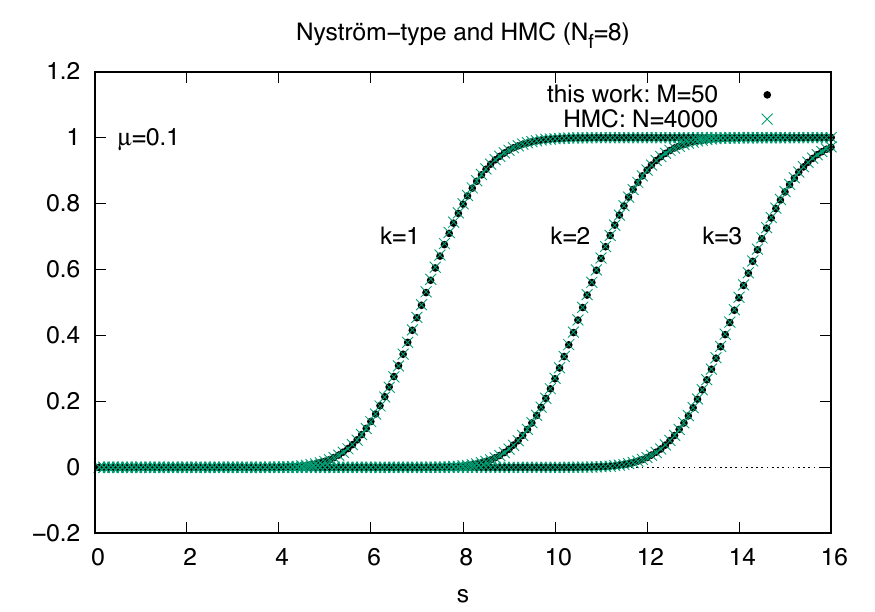}
  \end{center}
  \caption{ \label{fig:beta42222_confl2} Plot of $F_k(s)$ in $0\leq s \leq16$ for  $N_F=8$ in the complete confluent limit with $\mu_1=0.1$ and the topological charge $\nu=0$. Black dot: Nystr\"om-type discretization of order $M=50$. Green cross: hybrid Monte Carlo simulation with the random matrix rank $N=4000$.}
 \end{figure}

By applying the explicit expressions in Appendix \ref{app:trace}, we can evaluate $E_k(s)$'s numerically.
Plots for $F_k(s)$ in $0\leq s\leq6$ are depicted in Fig.~\ref{fig:beta42222_confl2}, and we find a good agreement with the computations of the hybrid Monte Carlo simulation with $N=4000$.

\section{Application: chiral condensate from lattice data}

As an application of our RMT results, we use the Dirac eigenvalues
of the SU(2) gauge theory with $N_F=8$ quarks in the fundamental representation.
A partial analysis of this system has been presented in
\cite{Huang:2015vkr}, where the Monte Carlo method is used to generate the RMT data.
Full analyses using the current RMT result will appear elsewhere \cite{lattice:full}.
As stated in the Introduction, we should use the chiral GSE with $N_F=4$, because due to the taste breaking effect, the 4-fold degeneracy
for the staggered fermions is totally broken 
so the number of lightest flavors is in fact $N_F=2$.  Furthermore, 
the pseudo-reality of the SU(2) gauge group yields an additional 2-fold degeneracy 
yields an additional 2-fold degeneracy, 
by which $N_F=2$ is promoted to $N_F=4$.

The microscopic eigenvalue density is related to the Dirac spectrum
through
\begin{align}
 \zeta_i&= \lambda_i V \Sigma, & \mu_f &= m_f V\Sigma,
\end{align}
where $\lambda_i$ denotes the eigenvalue of the Dirac operator,
$\Sigma$ the chiral condensate, $V$ the 4-volume, and $m_f$ the quark masses.  
We relate the smallest Dirac eigenvalue distribution from lattice
simulation through\footnote{This equation is not valid if the lattice
simulation is in the symmetric phase of the chiral symmetry, to 
which standard chiral RMT may not apply.
}
\begin{equation}
 p_1^{\mathrm{RMT}}(\zeta_1; \mu)
  \Big|_{\zeta_1=\lambda_1 V\Sigma,\, \mu=m_f V\Sigma}
= p_1^{\mathrm{latt.}}(\hat{\lambda}_1; \hat{m}_f). 
\label{eq:rmt-lattice}
\end{equation}
The parameters $\hat{V}$, $\hat{\Sigma}$ and $\hat{m}_f$ are the dimensionless
4-volume, the chiral condensate and
the fermion mass of the SU(2) gauge theory in the lattice unit, respectively.
Dimensionful quantities are
$\lambda_1= \hat{\lambda}_1/a$,
$V=a^4\hat{V}$, $\Sigma = \hat{\Sigma}/a^3$, and $m_f= \hat{m}_f/a$,
where $a$ is the lattice spacing.
The distribution of the smallest eigenvalue
$p_1^{\mathrm{latt.}}(\hat{\lambda}_1; \hat{m}_f)$
is determined from lattice simulation and its normalization is fixed
by
\begin{equation}
 \int_0^\infty d\hat{\lambda}\, p_1^{\mathrm{latt.}}(\hat{\lambda};\hat{m}_f)=1.
\end{equation}

As the sole undetermined quantity in eq.~(\ref{eq:rmt-lattice}) is the chiral condensate,
we can use this relation to best-fit the value of $\hat{\Sigma}$.
If the fit does not work, that is, if eq.~(\ref{eq:rmt-lattice})
is not numerically satisfied
by any choice of $\hat{\Sigma}$,
it implies that the chiral symmetry is restored
and the RM description is not applicable.
Note that $\zeta_1$ and $\mu$ are dimensionless
so they are directly related to quantities in the lattice unit:
$\zeta_1=\lambda_1 V \Sigma = \hat{\lambda}_1 \hat{V}\hat{\Sigma}$
and $\mu=m_f V \Sigma = \hat{m}_f \hat{V} \hat{\Sigma}$.
An integrated version of eq.~(\ref{eq:rmt-lattice}) is
\begin{align}
F_1(s)
 =\int_0^s d \zeta_1  p_1^{\mathrm{RMT}}(\zeta_1; \mu)
  \Big|_{\zeta_1=\lambda_1 V\Sigma, \mu=m_f V\Sigma}
 = 
  \int_0^{\hat{s}} d\hat{\lambda}_1 p_1^{\mathrm{latt.}}(\hat{\lambda}_1; \hat{m}_f)
 \quad (\equiv I(\hat{s}) \ ),
 \label{eq:integrated-dist}
\end{align}
where $s=\hat{s} \hat{V} \hat{\Sigma}$.
We use $I(\hat{s})$ in the fitting process.

Our lattice setting is the following.
We have three different lattice sizes, $(T/a) \times (L/a)^3= 8\times 8^3$,
$12 \times 12^3$ and $16 \times 16^3$.
In this paper, we use fermion mass $\hat{m}_f= am_f=0.010$.
We use several values of the bare gauge coupling $\beta=4/g^2$, for which we use
$\beta=1.1$--$1.475$.
These values are almost the same ones as used in \cite{Huang:2015vkr}.
See Table \ref{tab:lattice_result} in Appendix~\ref{app:lattice_data} for the details of the lattice data.
The topological charge $\nu$ is calculated with the APE smeared
\cite{Albanese:1987ds}
configuration with order-$a$ improved (i.e., ``clover'')
field strength.
Note that this gluonic definition does not give an integer value on a lattice.
The obtained values, however, cluster around integer values so that we can
identify configurations with $\nu=0$.
Eigenvalues and topological charges are calculated for every 10 trajectories.  

The details of our fitting procedure is as follows:
We divide a given lattice eigenvalue distribution into
$N_{\mathrm{bin}}=25$ bins,
whose support covers from 0 to 1.3 times the largest value in
the distribution.
In addition to the average value and the error in each bin,
we estimate the correlation matrix $C$ between bins
by using the jackknife method.
Since a naive estimation of the correlation matrix causes unstable fitting,
we use an improved estimation of the inverse, $C^{-1}_{\mathrm{imp.}}$.  
See appendix \ref{app:correlation_matrix} for the details.
The value of the chiral condensate $\hat{\Sigma}$ is determined
by minimizing the correlated $\chi$ squared:
\begin{equation}
 \chi^2(\hat{\Sigma})
  = \sum_{i,j=1}^{N_{\mathrm{bin}}}
  \left[I(\hat{s}_i) - I^{\mathrm{RMT}}(s_i;\hat{\Sigma})\right]
  \left(C_{\mathrm{imp}}^{-1} \right)_{ij}
  \left[I(\hat{s}_j) - I^{\mathrm{RMT}}(s_j;\hat{\Sigma})\right],
\end{equation}
where
\begin{equation}
 I^{\mathrm{RMT}}(s_i;\hat{\Sigma})
  =F_1(s_i),\qquad
  \text{with}\qquad
  s_i=\hat{s_i}\hat{V}\hat{\Sigma},
  \qquad
  \mu=\hat{V}\hat{\Sigma}\hat{m}_f.
\end{equation}
To estimate $p_1^{\mathrm{RMT}}(\zeta_1,\mu)$ with arbitrary
$\zeta_1$ and $\mu$,
which is needed to calculate $ I^{\mathrm{RMT}}(s_i;\hat{\Sigma})$
for a given $\hat{\Sigma}$,
we use interpolations in both $\zeta_1$ and $\mu$.
We first interpolate in $\mu$ and then in $\zeta_1$, 
with the 4-point interpolation is used for both.
Near the boundary of the available points where the 4-point interpolation
is not possible,
an interpolation with 3 points or an
extrapolation with 2 points is used as well.

\begin{figure}[htbp]
\hspace*{0cm}
 \begin{minipage}{0.50\hsize}
  \begin{center}
 \includegraphics[bb=0 0 360 223,width=100mm]{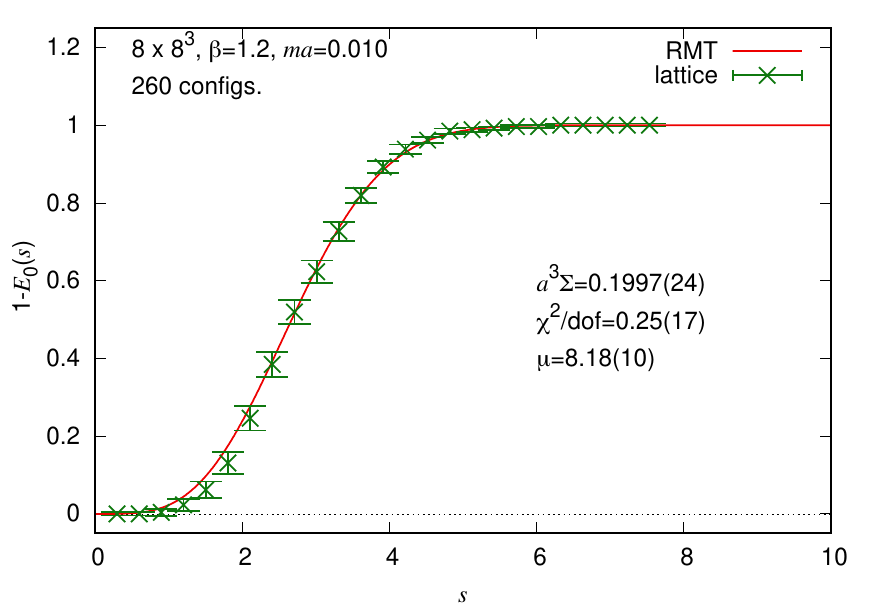}
  \end{center}
 \end{minipage}
 \begin{minipage}{0.50\hsize}
 \begin{center}
 \includegraphics[bb=0 0 360 223,width=100mm]{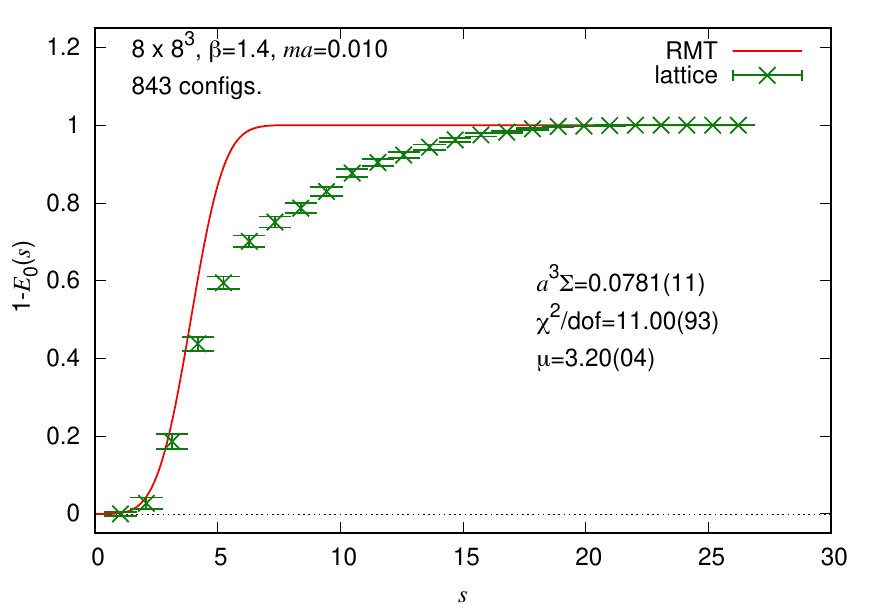}
 \end{center}
 \end{minipage}
 \caption{\label{fig:fit_examples}Typical example of a good fit (left) and a bad fit (right).
The horizontal scale for the RMT curve is determined by the the best
 value of the chiral condensate, which is denoted in the plot.
 }
\end{figure}

Fig.~\ref{fig:fit_examples} is a typical example of
a good fit (indicating the chirally broken phase)
and
a bad fit (chirally symmetric phase).
In the broken phase, the RMT well describes the smallest eigenvalue
distribution from the lattice data, with a reasonably small value of $\chi$ squared.
On the other hand, in the broken phase, the RMT curve can by no means describe
the lattice data.
In the figure, the plotted curve is the result with the best
value of $\hat{\Sigma}=a^3\Sigma$.
The value of $\chi$ squared, however,  indicates
that the quality of the fit is poor in the right panel
and the RMT result is rejected as fitting ansatz.

\begin{figure}[htbp]
\begin{center}
\hspace{3cm}
 \includegraphics[bb=0 0 360 223,width=120mm]{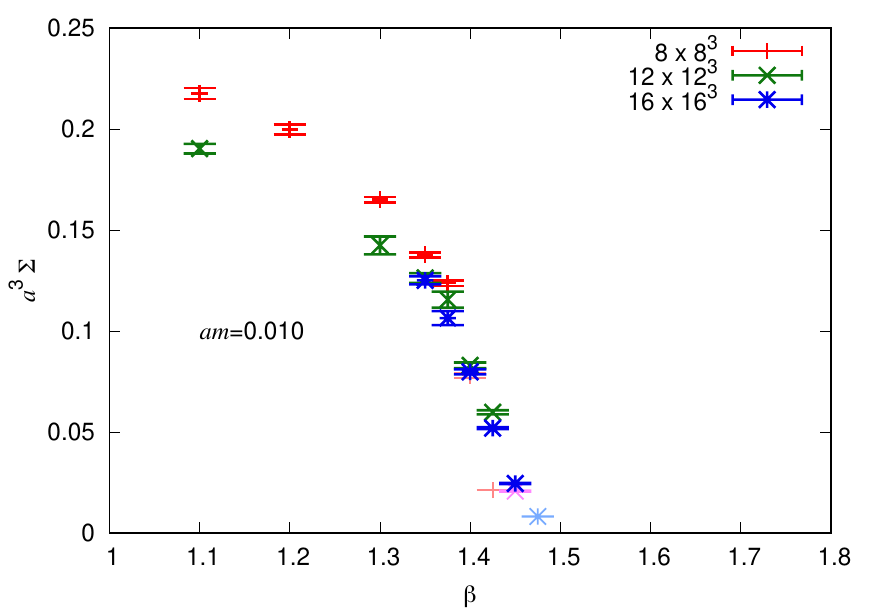}
 \end{center}
 \caption{ \label{fig:sigma_vs_beta}Chiral condensate versus bare coupling $\beta=4/g_0^2$. The
 pale colored symbols have poor values of $\chi^2$/d.o.f value ($>1$).}
\end{figure}

It is interesting to note that even though the fit result is unreliable in the
symmetric phase, the obtained value of the chiral condensate is small and
consistent with zero, as should be in the symmetric phase.
This is clearly seen in Fig.~\ref{fig:sigma_vs_beta}.
We observe that the larger the bare coupling $\beta=4/g^2$ is,
the smaller the obtained chiral condensate becomes and eventually the
fit becomes unreliable near the vanishing of the chiral condensate at around
$\beta=1.4$--$1.5$.
In this Figure, the unreliable data points, for which $\chi^2$ par degrees
of freedom exceeds 1, are plotted with pale colored symbols.
Such behavior is also reported in \cite{Huang:2015vkr}, where
the HMC with $N=400$ is used to obtain the RMT result.
\vspace{-0cm}

\section{Conclusions and discussions}
We have numerically evaluated the $k^{\text{\tiny th}}$ smallest eigenvalue distributions of
chiral random matrix ensembles with multiple flavors using the Nystr\"om-type method applied to the Fredholm determinant and Pfaffian describing the Janossy densities.
Adopting the compact determinant formulas (\ref{eq:det_formula}) and (\ref{eq:Pfaffian_formula}) for the Fredholm determinant for the Janossy densities, we performed numerical computations for the chiral GUE and GSE
in the asymptotic limit (\ref{eq:asymptotic_limit}).
One of our goals of these analyses is an application to the two-color QCD with $N_F$ fundamental staggered flavors.
For the system of $N_F=8$ flavors in the fundamental representation of SU(2),  
the distribution of eigenvalues of the Dirac operators is being
studied through the lattice simulation \cite{Huang:2015vkr}.

In the simulation we used, 
the taste symmetry of the staggered fermions is completely broken
due to the finite lattice spacing, so
that the remaining flavor symmetry is merely $N_F=2$.  In addition to this
flavor symmetry, 
due to the pseudo-reality of the fundamental representation of the SU(2) gauge group,
all the eigenvalues of the Dirac operator are doubly degenerated.
As a result, the distribution of the Dirac eigenvalues can fit with the chiral GSE with quadruply degenerated masses $N_F=4$ 
in the broken phase $\Sigma\ne 0$.

As shown in the left panel of Fig.~\ref{fig:fit_examples},
we observed that the fitting with the chiral GSE works out very nicely in the broken phase.
As the bare coupling $\beta=4/g^2$ grows the chiral condensate
becomes smaller and eventually the fitting becomes unreliable at around
$\beta=1.4$--$1.5$ (Fig.~\ref{fig:sigma_vs_beta}).
This implies that the chiral condensate vanishes
and the symmetry is restored at $\beta \gtrsim 1.45$.
A detailed analysis with more lattice data is currently ongoing \cite{lattice:full}.
We note that even with large
values of the scaled quark masses $\mu_f$,
fitting with the quenched chiral GSE is valid as long as the magnitude of the eigenvalue
is much smaller than $F_\pi^2/(\Sigma\sqrt{V})$. 
Although the value of $F_\pi$, the pion decay constant, is not available from the
current lattice data,
it is natural to assume that the smallest of the Dirac eigenvalues satisfies this
condition in the broken phase.
The Banks-Casher relation tells us that the smallest eigenvalue is small
enough to give a non-zero eigenvalue density around the origin.

Finally we will list some directions for the future research.
Firstly, the numerical computations developed in this article could also be applied to the two-color QCD with $N_F=8$--$12$ fundamental flavors. Among such systems, the existence of the conformal window is strongly expected, and
the technology of their lattice simulations is developping remarkably in recent years.
We anticipate that the RMT analysis of the spectral statistics of the Dirac operators
would discriminate the (near-)conformality of the QCD-like systems and unveil some novel aspects of the conformal window.

The Fredholm Pfaffian for the Janossy density of the chiral Gaussian orthogonal ensemble will deserve 
a future study direction;
the chiral GOE describes the distributions of the Dirac eigenvalues for QCD-like systems with staggered fermions in adjoint representation of SU($N_c$) \cite{Verbaarschot:1994qf}. It is known that the exponential convergence of the Nystr\"om-type discretization of the Fredholm Pfaffian for the orthogonal ensemble is not guaranteed due to the infinite oscillations
originating from the discontinuity of the quaternion kernel elements. 
Even though such hard problem resides, we may still be able to apply the Nystr\"om-type discretization for the practical purpose if the error can be estimated appropriately, and use it to estimate the value of the chiral condensate $\Sigma$ for the adjoint QCD-like system.\footnote{
The analytical computations of the smallest eigenvalue distribution ($k=1$) for the chiral GOE are found in \cite{Akemann:2014cna,Wirtz:2015oma}.
}

In \cite{FW,WBF}, an exact analysis of the Janossy density for the unitary ensemble is done on a basis of the Painlev\'e  II
transcendent and its associated isomonodromic system.
Generalization of such an exact analysis to the symplectic and orthogonal ensembles could be an interesting direction
yet to be studied, and it can be compared with our numerical results.

Recent years, the $(0+1)$-dimensional fermionic model with all-to-all random interactions referred to  as Sachdev-Ye-Kitaev (SYK) model \cite{Sachdev:1992fk,Kitaev} has been studied very actively in the context of the nonequilibrium quantum many-body systems and its application to the AdS/CFT correspondence (see references in a recent review article \cite{Rosenhaus:2018dtp}).  The level statistics of the SYK model was numerically examined, 
and good agreements with the RMT have been observed.
It would be interesting to explore how the Fredholm determinant or Pfaffian expression for the Janossy density
of the chiral random matrices
appears in the level statistics of the supersymmetric SYK Hamiltonian
\cite{Fu:2016,Garcia:2018}.

\acknowledgments{
The authors would like to thank Peter J. Forrester, C. -J. David Lin, and Taro Nagao for fruitful discussions and valuable comments.

The authors thank Asia Pacific Center for Theoretical Physics (APCTP)
where this work was initiated during the Workshop 2017 on 
``Discrete Approaches to the Dynamics of Fields and Space-Time''. 
The authors also thank the Yukawa Institute for Theoretical Physics (YITP) at Kyoto University, for the 
discussions during the Workshop YITP-S-17-04 on ``The 40th Shikoku Seminar on Particle and Nuclear Physics''
which was useful to complete this work. 

The research of H.F. is supported by the
Grant-in-Aid for Challenging Research (Exploratory) [\# 17K18781],
Grant-in-Aid for Scientific Research(C)  [\# 17K05239], [\# 18K03281], and 
Grant-in-Aid for Scientific Research(B)  [\# 16H03927]  from the Japan Ministry of Education, Culture, Sports, Science and Technology.
The research of I.K. is supported by MEXT as ``Priority Issue 9 to be Tackled by Using Post-K Computer'' 
(Elucidation of the Fundamental Laws and Evolution of the Universe) and JICFuS.
The research of S.M.N. is supported by
Grants-in-Aid for Scientific Research(C)  [\# 25400259] and [\# 17K05416].
National Centre for High-performance Computing and
National Chiao-Tung University (NCTU) HPC facility in Taiwan were
used for lattice simulation.
}

\appendix
\section{Quaternionic kernels for the chiral Gaussian symplectic ensemble}\label{app:kernel}
In this appendix, the explicit forms of the spectral kernel for the chiral GSE are summarized 
for quadruply degenerated masses $N_F=4\alpha$ and doubly degenerated masses $N_F=2\alpha$.

\subsection{Spectral kernel for quadruply degenerated masses}\label{app:kernel_quad}
The scaled correlation function of the $\beta=4$ chiral RMT with quadruply degenerated masses $N_F=4\alpha$ in the scaling limit (\ref{eq:asymptotic_limit})
is given in \cite{Nagao:2000cb}.
\begin{align}
&R_{\beta=4,\nu}^{(p)}(\zeta_1,\ldots,\zeta_p;\{\mu_a\})
=\frac{\mathrm{Pf}\left[Z\left(
\begin{array}{cc}
\left[K_{++}(\zeta_i,\zeta_j)\right]_{i,j=1,\ldots,p} & 
\left[K_{+-}(\mu_a,\zeta_j)\right]_{\substack{a=1,\ldots,\alpha\\j=1,\ldots,p}}
\\
\left[K_{-+}(\zeta_i,\mu_b)\right]_{\substack{i=1,\ldots,p\\b=1,\ldots,\alpha}}
&
\left[K_{--}(\mu_a,\mu_b)\right]_{a,b=1,\ldots,\alpha}
\end{array}
\right)\right]}{\mathrm{Pf}\left[Z\left[K_{--}(\mu_a,\mu_b)\right]_{a,b=1,\ldots,\alpha}\right]},\\
&K_{AB}(\zeta,\zeta')=\left[
\begin{array}{cc}
-S_{AB}(\zeta,\zeta') &-I_{AB}(\zeta,\zeta') \\
D_{AB}(\zeta,\zeta')  & -S_{BA}(\zeta',\zeta) 
\end{array}
\right],\quad Z=\mathrm{i}\sigma_2\otimes\mathbb{I},
\nonumber
\end{align}
where elements of block matrices are
\begin{align}
&S_{++}(\zeta,\zeta')=
2\int_0^1du\int_{0}^{1}dv\sqrt{\zeta\zeta'}\zeta v^2
\left(
J_{2\nu}(2v\zeta)uJ_{2\nu+1}(2uv\zeta')-J_{2\nu}(2uv\zeta)J_{2\nu+1}(2v\zeta')
\right),\nonumber \\
&S_{+-}(\zeta,\eta')=
(-1)^{\nu+1}2\int_0^1du\int_{0}^{1}dv\sqrt{\zeta\eta'}\zeta v^2
\left(
J_{2\nu}(2v\zeta)uI_{2\nu+1}(2uv\eta')-J_{2\nu}(2uv\zeta)I_{2\nu+1}(2v\eta')
\right),\nonumber \\
&S_{-+}(\eta,\zeta')=
(-1)^{\nu+1}2\int_0^1du\int_{0}^{1}dv\sqrt{\eta\zeta'}\eta v^2
\left(
I_{2\nu}(2v\eta)uJ_{2\nu+1}(2uv\zeta')-I_{2\nu}(2uv\eta)J_{2\nu+1}(2v\zeta')
\right),\nonumber \\
&S_{--}(\eta,\eta')=
(-1)^{2\nu}2\int_0^1du\int_{0}^{1}dv\sqrt{\eta\eta'}\eta v^2
\left(
I_{2\nu}(2v\eta)uI_{2\nu+1}(2uv\eta')-I_{2\nu}(2uv\eta)I_{2\nu+1}(2v\eta')
\right),\label{eq:S}
\\
&D_{++}(\zeta,\zeta')=2\int_0^1du\int_{0}^{1}dv\sqrt{\zeta\zeta'}
v^3u\left(
J_{2\nu+1}(2v\zeta)J_{2\nu+1}(2uv\zeta')-J_{2\nu+1}(2uv\zeta)J_{2\nu+1}(2v\zeta')
\right), \nonumber \\
&D_{+-}(\zeta,\eta')=(-1)^{\nu+1}2\int_0^1du\int_{0}^{1}dv\sqrt{\zeta\eta'}
v^3u\left(
J_{2\nu+1}(2v\zeta)I_{2\nu+1}(2uv\eta')-J_{2\nu+1}(2uv\zeta)I_{2\nu+1}(2v\eta')
\right), \nonumber \\
&D_{-+}(\eta,\zeta')=-D_{+-}(\zeta',\eta),\nonumber \\
&D_{--}(\eta,\eta')=(-1)^{2\nu}2\int_0^1du\int_{0}^{1}dv\sqrt{\eta\eta'}
v^3u\left(
I_{2\nu+1}(2v\eta)I_{2\nu+1}(2uv\eta')-I_{2\nu+1}(2uv\eta)I_{2\nu+1}(2v\eta')
\right), \label{eq:D}
\end{align}
\begin{align}
&I_{++}(\zeta,\zeta')=2\int_0^1du\int_{0}^{1}dv\sqrt{\zeta\zeta'}
\zeta\zeta'v\left(
J_{2\nu}(2v\zeta)J_{2\nu}(2uv\zeta')-J_{2\nu}(2uv\zeta)J_{2\nu}(2v\zeta')
\right),\nonumber \\
&I_{+-}(\zeta,\eta')=(-1)^{\nu+1}2\int_0^1du\int_{0}^{1}dv\sqrt{\zeta\eta'}
\zeta\eta'v\left(
J_{2\nu}(2v\zeta)I_{2\nu}(2uv\eta')-J_{2\nu}(2uv\zeta)I_{2\nu}(2v\eta')
\right),\nonumber \\
&I_{-+}(\zeta,\eta')=-I_{+-}(\eta',\zeta),\nonumber \\
&I_{--}(\eta,\eta')=(-1)^{2\nu}2\int_0^1du\int_{0}^{1}dv\sqrt{\eta\eta'}
\eta\eta'v\left(
I_{2\nu}(2v\eta)I_{2\nu}(2uv\zeta')-I_{2\nu}(2uv\eta)I_{2\nu}(2v\eta')
\right).\label{eq:I}
\end{align}

\subsection{Spectral kernel for doubly degenerated masses}\label{app:kernel_odd}
The $p$-level correlation function for 
 $\beta=4$ chiral RMT with $N_F=2\alpha$ doubly degenerated masses in the scaling limit (\ref{eq:asymptotic_limit})
is given in \cite{Nagao:2000cb}.
\begin{align}
R_{\beta=4,\nu}^{(p)}(\zeta_1,\ldots,\zeta_p;\{\mu_a\})
=\frac{\mathrm{Pf}\left[ZK^{(p)}\right]}{\mathrm{Pf}\left[ZK^{(0)}\right]}.
\end{align}
For even $\alpha$, the kernel $ZK^{(p)}$ is given as follows:
\begin{align}
&ZK^{(p)}=
\left(
\begin{array}{ccc}
[I_{--}(\mu_a,\mu_b)]_{a,b=1,\ldots,\alpha} 
& 
[I_{-+}(\mu_a,\zeta_i)]_{
\substack{a=1,\ldots,\alpha \\
j=1,\ldots,p}} 
& 
\left[S_{-+}(\mu_a,\zeta_j)\right]_{
\substack{a=1,\ldots,\alpha\\
j=1,\ldots,p}} 
\\
-[I_{-+}^{\mathrm{T}}(\mu_b,\zeta_i)]_{
\substack{b=1,\ldots,\alpha \\
i=1,\ldots,p}} &
[I_{++}(\zeta_i,\zeta_j)]_{i,j=1,\ldots,p} 
&
[S_{++}(\zeta_i,\zeta_j)]_{i,j=1,\ldots,p} 
\\
-[S_{-+}^{\mathrm{T}}(\mu_b,\zeta_i)]_{
\substack{b=1,\ldots,\alpha \\
i=1,\ldots,p}}
& 
-[S^{\mathrm{T}}_{++}(\zeta_j,\zeta_i)]_{i,j=1,\ldots,p}  
&
[D_{++}(\zeta_i,\zeta_j)]_{i,j=1,\ldots,p}
\end{array}
\right),
\label{eq:even_R}
\end{align}
where $S_{AB}$'s, $D_{AB}$'s, and $I_{AB}$'s are the same as $N_F=4\alpha$ in Appendix \ref{app:kernel_quad}, and T stands for the transposition of the block matrix.

For odd $\alpha$, the kernel $ZK^{(p)}$ is given as follows:
\begin{align}
&ZK^{(p)}
\label{eq:odd_R}
\\
&
=\left(
\begin{array}{cccc}
[I_{--}(\mu_a,\mu_b)]_{a,b=1,\ldots,\alpha} 
&
[Q_{-}(\mu_a)]_{a=1,\ldots,\alpha}
&
[I_{-+}(\mu_a,\zeta_j)]_{
\substack{a=1,\ldots,\alpha \\
j=1,\ldots,M}}
&
\left[S_{-+}(\mu_a,\zeta_j)\right]_{
\substack{a=1,\ldots,\alpha\\
j=1,\ldots,p}} 
\\
-[Q_{-}^{\mathrm{T}}(\mu_b)]_{b=1,\ldots,\alpha}
&
0
&
-[Q_{+}^{\mathrm{T}}(\zeta_j)]_{j=1,\ldots,p}
&
-[P^{\mathrm{T}}_+(\zeta_j)]_{j=1,\ldots,p} 
\\
-[I^{\mathrm{T}}_{-+}(\mu_b,\zeta_i)]_{
\substack{i=1,\ldots,p \\
b=1,\ldots,\alpha}}
&
[Q_{+}(\zeta_i)]_{i=1,\ldots,p}
&
[I_{++}(\zeta_i,\zeta_j)]_{i,j=1,\ldots,p}
&
[S_{++}(\zeta_i,\zeta_j)]_{i,j=1,\ldots,p} 
\\
-[S_{-+}^{\mathrm{T}}(\mu_b,\zeta_i)]_{
\substack{i=1,\ldots,p \\ b=1,\ldots,\alpha 
}}
&
\left[P_{+}(\zeta_i)\right]_{i=1,\ldots,p}
&
-[S^{\mathrm{T}}_{++}(\zeta_j,\zeta_i)]_{i,j=1,\ldots,p}
&
[D_{++}(\zeta_i,\zeta_j)]_{i,j=1,\ldots,p}
\end{array}
\right).
\nonumber
\end{align}
where $S_{AB}$'s, $D_{AB}$'s, and $I_{AB}$'s are the same as $N_F=4\alpha$ in Appendix \ref{app:kernel_quad}, and
\begin{align}
&Q_+(\zeta)=2\sqrt{\zeta}\zeta\int_{0}^1dv\,J_{2\nu}(2v\zeta), \quad
Q_-(\eta)=(-1)^{\nu+1}2\sqrt{\eta}\eta\int_{0}^1dv\,I_{2\nu}(2v\eta), \label{eq:Q}\\
&P_+(\zeta)=2\sqrt{\zeta}\int_{0}^1dv\,vJ_{2\nu+1}(2v\zeta).\label{eq:P}
\end{align}

\section{Confluent limits of the correlation function}\label{app:confluent}
\subsection{Chiral Gaussian unitary ensemble}\label{app:confluent_GUE}
Let $Z_{\beta=2,\nu}(x_1,\ldots,x_n)$ be the partition function which is obtained as the scaling limit (\ref{eq:asymptotic_limit})
of the chiral Gaussian unitary ensemble with $N_f=2n$ mass parameters $x_a=m_a/\Delta$
 \cite{Damgaard:1997ye,Wilke:1997gf,Nishigaki:1998is}.
\begin{align}
Z_{\beta=2,\nu}(x_1,\ldots,x_n)=\frac{\det\left[x_a^{b-1}I_{\nu+b-1}(x_a)\right]_{a,b=1}^n}{\prod_{a>b}(x_a^2-x_b^2)}.
\end{align}
To consider the confluent limit $x_i\to x_j$ of this partition function \cite{Leutwyler:1992yt}, we will use the l' H\^opital's rule given as follows. 

Let  $f,g$ be differentiable functions  on an interval $I\in\mathbb{R}$. Assume that for $c\in I$, 
(1) $\lim_{x\to c}f(x)=\lim_{x\to c}g(x)=0\, \mathrm{or}\, \infty$, (2)  $\lim_{x\to c}f'(x)/g'(x)$ exists, (3) $g'(x)\neq 0$ for $x\in I\setminus \{c\}$,
then
\begin{align}
\lim_{x\to c}\frac{f(x)}{g(x)}=\lim_{x\to c}\frac{f'(x)}{g'(x)}.
\label{lhopital}
\end{align}

One finds that the confluent limit $x_i\to x_1=x$ ($i=1,\ldots,n$) of the partition function $Z_{\beta=2,\nu}(x_1,\ldots,x_n)$
by adopting the l' H\^opital's rule (\ref{lhopital}) repeatedly.
\begin{align}
&\lim_{x_{n}\to x_1}\lim_{x_{n-1}\to x_1}\cdots \lim_{x_2\to x_1}Z_{\beta=2,\nu}(x_1,\ldots,x_n)
\nonumber \\
&=\lim_{x_{n}\to x_1}\lim_{x_{n-1}\to x_1}\cdots \lim_{x_2\to x_1}\frac{
\left|\begin{array}{cccc}
I_0(x_1) & x_1I_1(x_1) &\cdots & x_{1}^{n-1}I_{n-1}(x_1) \\
I_0(x_2) & x_2I_1(x_2) &\cdots & x_{2}^{n-1}I_{n-1}(x_2) \\
\vdots & \vdots & \vdots &\vdots  \\
I_0(x_{n}) & x_nI_1(x_n) &\cdots & x_{n}^{n-1}I_{n-1}(x_n) 
\end{array}
\right|
}{(x_1^2-x_2^2)(x_1^2-x_2^2)\cdots (x_{n-1}^2-x_n^2)}
\nonumber \\
&=\left(
\frac{\frac{1}{2}\frac{1}{2^2}\frac{1}{2^3}\cdots \frac{1}{2^{n-1}}}{1! 2! 3!\cdots (n-1)!}
\right)\left|\begin{array}{cccc}
I_0(x_1) & x_1I_1(x_1) &\cdots & x_{1}^{n-1}I_{n-1}(x_1) \\
x_1^{-1}I_{-1}(x_1) & I_0(x_1) &\cdots & x_{1}^{n-2}I_{n-2}(x_1) \\
x_1^{-2}I_{-2}(x_1) & x_1^{-1}I_{-1}(x_1) &\cdots & x_{1}^{n-3}I_{n-3}(x_1) \\
\vdots & \vdots & \vdots &\vdots  \\
x_1^{-(n-1)}I_{-(n-1)}(x_{1}) & x_1^{-(n-2)}I_{-(n-2)}(x_1) &\cdots & x_{1}^{0}I_{0}(x_1) 
\end{array}
\right|
\nonumber \\
&=\frac{\frac{1}{2}\frac{1}{2^2}\frac{1}{2^3}\cdots \frac{1}{2^{n-1}}}{1! 2! 3!\cdots (n-1)!}
\det\left[x_1^{a-b}I_{a-b}(x)\right]_{a,b=1}^n,
\end{align}
where the following formula of the Bessel function is adopted
\begin{align}
2\frac{\partial}{\partial x^2}\left(x^kI_k(x)\right)=x^{k-1}I_{k-1}(x).
\label{bessel_formula}
\end{align}

Next we will consider the scalar kernel $K_s(\zeta,\zeta',\mu_1,\mu_2,\ldots,\mu_{\alpha})$ for the  chiral GUE with $\nu=0$ \cite{Damgaard:1997ye,Wilke:1997gf,Nishigaki:1998is}.
\begin{align}
&K_s(\zeta,\zeta',\mu_1,\mu_2,\ldots,\mu_{\alpha})
\nonumber \\
&=
\frac{\sqrt{\zeta\zeta'}}{(\zeta^2-\zeta^{'2})\prod_{k=1}^{\alpha}\sqrt{(\zeta^2+\mu_k^2)(\zeta^{'2}+\mu_k^2)}}
\frac{\left|
\begin{array}{cccc}
J_{0}(\zeta) & \zeta J_1(\zeta) & \cdots &\zeta^{\alpha+1}J_{\alpha+1}(\zeta) \\
J_{0}(\zeta') & \zeta J_1(\zeta') & \cdots &\zeta^{'\alpha+1}J_{\alpha+1}(\zeta') \\
I_{0}(\mu_1) & \zeta (-\mu_1)I_1(\mu_1) & \cdots &\mu_1^{\alpha+1}J_{\alpha+1}(\mu_1) \\
\vdots & \vdots & \cdots &\vdots \\
I_{0}(\mu_{\alpha}) & \zeta (-\mu_{\alpha})I_1(\mu_{\alpha}) & \cdots &\mu_{\alpha}^{\alpha+1}J_{\alpha+1}(\mu_{\alpha}) 
\end{array}
\right|}{\det\left[(-\mu_k)^{\ell}I_{\ell}(\mu_k)\right]_{k,\ell=1}^{\alpha}}.
\label{eq:massive_correlator}
\end{align}
The confluent limit of $K_s(\zeta,\zeta',\mu_1,\mu_2,\ldots,\mu_{\alpha})$
is also obtained in the same way as the partition function $Z_{\beta=2,\nu}(x_1,\ldots,x_n)$ considered above.
For our notational convenience, we introduce 
\begin{align}
A_k(x)=x^kI_k(x),\quad B_k(x)=(-x)^kJ_k(x),
\end{align}
and $A_k$'s obey
\begin{align}
\frac{d}{dx^2}A_k(x)=\frac{1}{2}A_{k-1}.
\end{align}

Adopting such notation, one can express the complete confluent limit  ($\mu_i\to \mu_1=\mu$ for $i=2,\ldots, \alpha$) 
of the determinant factor in the scalar kernel $K_s(\zeta,\zeta',\mu_1,\mu_2,\ldots,\mu_{\alpha})$.
\begin{align}
\lim_{\mu_2\to \mu_1}\cdots \lim_{\mu_{\alpha}\to \mu_1}
\frac{\left|
\begin{array}{cccc}
B_{0}(\zeta) & B_1(\zeta) & \cdots &B_{\alpha+1}(\zeta) \\
B_{0}(\zeta') & B_1(\zeta') & \cdots &B_{\alpha+1}(\zeta') \\
A_{0}(\mu_1) & A_1(\mu_1) & \cdots &A_{\alpha+1}(\mu_1) \\
A_{0}(\mu_2) & A_1(\mu_2) & \cdots &A_{\alpha+1}(\mu_2) \\
\vdots & \vdots & \cdots &\vdots \\
A_{0}(\mu_{\alpha}) &A_1(\mu_{\alpha}) & \cdots & A_{\alpha+1}(\mu_{\alpha}) 
\end{array}
\right|}{\det\left[A_{\ell}(\mu_k)\right]_{k,\ell=1}^{\alpha}}
=\frac{\left|
\begin{array}{cccc}
B_{0}(\zeta) & B_1(\zeta) & \cdots &B_{\alpha+1}(\zeta) \\
B_{0}(\zeta') & B_1(\zeta') & \cdots &B_{\alpha+1}(\zeta') \\
A_{0}(\mu) & A_1(\mu) & \cdots &A_{\alpha+1}(\mu) \\
A_{-1}(\mu) & A_0(\mu) & \cdots &A_{\alpha}(\mu) \\
\vdots & \vdots & \cdots &\vdots \\
A_{-\alpha+1}(\mu) &A_{-\alpha+2}(\mu) & \cdots & A_{0}(\mu) 
\end{array}
\right|}{\det\left[A_{k-\ell}(\mu)\right]_{k,\ell=1}^{\alpha}}.
\end{align}
Completed by the confluent limit of remaining factors in $K_s(\zeta,\zeta',\mu_1,\mu_2,\ldots,\mu_{\alpha})$, one obtains the confluent limit of the spectral kernel for the chiral GUE.

\subsection{Chiral Gaussian symplectic ensemble}\label{app:confluent_GSE}
The partition function $Z_{\beta=4,\nu}(\{\mu_a\})$
for the scaling limit (\ref{eq:asymptotic_limit}) of the chiral GSE ($\beta=4$) with $2\alpha$ ($\alpha$: even) flavors of the doubly degenerated masses
is given as follows \cite{Nagao:2000cb}.
\begin{align}
Z_{\beta=4,\nu}(\{\mu_a\})=c_{\alpha}\left(\prod_{i=1}^{\alpha}\mu_i^{2\nu}\right)\frac{\mathrm{Pf}(Zf)}{\Delta(\mu_1^2,\ldots,\mu_{\alpha}^2)},
\end{align}
where 
\begin{align}
&c_{\alpha}=(-1)^{\frac{\alpha(\alpha+1)}{2}}\prod_{k=0}^{\alpha-1}(2k+1)!,
\quad \Delta(\mu_1^2,\ldots,\mu_{\alpha}^2)=\prod_{i>j}(\mu_i^2-\mu_j^2). \nonumber\\
&f_{ij}=f(\mu_i,\mu_j)=\int_{0}^{1}dt\,t\frac{I_{2\nu}(2t\mu_i)}{\mu_i^{\nu}}\int_0^1du\frac{I_{2\nu}(2tu\mu_j)}{\mu_j^{2\nu}}-(i\leftrightarrow j).
\end{align}
The complete confluent limit  $\mu_i\to \mu_1=\mu$ of the partition function $Z_{\beta=4,\nu}(\mu_1^{\otimes 2},\ldots,\mu_n^{\otimes 2})$ yields
\begin{align}
\lim_{\mu_{2,\ldots,n}\to\mu_1=\mu}Z_{\beta=4,\nu}(\mu_1^{\otimes 2},\ldots,\mu_n^{\otimes 2})
=\frac{\frac{1}{2}\frac{1}{2^2}\cdots\frac{1}{2^{n-1}}}{1!2!\cdots (n-1)!}c_4\cdot\mathrm{Pf}\left(Z[f^{(i,j)}(\mu,\mu)]_{i,j=0}^{n-1}\right),
\end{align}
where 
\begin{align}
f^{(k,\ell)}(\mu_i,\mu_j)=\left(\frac{\partial}{\partial \mu_i^2}\right)^k\left(\frac{\partial}{\partial \mu_j^2}\right)^{\ell}f(\mu_i,\mu_j).
\end{align}

The complete confluent limit  ($\mu_i\to \mu_1=\mu$ for $i=2,\ldots, \alpha$)  of the correlation function (\ref{eq:even_R}) 
in \cite{Nagao:2000cb}
is also obtained in the same way.
In the limit $\mu_{2,\ldots,n}\to\mu_1=\mu$, matrix elements in eq.~(\ref{eq:even_R}) are replaced in the following way. (For simplicity, we consider the case of $\nu=0$.)
\begin{align}
I_{--}(\mu_i,\mu_j)\,\rightarrow\,&I_{--}^{(i-1,j-1)}(\mu,\mu)\nonumber \\
&=\mu^2\int_0^1dt\int_0^1du\,t\Bigl[(2t)^{2(i-1)}(2tu)^{2(j-1)}A_{-i+1}(2t\mu)A_{-j+1}(2tu\mu)
\nonumber \\
&\qquad\qquad\qquad\qquad
-(2t)^{2(j-1)}(2tu)^{2(i-1)}A_{-j+1}(2t\mu)A_{-i+1}(2tu\mu)\Bigr],
\nonumber \\
I_{-+}(\mu_i,\zeta_{\ell})\,\rightarrow\,&I_{-+}^{(i-1,0)}(\mu,\zeta_{\ell})
=\mu\int_0^1dt\int_0^1du\,t\Bigl[
(2t)^{2(i-1)}A_{-i+1}(2t\mu)J_0(2tu\zeta_{\ell})
\nonumber \\
&\qquad\qquad\qquad\qquad\qquad\qquad\qquad
-(2tu)^{2(i-1)}A_{-i+1}(2tu\mu)J_0(2t\zeta_{\ell})
\Bigr],
\nonumber\\
S_{-+}(\mu_i,\zeta_{\ell})\,\rightarrow\,&S_{-+}^{(i-1,0)}(\mu,\zeta_{\ell})
=\mu\int_0^1dt\int_0^1du\,t^2\Bigl[
(2t)^{2(i-1)}uA_{-i+1}(2t\mu)J_1(2tu\zeta_{\ell})
\nonumber \\
&\qquad\qquad\qquad\qquad\qquad\qquad\qquad
-(2tu)^{2(i-1)}A_{-i+1}(2tu\mu)J_1(2t\zeta_{\ell})
\Bigr].
\end{align}
The other matrix elements are kept untouched, because they are not dependent on mass parameters $\mu_i$'s.

Lastly, for the odd $\alpha$ case in addition to the above replacements, the matrix elements $Q_-$'s in eq.~(\ref{eq:odd_R}) are also
replaced by
\begin{align}
Q_{-}(\mu_j)\,\rightarrow\,Q_-^{(j-1)}(\mu)=\frac{\partial^{i-1}}{\partial (\mu^2)^{j-1}}Q_{-}(\mu)
=\int_0^1dv\,(2v)^{2(j-1)-1}\frac{d^{j-1}}{d(x^2)^{j-1}}(xA_0(x))\Big|_{x=2v\mu}.
\end{align}

\newpage
\subsubsection{Quadruply degenerated kernels in the confluent limit}\label{app:quad}
For the chiral GSE with $N_F=4\alpha$ quadruply degenerated masses, we can use the spectral kernel given in Appendix \ref{app:kernel_quad}.
In the complete confluent limit,  some matrix elements in eqs.~(\ref{eq:S})--(\ref{eq:I}) are replaced as follows.
(We also choose $\nu=0$ for simplicity.)
\begin{align}
&S_{+-}(\zeta,\mu_a)
\nonumber \\
&\rightarrow\,S_{+-}^{(0,a-1)}(\zeta,\mu)
\nonumber \\
&=2\int_0^1du\int_{0}^1dv\,\sqrt{\zeta\mu}\zeta v^2\Bigl[
(2uv)^{2(a-1)}uJ_0(2v\zeta)\frac{d^{a-1}}{d(x^2)^{a-1}}(x^{-1}A_{1}(x))\Big|_{x=2uv\mu}
\nonumber \\
&\qquad\qquad\qquad\qquad\qquad\qquad
-(2v)^{2(a-1)}J_0(2uv\zeta)\frac{d^{a-1}}{d(x^2)^{a-1}}(x^{-1}A_{1}(x))\Big|_{x=2v\mu}
\Bigr],
\nonumber\\
&S_{-+}(\mu_a,\zeta)
\nonumber \\
&\rightarrow\,S_{-+}^{(0,a-1)}(\zeta,\mu)
\nonumber \\
&=2\int_0^1du\int_{0}^1dv\,\sqrt{\zeta\mu}\mu v^2\Bigl[
(2v)^{2(a-1)}uA_{-a+1}(2v\mu)J_{1}(2uv\mu)
\nonumber \\
&\qquad\qquad\qquad\qquad\qquad\qquad
-(2uv)^{2(a-1)}A_{-a+1}(2uv\mu)J_{1}(2v\mu)
\Bigr],
\nonumber \\
&S_{--}(\mu_a,\mu_b)
\nonumber \\
&\rightarrow\,S_{--}^{(a-1,b-1)}(\mu,\mu)
\nonumber \\
&
=-2\int_0^1du\int_{0}^1dv\,\mu^2 v^2\Bigl[
(2v)^{2(a-1)}(2uv)^{2(b-1)}uA_{-a+1}(2v\mu)\frac{d^{b-1}}{d(x^2)^{b-1}}(x^{-1}A_{1}(x))\Big|_{x=2uv\mu}
\nonumber \\
&\qquad\qquad\qquad\qquad\qquad
-(2v)^{2(b-1)}(2uv)^{2(a-1)}A_{-a+1}(2uv\mu)\frac{d^{b-1}}{d(x^2)^{b-1}}(x^{-1}A_{1}(x))\Big|_{x=2v\mu}
\Bigr],
\label{eq:S_confl}
\\
&D_{+-}(\zeta,\mu_a)
\nonumber \\
&\rightarrow\,D_{+-}^{(0,a-1)}(\zeta,\mu)
\nonumber \\
&=-2\int_0^1du\int_{0}^1dv\,\sqrt{\zeta\mu} v^3u\Bigl[
(2uv)^{2(a-1)}J_1(2v\zeta)\frac{d^{a-1}}{d(x^2)^{a-1}}(x^{-1}A_{1}(x))\Big|_{x=2uv\mu}
\nonumber \\
&\qquad\qquad\qquad\qquad\qquad\qquad
-(2v)^{2(a-1)}J_1(2uv\zeta)\frac{d^{a-1}}{d(x^2)^{a-1}}(x^{-1}A_{1}(x))\Big|_{x=2v\mu}
\Bigr],\nonumber \\
&D_{-+}(\mu_a,\zeta')\,\rightarrow\,-D_{-+}^{(0,a-1)}(\zeta',\mu),\nonumber \\
&D_{--}(\mu_a,\mu_b)
\nonumber \\
&\rightarrow\,D_{--}^{(a-1,b-1)}(\zeta,\mu)
\nonumber \\
&=2\int_0^1du\int_{0}^1dv\,\mu v^3u\Bigl[
(2v)^{2(a-1)}(2uv)^{2(b-1)}A_{2-a}(2v\mu)\frac{d^{b-1}}{d(x^2)^{b-1}}(x^{-1}A_{1}(x))\Big|_{x=2uv\mu}
\nonumber \\
&\qquad\qquad\qquad\qquad\qquad
-(2uv)^{2(a-1)}(2v)^{2(b-1)}A_{2-a}(2uv\mu)\frac{d^{b-1}}{d(x^2)^{b-1}}(x^{-1}A_{1}(x))\Big|_{x=2v\mu}\Bigr],
\label{eq:D_confl}
\end{align}
\begin{align}
&I_{+-}(\zeta,\mu_a)
\nonumber \\
&\rightarrow\,I_{+-}^{(0,a-1)}(\zeta,\mu)
=2\int_0^1du\int_{0}^1dv\,\sqrt{\zeta\mu}\zeta\mu v\Bigl[
(2uv)^{2(a-1)}J_0(2v\zeta)A_{1-a}(2uv\mu)
\nonumber \\
&\qquad\qquad\qquad\qquad\qquad\qquad\qquad\qquad\qquad\qquad
-(2v)^{2(a-1)}J_0(2uv\zeta)A_{1-a}(2v\mu)
\Bigr],\nonumber \\
&I_{-+}(\mu,\zeta')\,\rightarrow\,-I_{+-}^{(0,a-1)}(\zeta',\mu),\nonumber \\
&I_{--}(\mu_a,\mu_b)
\nonumber \\
&\rightarrow\,I_{--}^{(a-1,b-1)}(\mu,\mu)
=-2\int_0^1du\int_{0}^1dv\,\mu^3 v\Bigl[
(2v)^{2(a-1)}(2uv)^{2(b-1)}A_{1-a}(2v\mu)A_{1-b}(2uv\mu)
\nonumber \\
&\qquad\qquad\qquad\qquad\qquad\qquad\qquad\qquad\qquad
-(2uv)^{2(a-1)}(2v)^{2(b-1)}A_{1-a}(2uv\mu)A_{1-b}(2v\mu)
\Bigr],
\label{eq:I_confl}
\end{align}
where 
\begin{align}
A_k(x)=x^{k}I_k(x),
\quad
2\frac{dI_k(x)}{dx}=x^{-1}I_{k-1}(x)-kx^{-2}I_k(x).
\end{align}

\section{Janossy density} \label{app:Janossy}
\subsection{Janossy density for the determinantal random point process}\label{app:RPP}
\begin{figure}[h] 
\begin{center}
\hspace{1cm}
 \includegraphics[bb=0 0 360 223,width=100mm]{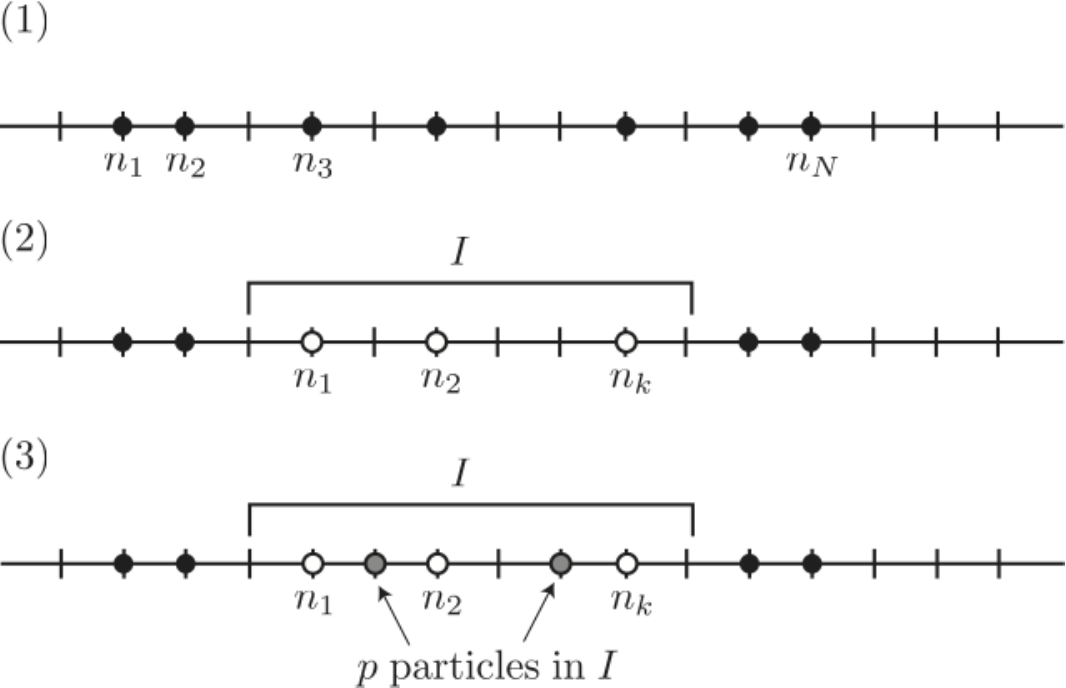}
\caption{\label{fig:Janossy}
Distribution of particles. (1) there is a particle at each of the points $n_i$ ($i=1,\ldots,N$). (2) there are exactly $k$ particles in $I$, one in each of $k$ designated points $n_i$  ($i=1,\ldots,k$). (3)  there are exactly $p$ particles in $I$ except for $k$ designated points $n_i$  ($i=1,\ldots,k$). }
\end{center}
\end{figure}

Below we shall overview the definition of the Janossy density for the determinantal random point process \cite{Soshnikov,Lyons,BHKP}.
Consider an ensemble of $N$ particles on $\mathbb{Z}$ with the joint distribution (see (1) in Fig.\ref{fig:Janossy}) given by
\begin{align}
p(n_1,\ldots,n_N)=\frac{1}{N!}\det\left[K(n_i,n_j)\right]_{i,j=1}^N,\quad n_i\in \mathbb{Z},
\end{align}
with the kernel $\mathbf{K}=\left[K(n,m)\right]_{n,m\in\mathbb{Z}}$ obeying the projective condition:
\begin{align}
\mathbf{K}\cdot \mathbf{K}=\mathbf{K},\quad \mathrm{tr}\,\mathbf{K}=N.
\end{align}
Then the $k$-point function $R_k(n_1,\ldots,n_k)$ is given  by
\begin{align}
R_k(n_1,\ldots,n_k)=\det\left[K(n_i,n_j)\right]_{i,j=1}^k.
\end{align}

Consider the probability $J_{k,I}(n_1,\ldots,n_k)$ of finding no particle in an interval $I\subset \mathbb{Z}$ except for $k$ designated point.
 (See (2) in Fig.\ref{fig:Janossy}.) 
$J_{k,I}(n_1,\ldots,n_k)$ is 
called Janossy density \cite{Macchi}, which is
given by the restricted kernel $\mathbf{K}_I=\left[K(n,m)\right]_{n,m\in I}$ on $I$ for the determinantal point process.
\begin{align}
J_{k,I}(n_1,\ldots,n_k)&=\det(\mathbb{I}-\mathbf{K}_I)\cdot \det\left[
\langle n_i|\mathbf{K}_I(\mathbb{I}-\mathbf{K}_I)^{-1}|n_j\rangle
\right]_{i,j=1}^k
\nonumber \\
&=(-1)^k\det
\left|
\begin{array}{cc}
-\left[\langle n_i|\mathbf{K}_I|n_j\rangle\right]_{i,j=1,\ldots,k} &
-\left[\langle n|\mathbf{K}_I|n_j\rangle\right]_{j=1,\ldots,k;n\in I}\\
-\left[\langle n_i|\mathbf{K}_I|m\rangle\right]_{i=1,\ldots,k; m\in I}&
\left[\langle n|(\mathbb{I}-\mathbf{K}_I)|m\rangle\right]_{n,m\in I} 
\end{array}
\right|.
\label{Janossy_prob0}
\end{align}
Here we denote the restricted kernel by $K_I(n,m)=\langle n|\mathbf{K}_I|m\rangle$ 
with the orthonormal complete basis $\{|n\rangle\,|\, n\in I\}$
and its dual $\{\langle n|\,|\, n\in I\}$.
The first line of (\ref{Janossy_prob0}) is quoted e.g.~from \cite{BO} ($\pi(X)$ on page 341),
and the second line is by the identity
$\det D \cdot \det\left(A-CD^{-1}B\right)=\det\left|\begin{array}{cc} A & B \\ C & D\end{array}\right|$.

Generalization to the probability $J_{p,k,I}(n_1,\ldots,n_k)$ of finding exactly $p$ particles in $I$ except for $k$ designated points is straightforward (see (3) in Fig.\ref{fig:Janossy}).
Just as in the case of the ordinary gap probability ($k=0$),
we merely introduce the spectral parameter $z$ so that
$J_{p,k,I}(n_1,\ldots,n_k)$ is given by
\begin{align}
J_{p,k,I}(n_1,\ldots,n_k)=\frac{1}{p!}\left(-\partial_{z}\right)^p
\det(\mathbb{I}-z\mathbf{K}_I)\cdot \det\left[
\langle n_i|\mathbf{K}_I(\mathbb{I}-z\mathbf{K}_I)^{-1}|n_j\rangle
\right]_{i,j=1}^k\bigg|_{z=1}.
\label{Janossy_prob_p}
\end{align}

For the continuous determinantal random point process on $X\subset \mathbb{R}$ with the measure $\mu$, 
the Janossy density $J_{k,I}(x_1,\ldots, x_k)\mu(dx_1)\cdots \mu(dx_k)$ for the distribution of the particles in a subset $I\subset X$
is defined as the probability density of finding exactly $k$ particles in $I$ and one at each of the $k$ infinitesimal intervals $(x_i,x_i+dx_i)\subset I$. 
$J_{k,I}(x_1,\ldots, x_k)$ is given by the Fredholm determinant $\det(\mathbb{I}-\mathbf{K}_I)$
and the determinant of $\mathbf{L}_I:=\mathbf{K}_I(\mathbb{I}-\mathbf{K}_I)^{-1}$ such that 
\begin{align}
J_{k,I}(x_1,\ldots, x_k)
=\det(\mathbb{I}-\mathbf{K}_I)\cdot \det\left[\mathbf{L}_I(x_i,x_j)
\right]_{i,j=1}^k.
\label{Janossy_density_cont}
\end{align}

\subsection{Massive chiral Gaussian ensemble with $N_F=\beta n$ fermions and the Janossy density}\label{app:Janossy2}

Consider a block diagonal Hermitian matrix $H$ of Dyson index $\beta=1,2,4$:
\begin{align}
H=
\left(
\begin{array}{cc}
0 & W \\
W^{\dagger} & 0
\end{array}
\right),\quad W\in F^{N\times (N+\nu)},\ \ 
F=\mathbb{R}, \mathbb{C}, \mathbb{H}.
\end{align}
The partition function $Z_{N,\beta,\nu}(\{m_a\})$ of the massive chiral Gaussian ensemble with $N_F=\beta n$ fermions
is given by
\begin{align}
Z_{N,\beta,\nu}(\{m_a\})&=\int dH\,\mathrm{e}^{-\beta\mathrm{tr} H^2}\prod_{a=1}^{n}(H+\mathrm{i}m_a)^{\beta}
\nonumber \\
&=\int_{0}^{\infty}\prod_{i=1}^N
\left(
dx_i \,x_i^{\frac{\beta(\nu+1)}{2}-1}\mathrm{e}^{-\beta x_i}\prod_{a=1}^{n}\left|x_i+m_a^2\right|^{\beta}
\right)\prod_{i>j}^N\left| x_i-x_j\right|^{\beta}.
\end{align} 
It can further be rewritten as an $N+n$ eigenvalue integral in the following form (up to $m$-dependent prefactor $C_{N,\beta,\nu}(\{m_a\})$):
\begin{align}
Z_{N,\beta,\nu}(\{m_a\})=\frac{1}{C_{N,\beta,\nu}(\{m_a\})}
\int_{-\infty}^{\infty}
&\prod_{i=1}^{N+n}
\left(
dx_i \,x_i^{\frac{\beta(\nu+1)}{2}-1}\mathrm{e}^{-\beta x_i}
\right)
\prod_{i>j}^{N+n}\left| x_i-x_j\right|^{\beta}
\nonumber \\
&\cdot\prod_{\ell=1}^N\theta(x_\ell)
\prod_{k=N+1}^{N+n}\delta\left(x_k-(-m_{k-N}^2)\right),
\label{partition_function_Janossy}
\end{align}
where $\theta(x)$ stands for the Heaviside function.

\begin{figure}[h] 
\begin{center}
\hspace*{0cm}
 \includegraphics[bb=0 0 360 223,width=100mm]{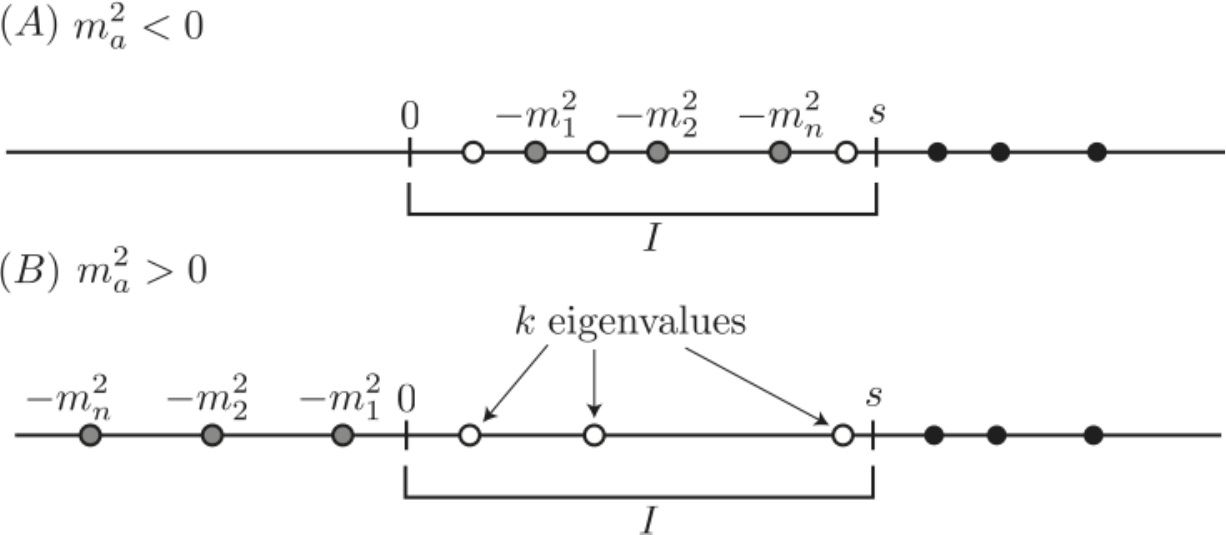}
\caption{\label{fig:Janossy2}
The determinantal random point process for $x_i$ $i=1,\ldots,N+n$ with designated points at $-m_{a}^2$ ($a=1,\ldots,n$). 
(A) For $m_a^2<0$: the Janossy density $J_{k,n,[0,s]}(\{-m_a^2\})$.  
(B) For $m_a^2>0$: the probability distribution $E(k;[0,s];\{m_a\})$ as an analytic continuation of the Janossy density. }
\end{center}
\end{figure}

This partition function is regarded as that of the determinantal random point process for $x_i$ ($i=1,\ldots,N+n$) with designated points at $-m_{a}^2$ ($a=1,\ldots,n$).
In the case of $m_{a}^2<0$, 
the Janossy density $J_{k,n,[0,s]}(-m_1^2,\ldots, -m_n^2)$ on the interval $I=[0,s]$ ($s>0$) 
for the above massive chiral Gaussian ensemble 
is found by adopting the spectral kernel $K(z_i,z_j)$ \cite{Mehta,Nagao,Forrester_book,Nishigaki} to eq.~(\ref{Janossy_density_cont}). 
(See (A) in Fig.\ref{fig:Janossy2}.)

Applying an analytic continuation with respect to the mass parameters $m_a$'s, one finds the joint probability $E(k;[0,s];\{m_a\})$ 
in eq.~(\ref{eq:E_k}) as the Janossy density $J_{k,n,[0,s]}(\{-m_a^2\})$ with $m_{a}^2>0$. (See (B) in Fig.\ref{fig:Janossy2}.)

\section{Probability distribution of the $k^{\text{\tiny th}}$ smallest eigenvalue}\label{app:trace}
The probability $E_k(s)=E(k;[0,s];\{-m_{a}^2\})$ 
of finding exactly $k$ eigenvalues in the interval $[0,s]$ is given by the $k^{\text{\tiny th}}$ derivative of the Fredholm determinant $\tau(z;[0,s];\{-m_{a}^2\})$ by the parameter $z$ such that
\begin{align}
E_k(s)=\frac{1}{k!}(-\partial_{z})^k\tau(z;[0,s];\{-y_a\})\Big|_{z=1}.
\end{align}
The Fredholm determinant and Pfaffian in eqs.~(\ref{eq:det_formula}) and (\ref{eq:Pfaffian_formula}) are represented by 
\begin{align}
\tau(z;[0,s];\{-m_{a}^2\})=\det \left|
\begin{array}{cc}
-\kappa &-\sqrt{z}\bm{k}^{\mathrm{T}} \\
 -\sqrt{z}\bm{k} &
\hat{\mathbb{I}}-z\bm{K}
\end{array}
\right|\Bigg/\det(-\kappa),
\label{fredholm_chGSE}
\end{align}
where det stands for determinant and quaternionic determinant for unitary and symplectic ensembles, respectively.
The Taylor expansion of $\tau(z;[0,s];\{-m_{a}^2\})$ in eq.~(\ref{fredholm_chGSE}) around $z=1$ is found as combinations of the functional traces $T_n$'s  as follows. (The same expansions for the quenched ($\alpha=0$)
ensembles are given in \cite{Nishigaki:2016nka}.)
\begin{align}
&E_0(s)=\tau(z=1;[0,s];\{-y_a\}),\quad E_1(s)=E_0(s)\bar{T}_1,\quad E_2(s)=\frac{E_0(s)}{2!}\left(\bar{T}_1^2-\bar{T}_2\right),
\nonumber\\
&E_3(s)=\frac{E_0(s)}{3!}\left(\bar{T}_1^3-3\bar{T}_1\bar{T}_2+\bar{T}_3\right),\quad
E_4(s)=\frac{E_0(s)}{4!}\left(\bar{T}_1^4-6\bar{T}_1^2\bar{T}_2+3\bar{T}_2^2+4\bar{T}_1\bar{T}_3-\bar{T}_4\right),
\nonumber \\
&E_5(s)=\frac{E_0(s)}{5!}\left(\bar{T}_1^5-10\bar{T}_1^3\bar{T}_2+10\bar{T}_1^2\bar{T}_3+15\bar{T}_1\bar{T}_2^2-5\bar{T}_1\bar{T}_4-10\bar{T}_2\bar{T}_3+\bar{T}_5\right),
\nonumber \\
&E_6(s)=\frac{E_0(s)}{6!}\bigl(
\bar{T}_1^6-15\bar{T}_1^4\bar{T}_2+20\bar{T}_1^3\bar{T}_3+45\bar{T}_1^2\bar{T}_2^2-15\bar{T}_1^2\bar{T}_4-60\bar{T}_1\bar{T}_2\bar{T}_3-15\bar{T}_2^3
\nonumber \\
&\qquad\qquad\qquad\quad
+6\bar{T}_1\bar{T}_5+15\bar{T}_2\bar{T}_4+10\bar{T}_2^3-\bar{T}_6
\bigr),
\nonumber \\
&E_7(s)=\frac{E_0(s)}{7!}\bigl(\bar{T}_1^7-21\bar{T}_1^5\bar{T}_2+35\bar{T}_1^4\bar{T}_3+105\bar{T}_1^3\bar{T}_1^2-35\bar{T}_1^3\bar{T}_4-210\bar{T}_1^2\bar{T}_2\bar{T}_3-105\bar{T}_1\bar{T}_2^3
\nonumber \\
&\qquad\qquad\qquad\quad
+21\bar{T}_1^2\bar{T}_5+105\bar{T}_1\bar{T}_2\bar{T}_4+70\bar{T}_1\bar{T}_3^2+105\bar{T}_2^2\bar{T}_3-7\bar{T}_1\bar{T}_6-21\bar{T}_2\bar{T}_5
\nonumber \\
&\qquad\qquad\qquad\quad
-35\bar{T}_3\bar{T}_4+\bar{T}_7
\bigr).
\end{align}
The functional traces consist of operators $\mathcal{K}^{(n)}$'s are given by
\begin{align}
&\mathcal{K}(z)=\left(\begin{array}{cc}
-\kappa &-\sqrt{z}\bm{k}^{\mathrm{T}} \\
 -\sqrt{z}\bm{k} &
\mathbb{I}-z\bm{K} 
\end{array}
\right),\quad 
\mathcal{K}^{(0)}=
\mathcal{K}(z=1)=
\left(\begin{array}{cc}
-\kappa &-\bm{k}^{\mathrm{T}} \\
-\bm{k} &
\mathbb{I}-\bm{K} 
\end{array}
\right),
\nonumber \\
&\mathcal{K}^{(1)}=-\frac{\partial}{\partial z}
\mathcal{K}(z)\Big|_{z=1}
=\left(\begin{array}{cc}
0 & \frac{1}{2}\bm{k}^{\mathrm{T}} \\
\frac{1}{2}\bm{k}& \bm{K}
\end{array}
\right),
\nonumber \\
&\mathcal{K}^{(n)}=\frac{\partial^n}{\partial z^n}\mathcal{K}(z)
\bigg|_{z=1}
=(-1)^{n-1}\frac{(2n-3)!}{2^n}\left(\begin{array}{cc}
0 & \frac{1}{2}\bm{k}^{\mathrm{T}} \\
\frac{1}{2}\bm{k} & 0
\end{array}
\right),\quad (n\ge 2).
\end{align}

\newpage
Using these operators, one can show $\bar{T}_n$ ($k=1,\ldots,7$) as follows.
\begin{align}
\bar{T}_1&=\frac{1}{2}\mathrm{tr}\left[\mathcal{K}^{(1)}\cdot\mathcal{K}^{(0)-1}\right],
\quad 
\bar{T}_2=\frac{1}{2}\mathrm{tr}\left[\left(\mathcal{K}^{(1)}\cdot\mathcal{K}^{(0)-1}\right)^2\right]
+\frac{1}{2}\mathrm{tr}\left[\mathcal{K}^{(2)}\cdot\mathcal{K}^{(0)-1}\right],
\nonumber \\
\bar{T}_3&=2!\cdot\frac{1}{2}\mathrm{tr}\left[\left(\mathcal{K}^{(1)}\cdot\mathcal{K}^{(0)-1}\right)^3\right]+3\cdot\frac{1}{2}\mathrm{tr}\left[\mathcal{K}^{(1)}\cdot\mathcal{K}^{(0)-1}\cdot\mathcal{K}^{(2)}\cdot\mathcal{K}^{(0)-1}\right]
+\frac{1}{2}\mathrm{tr}\left[\mathcal{K}^{(3)}\cdot\mathcal{K}^{(0)-1}\right],
\nonumber \\
\bar{T}_4&=3!\cdot\frac{1}{2}\mathrm{tr}\left[\left(\mathcal{K}^{(1)}\cdot\mathcal{K}^{(0)-1}\right)^4\right]
+12\cdot\frac{1}{2}\mathrm{tr}\left[\left(\mathcal{K}^{(1)}\cdot\mathcal{K}^{(0)-1}\right)^2\cdot\mathcal{K}^{(2)}\cdot\mathcal{K}^{(0)-1}\right]
\nonumber \\
&\quad
+3\cdot\frac{1}{2}\mathrm{tr}\left[\left(\mathcal{K}^{(2)}\cdot\mathcal{K}^{(0)-1}\right)^2\right]
+4\cdot\frac{1}{2}\mathrm{tr}\left[\mathcal{K}^{(1)}\cdot\mathcal{K}^{(0)-1}\cdot\mathcal{K}^{(3)}\cdot\mathcal{K}^{(0)-1}\right]
+\frac{1}{2}\mathrm{tr}\left[\mathcal{K}^{(4)}\cdot\mathcal{K}^{(0)-1}\right],
\nonumber 
\\
\bar{T}_5&=4!\cdot\frac{1}{2}\mathrm{tr}\left[\left(\mathcal{K}^{(1)}\cdot\mathcal{K}^{(0)-1}\right)^5\right]
+60\cdot\frac{1}{2}\mathrm{tr}\left[\left(\mathcal{K}^{(1)}\cdot\mathcal{K}^{(0)-1}\right)^3\cdot\mathcal{K}^{(2)}\cdot\mathcal{K}^{(0)-1}\right]
\nonumber \\
&\quad +20\cdot\frac{1}{2}\mathrm{tr}\left[\left(\mathcal{K}^{(1)}\cdot\mathcal{K}^{(0)-1}\right)^2\cdot\mathcal{K}^{(3)}\cdot\mathcal{K}^{(0)-1}\right]
+30\cdot\frac{1}{2}\mathrm{tr}\left[\mathcal{K}^{(1)}\cdot\mathcal{K}^{(0)-1}\cdot\left(\mathcal{K}^{(2)}\cdot\mathcal{K}^{(0)-1}\right)^2\right]
\nonumber \\
&\quad +5\cdot\frac{1}{2}\mathrm{tr}\left[\mathcal{K}^{(1)}\cdot\mathcal{K}^{(0)-1}\cdot\mathcal{K}^{(4)}\cdot\mathcal{K}^{(0)-1}\right]
+10\cdot\frac{1}{2}\mathrm{tr}\left[\mathcal{K}^{(2)}\cdot\mathcal{K}^{(0)-1}\cdot\mathcal{K}^{(3)}\cdot\mathcal{K}^{(0)-1}\right]
\nonumber \\
&\quad
+\frac{1}{2}\mathrm{tr}\left[\mathcal{K}^{(5)}\cdot\mathcal{K}^{(0)-1}\right],
\nonumber 
\\
\bar{T}_6&=5!\cdot\frac{1}{2}\mathrm{tr}\left[\left(\mathcal{K}^{(1)}\cdot\mathcal{K}^{(0)-1}\right)^6\right]
+360\cdot\frac{1}{2}\mathrm{tr}\left[\left(\mathcal{K}^{(1)}\cdot\mathcal{K}^{(0)-1}\right)^4\cdot\mathcal{K}^{(2)}\cdot\mathcal{K}^{(0)-1}\right]
\nonumber \\
&\quad +120\cdot\frac{1}{2}\mathrm{tr}\left[\left(\mathcal{K}^{(1)}\cdot\mathcal{K}^{(0)-1}\right)^3\cdot\mathcal{K}^{(3)}\cdot\mathcal{K}^{(0)-1}\right]
\nonumber \\
&\quad
+180\cdot\frac{1}{2}\mathrm{tr}\left[\left(\mathcal{K}^{(1)}\cdot\mathcal{K}^{(0)-1}\right)^2\cdot\left(\mathcal{K}^{(2)}\cdot\mathcal{K}^{(0)-1}\right)^2\right]
\nonumber \\
&\quad
+90\cdot\frac{1}{2}\mathrm{tr}\left[\mathcal{K}^{(1)}\cdot\mathcal{K}^{(0)-1}\cdot\mathcal{K}^{(2)}\cdot\mathcal{K}^{(0)-1}\cdot\mathcal{K}^{(1)}\cdot\mathcal{K}^{(0)-1}\cdot\mathcal{K}^{(2)}\cdot\mathcal{K}^{(0)-1}\right]
\nonumber \\
 &\quad
 +30\cdot\frac{1}{2}\mathrm{tr}\left[\left(\mathcal{K}^{(1)}\cdot\mathcal{K}^{(0)-1}\right)^2\cdot\mathcal{K}^{(4)}\cdot\mathcal{K}^{(0)-1}\right]
\nonumber \\
 &\quad
 +120\cdot\frac{1}{2}\mathrm{tr}\left[\mathcal{K}^{(1)}\cdot\mathcal{K}^{(0)-1}\cdot\mathcal{K}^{(2)}\cdot\mathcal{K}^{(0)-1}\cdot\mathcal{K}^{(3)}\cdot\mathcal{K}^{(0)-1}\right]
\nonumber \\
 &\quad
 +30\cdot\frac{1}{2}\mathrm{tr}\left[\left(\mathcal{K}^{(2)}\cdot\mathcal{K}^{(0)-1}\right)^3\right]
\nonumber \\
 &\quad
 +6\cdot\frac{1}{2}\mathrm{tr}\left[\mathcal{K}^{(1)}\cdot\mathcal{K}^{(0)-1}\cdot\mathcal{K}^{(5)}\cdot\mathcal{K}^{(0)-1}\right]
+15\cdot\frac{1}{2}\mathrm{tr}\left[\mathcal{K}^{(2)}\cdot\mathcal{K}^{(0)-1}\cdot\mathcal{K}^{(4)}\cdot\mathcal{K}^{(0)-1}\right]
\nonumber \\
 &\quad
 +10\cdot\frac{1}{2}\mathrm{tr}\left[\left(\mathcal{K}^{(3)}\cdot\mathcal{K}^{(0)-1}\right)^2\right]
+\frac{1}{2}\mathrm{tr}\left[\mathcal{K}^{(6)}\cdot\mathcal{K}^{(0)-1}\right],
\nonumber 
\end{align}
\newpage
\begin{align}
\bar{T}_7&=6!\cdot\frac{1}{2}\mathrm{tr}\left[\left(\mathcal{K}^{(1)}\cdot\mathcal{K}^{(0)-1}\right)^7\right]
+2520\cdot\frac{1}{2}\mathrm{tr}\left[\left(\mathcal{K}^{(1)}\cdot\mathcal{K}^{(0)-1}\right)^5\cdot\mathcal{K}^{(2)}\cdot\mathcal{K}^{(0)-1}\right]
\nonumber \\
 &\quad
 +840\cdot\frac{1}{2}\mathrm{tr}\left[\left(\mathcal{K}^{(1)}\cdot\mathcal{K}^{(0)-1}\right)^4\cdot\mathcal{K}^{(3)}\cdot\mathcal{K}^{(0)-1}\right]
\nonumber \\
&\quad
+1260\cdot\frac{1}{2}\mathrm{tr}\left[\left(\mathcal{K}^{(1)}\cdot\mathcal{K}^{(0)-1}\right)^3\cdot\left(\mathcal{K}^{(2)}\cdot\mathcal{K}^{(0)-1}\right)^2\right]
\nonumber \\
&\quad
+1260\cdot\frac{1}{2}\mathrm{tr}\left[\left(\mathcal{K}^{(1)}\cdot\mathcal{K}^{(0)-1}\right)^2\cdot\mathcal{K}^{(2)}\cdot\mathcal{K}^{(0)-1}\cdot\mathcal{K}^{(1)}\cdot\mathcal{K}^{(0)-1}\cdot\mathcal{K}^{(2)}\cdot\mathcal{K}^{(0)-1}\right]
\nonumber \\
 &\quad
 +210\cdot\frac{1}{2}\mathrm{tr}\left[\left(\mathcal{K}^{(1)}\cdot\mathcal{K}^{(0)-1}\right)^3\cdot\mathcal{K}^{(4)}\cdot\mathcal{K}^{(0)-1}\right]
\nonumber \\
 &\quad
 +840\cdot\frac{1}{2}\mathrm{tr}\left[\left(\mathcal{K}^{(1)}\cdot\mathcal{K}^{(0)-1}\right)^2\cdot\mathcal{K}^{(2)}\cdot\mathcal{K}^{(0)-1}\cdot\mathcal{K}^{(3)}\cdot\mathcal{K}^{(0)-1}\right]
\nonumber \\
 &\quad
 +420\cdot\frac{1}{2}\mathrm{tr}\left[\mathcal{K}^{(1)}\cdot\mathcal{K}^{(0)-1}\cdot\mathcal{K}^{(2)}\cdot\mathcal{K}^{(0)-1}\cdot\mathcal{K}^{(1)}\cdot\mathcal{K}^{(0)-1}\cdot\mathcal{K}^{(3)}\cdot\mathcal{K}^{(0)-1}\right]
\nonumber \\
 &\quad
 +630\cdot\frac{1}{2}\mathrm{tr}\left[\mathcal{K}^{(1)}\cdot\mathcal{K}^{(0)-1}\cdot\left(\mathcal{K}^{(2)}\cdot\mathcal{K}^{(0)-1}\right)^3\right]
\nonumber \\
&\quad
+42\cdot\frac{1}{2}\mathrm{tr}\left[\left(\mathcal{K}^{(1)}\cdot\mathcal{K}^{(0)-1}\right)^2\cdot\mathcal{K}^{(5)}\cdot\mathcal{K}^{(0)-1}\right]
\nonumber \\
 &\quad
 +210\cdot\frac{1}{2}\mathrm{tr}\left[\mathcal{K}^{(1)}\cdot\mathcal{K}^{(0)-1}\cdot\mathcal{K}^{(2)}\cdot\mathcal{K}^{(0)-1}\mathcal{K}^{(4)}\cdot\mathcal{K}^{(0)-1}\right]
\nonumber \\
 &\quad
 +140\cdot\frac{1}{2}\mathrm{tr}\left[\mathcal{K}^{(1)}\cdot\mathcal{K}^{(0)-1}\cdot\left(\mathcal{K}^{(3)}\cdot\mathcal{K}^{(0)-1}\right)^2\right]
\nonumber\\
 &\quad
 +210\cdot\frac{1}{2}\mathrm{tr}\left[\left(\mathcal{K}^{(2)}\cdot\mathcal{K}^{(0)-1}\right)^2\cdot\mathcal{K}^{(3)}\cdot\mathcal{K}^{(0)-1}\right]
\nonumber \\
 &\quad
 +7\cdot\frac{1}{2}\mathrm{tr}\left[\mathcal{K}^{(1)}\cdot\mathcal{K}^{(0)-1}\cdot\mathcal{K}^{(6)}\cdot\mathcal{K}^{(0)-1}\right]
\nonumber \\
 &\quad
 +21\cdot\frac{1}{2}\mathrm{tr}\left[\mathcal{K}^{(2)}\cdot\mathcal{K}^{(0)-1}\cdot\mathcal{K}^{(5)}\cdot\mathcal{K}^{(0)-1}\right]
\nonumber \\
&\quad+\frac{1}{2}\mathrm{tr}\left[\mathcal{K}^{(7)}\cdot\mathcal{K}^{(0)-1}\right].
\end{align}

The above expansion of the functional trace can also be considered as follows.
Rewriting the Fredholm determinant and Pfaffian given in eq. (\ref{fredholm_chGSE}) 
into the following form:\footnote{
The authors thank the anonymous referee for pointing out such expansion.}
\begin{align}
\det\left|
\begin{array}{cc}
-\kappa & -\sqrt{z}\bm{k}^{\mathrm{T}} \\
-\sqrt{z}\bm{k} & \hat{\mathbb{I}}-z\bm{K}
\end{array}
\right|/\det(-\kappa)
=\det\left(
\hat{\mathbb{I}}-z(\bm{K}-\bm{k}\kappa^{-1}\bm{k}^{\mathrm{T}})
\right),
\label{rewriting_det}
\nonumber
\end{align}
then one finds that $E_k(s)$'s in eq.(\ref{fredholm_chGSE}) are represented as 
the quenched model. 
Using the representation eq.(2.6) in \cite{Nishigaki:2016nka},
we obtain a little different expansion with the functional traces of the resolvents 
$T_n(s)=\mathrm{tr}\left(\tilde{\bm{K}}(\mathbb{I}-\tilde{\bm{K}})^{-1}\right)^n$ for the Fredholm determinant
and 
$T_n(s)=\mathrm{tr}\left(\tilde{\bm{K}}(\mathbb{I}-\tilde{\bm{K}})^{-1}\right)^n/2$ for the Fredholm Pfaffian with $\tilde{\bm{K}}=\bm{K}-\bm{k}\kappa^{-1}\bm{k}^{\mathrm{T}}$.

\section{Gauss-Legendre quadrature rule}\label{app:nystrom}
The quadrature rule is an efficient method to perform the numerical evaluation for the integral of the smooth function.
The quadrature formula for the integral over the interval is represented as \cite{wolfram_GL}
\begin{align}
Q_I(f)=\sum_{i=1}^mw_if(x_i)\approx \int_If(x)dx,
\end{align}
where $w_i$ and $x_i$ denote the weight and nodes, respectively,  determined by the prescription of the quadrature rule.
There are several kinds of quadrature rules. 
The most basic method is the Gauss-Legendre rule and more efficient one is the Clenshaw-Curtis rule.
In the following, we will summarize the Gauss-Legendre rule.

Let $I=[-1,1]$ and $M\in\mathbb{N}$.\\
\begin{enumerate}
\item The node $x_i^{[-1,1]}$ is given by the $i^{\rm th}$ zero of the Legendre polynomial $P_{M}(x)$.\\
\item The weight $w_i^{[-1,1]}$ is given by
\begin{align}
w_i^{[-1,1]}=\frac{2}{(1-x_i^2)^2P'_M(x_i)^2}.
\end{align}
\end{enumerate}
For some lower orders $M$, nodes and weights are listed in the following table \cite{wolfram_GL}.
\begin{table}[htb]
  \caption{Nodes and weights of the Gauss-Legendre rule}
  \begin{center}
  \begin{tabular}{|c||c|c|} \hline
    $M$ & $x_i^{[-1,1]}$ & $w_i^{[-1,1]}$ \\ \hline 
    1 & 0 & 2 \\ \hline
     2 & $\pm\sqrt{1/3}$ & 1 \\ \hline
    3 & 0 & 8/9 \\ \cline{2-3}
      & $\pm\sqrt{3/5}$ & 5/9 \\ \hline
    4  & $\pm\sqrt{\left(3-2\sqrt{6/5}\right)/7}$ & $\frac{18+\sqrt{30}}{36}$  \\ \cline{2-3}
    &   $\pm\sqrt{\left(3+2\sqrt{6/5}\right)/7}$  & $\frac{18+\sqrt{30}}{36}$  \\ \hline
        5  & 0 & 128/225  \\ \cline{2-3}
    &   $\pm\frac{1}{3}\sqrt{5-2\sqrt{10/7}}$  & $\frac{322+13\sqrt{70}}{900}$  \\ \cline{2-3}
          &  $\pm\frac{1}{3}\sqrt{5+2\sqrt{10/7}}$ & $\frac{322-13\sqrt{70}}{900}$  \\ \hline
  \end{tabular}
  \end{center}
\end{table}

The following proposition holds for the Gauss-Legendre quadrature rule.
\begin{prop}
The Gauss-Legendre quadrature rule of order $M$ is exact,
if $f(x)$ is an $(2M-1)^{\rm th}$ order (or lesser) polynomial of $x$.
\end{prop}

By a simple change of variable, one finds the quadrature formula for the interval $I=[a,b]$.
\begin{align}
\int_a^bdx\,f(x)=\frac{b-a}{2}\int_{-1}^1f\left(\frac{b-a}{2}x+\frac{a+b}{2}\right)
\approx\frac{b-a}{2}\sum_{i=1}^Mw_i^{[-1,1]}f\left(\frac{b-a}{2}x_i^{[-1,1]}+\frac{a+b}{2}\right).
\end{align}
In particular for $I=[0,s]$, the quadrature formula reduces to
\begin{align}
\int_0^sdx\,f(x)\approx\sum_{i=1}^M\frac{sw_i}{2}f\left(\frac{s}{2}(x_i+1)\right).
\end{align}
In particular for the numerical evaluation of the Fredholm determinant on $\tau(z;[0,s];\{\mu_a\})$, 
the nodes and weights for $I=[0,s]$ are chosen as
\begin{align}
\zeta_i=\frac{s}{2}(x_i^{[-1,1]}+1),\quad w_i=\frac{sw_i^{[-1,1]}}{2}.
\label{eq:node_weight}
\end{align}

\section{Details of the lattice result}
\label{app:lattice_data}
In Table \ref{tab:lattice_result}, we list the result of the fitting of
lattice data.

\begin{table}
\center
 \begin{tabular}{cc|rrrr}
  \multicolumn{1}{c}{lattice size} &
  \multicolumn{1}{c|}{$\beta$} &
  \multicolumn{1}{c}{$\hat{\Sigma}$} &
  \multicolumn{1}{c}{$\mu$} &
  \multicolumn{1}{c}{$\chi^2/\mathrm{d.o.f}$} &
  \multicolumn{1}{c}{num($\nu=0$)} \\
\hline
  $8^4$
   & 1.100 & 0.2176(27) & 8.91(11) &  0.30(17) & 230  \\
   & 1.200 & 0.1997(24) & 8.18(10) &  0.25(17) & 260  \\
   & 1.300 & 0.1651(14) & 6.76(05) &  0.45(20) & 314  \\
   & 1.350 & 0.1378(12) & 5.65(05) &  0.27(16) & 467  \\
   & 1.375 & 0.1238(13) & 5.07(05) &  0.52(20) & 407  \\
   & 1.400 & 0.0781(11) & 3.20(04) & 11.00(93) & 843  \\
   & 1.425 & 0.0215(01) & 0.880(4) &  2.85(47) & 2338 \\
\hline
  $12^4$
    & 1.100 & 0.1903(23) & 39.46(48) & 0.66(25) & 399 \\
    & 1.300 & 0.1425(44) & 29.54(91) & 0.20(15) & 63 \\
    & 1.350 & 0.1263(23) & 26.19(49) & 0.37(20) & 38 \\
    & 1.375 & 0.1156(39) & 23.98(81) & 0.13(13) & 45 \\
    & 1.400 & 0.0831(14) & 17.23(29) & 0.50(23) & 106 \\
    & 1.425 & 0.0598(10) & 12.41(21) & 0.37(20) & 206 \\
    & 1.450 & 0.0209(04) & 4.32(08)  & 6.74(73) & 600 \\
\hline
  $16^4$
    & 1.350 & 0.1252(20) & 82.0(1.3) & 0.34(16) & 105 \\
    & 1.375 & 0.1064(34) & 69.8(2.3) & 0.22(16) & 41 \\
    & 1.400 & 0.0799(12) & 52.35(80) & 0.37(21) & 155 \\
    & 1.425 & 0.0521(05) & 34.13(33) & 0.48(20) & 369 \\
    & 1.450 & 0.0246(02) & 16.14(17) & 0.83(26) & 561 \\
    & 1.475 & 0.0083(01) & 5.47(12)  & 1.32(30) & 248 \\
 \end{tabular}

 \caption{Fit result of $\hat{\Sigma}$, chiral condensate in the lattice
 unit, together with the corresponding value of $\mu$.
 The bare coupling constant is given  through $\beta=4/g^2$.
 The most right column is the number of
 configurations we used in each of the fitting, which belong to
 the topological charge $\nu=0$ sector.
 The reduced chi squared, $\chi^2/\mathrm{d.o.f}$, indicates the quality
 of the fitting.
 }
 \label{tab:lattice_result}
 
\end{table}

\section{Estimation of the correlation matrix}
\label{app:correlation_matrix}

An element of the correlation matrix is given 
\begin{align}
 C_{ij} = 
  \langle (y_i-\langle y_i\rangle )(y_j-\langle y_j\rangle) \rangle,
\end{align}
where $y_i=I(\hat{s}_i)$ with 
 $I(\hat{s}_i) = \int_0^{\hat{s}} d\hat{\lambda}_1 p_1^{\mathrm{latt.}}(\hat{\lambda}_1; \hat{m}_f)$ defined in eq.~(\ref{eq:integrated-dist})
and $\hat{s}_i$ is the upper end of
the $i$-th bin.
The bracket $\langle \cdot \rangle$
represents the average over lattice configurations which belong to $\nu=0$
sector.
Since the correlation matrix is an average of fluctuation,
one needs to use a resampling method like jackknife
or bootstrapping to estimate.  In this analysis, we use the jackknife method.

What we need in the fitting is not the correlation matrix
itself but its inverse.
As the estimate of $C$ contains some error, 
we need some care to invert it.
If the bin width is too fine, neighboring bins may give (almost) the 
same value which causes zero-mode (or almost zero-mode)
of the correlation matrix.
If eigenvalue of $C$ is too small, the relative error of the eigenvalue becomes
large, which makes estimation of $C^{-1}$ unreliable.
Note that the smallest eigenmode gives the largest contribution to the inverse.

We therefore employ the following steps.
First of all, some of the bins do not have eigenvalues of the Dirac
operator in it (the
largest several bins and sometimes the first bin(s)).
Let us suppose that $i$-th bin has no eigenvalue.  Then, $i$-th column/row
of the  correlation matrix, $C_{ij}$ and $C_{ji}$ for arbitrary $j$
becomes zero as $y_i$ is always 1 (or always 0).
This obviously reduces the rank of $C$.
We therefore replace the diagonal element $C_{ii}=0$ with the upper
bound of the
estimate, $1/n^3$, where
$n$ is number of independent configurations we
use\footnote{
This value is estimated by assuming
that 1 configuration has 1 eigenvalue in the bin,
and other $n-1$ configurations do not have any.
We also assume that $n$ is large enough.}.
The off-diagonal elements are kept zero.
After this modification of the correlation matrix, which is now denoted
as $C'$,
we still may have very small eigenvalues.
Numerically, we even may observe (small) negative eigenvalue of $C'$%
\footnote{
The correlation matrix must be positive semi-definite, but
with finite statistics and numerical precision, we may observe
negative eigenvalue.}.
We therefore truncate the correlation matrix by cutting
small eigenmodes in inverting the matrix to give an improved estimate of the inverse of the correlation matrix $C_{\mathrm{imp.}}^{-1}$.
The cutoff $c_{\mathrm{cut}}$ we use is
0.1 times smallest diagonal element, $c_{\mathrm{cut}} = 0.1/n^3$.
That is,
\begin{align}
 C'|i\rangle &= c_i |i\rangle,
&
 C_{\mathrm{imp.}}^{-1}
 &= \sum_{i \ \mathrm{s.t.}\ c_i > c_{\mathrm{cut}}} |i\rangle \frac{1}{c_i}\langle i|.
\end{align}

 \section{Hybrid Monte Carlo (HMC) for RMT}

A hybrid Monte Carlo simulation technique \cite{Duane:1987de}
is applicable to finite $N$ random matrix theory.

By introducing $\zeta_i=\sqrt{8N x_i}$ and $\mu_a = \sqrt{8N}\,m_a$ as
eq.~(\ref{eq:asymptotic_limit}),
the partition function (\ref{eq:partition_function}) becomes
\begin{equation}
 Z= C \int_0^\infty \cdots\int_0^\infty
  \prod_{i=1}^{N} d\zeta_i
  \, e^{-S},
\end{equation}
where $C$ represents irrelevant normalization factor and the action is
\begin{equation}
 S=\sum_{i=1}^N\left(
		     \beta\frac{\zeta_i^2}{8N}
		     - \frac{\beta(\nu+1)-1}{2}\ln(\zeta_i^2)
		     -\sum_{a=1}^{n_f}\ln\left(\zeta_i^2 + \mu_a^2\right)
		     - \beta \sum_{j<i} \ln\left|\zeta_i^2-\zeta_j^2\right|
		    \right).
\end{equation}
The dynamical variables here is the eigenvalue $\zeta_i$.
The Hamiltonian for the HMC is
\begin{equation}
 H= \sum_{i=1}^N\frac{p_i^2}{2} + S,
\end{equation}
where $p_i$ the conjugate momentum to $\zeta_i$.
It is straightforward to write down the equation of motions and apply
the HMC algorithm.  For the molecular dynamical time evolution,
we use a leapfrog integrator.

The only non-trivial part is ordering of the variables.
We assume that $0< \zeta_1 < \zeta_2 < \dots <\zeta_N$.
Since there is a divergence in the potential at $\zeta_i=0$ and
$\zeta_i=\zeta_j$ ($i\neq j$), 
if the initial configuration satisfies this ordering,
a smooth molecular dynamical evolution keeps the configuration satisfy the same
constraint.
Discrete time evolutions, however, can break the constraint
so that we use the so called retry trick.
We check whether the trial configuration satisfies the constraint
before the metropolis test.
If it does not, rerun the molecular dynamics with the same random momentum
but a finer time step, $\delta \tau \to \delta \tau/2$.
If the constraint is still broken after several reductions of the time step
(our limit is 6 times), the trial configuration is rejected.
For $\beta=4$, the frequency of the retry is order 0.01\%  and
we did not encounter rejections for this reason.
As $\beta$ becomes smaller, the effect of the potential barrier becomes
weaker.  In fact, more frequent retries are needed for $\beta=2$,
and some trial configurations are rejected in the end.
Note that $\beta=1$ and $\nu=0$, the potential
barrier at $\zeta_i=0$ disappears.

Here is some parameters we used in $\beta=4$ case.
The trajectory length between Metropolis test is $\tau=1$.
We keep the acceptance ratio rather high, typically 0.96--0.97, to reduce
the frequency of retries.
To avoid the auto correlation, we measure the smallest 10 $\zeta_i$ every 10 trajectories and all $\zeta_i$ every 500 trajectories.
In making the distribution in Figs
\ref{fig:beta44}, \ref{fig:beta44E_all}, \ref{fig:beta42}--\ref{fig:beta42222_confl2},
we check the integrated
auto correlation, which is $2\tau_{\mathrm{int}} \lesssim 1.2$
and used every 2 measurements.

The number of independent configurations used to plot Figs. \ref{fig:beta44}, \ref{fig:beta44E_all}, \ref{fig:beta42}--\ref{fig:beta42222_confl2}
in Sec. \ref{section3} are
tabulated in Table \ref{table:ensemble}.
\begin{table}[htb]
\center
 \begin{tabular}{c|rrr}
  \multicolumn{1}{c|}{$N$}
&  \multicolumn{1}{c}{$N_f=2$} &
   \multicolumn{1}{c}{$N_f=4$} &
   \multicolumn{1}{c}{$N_f=8$} \\
\hline
  250 &  2495000 & 1535950 & \\
 1000 &  245000 & 495000  & 245000 \\
 2000 &  245000 & 245000 & 245000 \\
 4000 &         &         & 145000 \\
 \end{tabular}
 \caption{\label{table:ensemble}Number of independent Monte Carlo configurations used in Sec. \ref{section3}.}
\end{table}

\section{Data of $k^{\mathrm{th}}$ smallest eigenvalue distributions for chiral GSE with $N_F=8$}
Numerical data of $F_k(s;\mu)$ $(k=1,2,3,4)$ for the chiral GSE with $N_F=8$ degenerate flavors,
in the range $0\leq s \leq 20$ and $0\leq \mu\leq 100$ are appended as a Mathematica Notebook
``\texttt{F1234\_chGSE\_NF8.nb}''.

\bibliographystyle{JHEP}
\bibliography{manuscript2019_v3}

\end{document}